\documentclass[10pt]{article}

\usepackage[margin=1in]{geometry}
\usepackage{amsmath,amssymb,amsthm,mathtools}
\usepackage{booktabs}
\usepackage{tabularx}
\usepackage{float}
\usepackage{pgfplots}
\usepgfplotslibrary{groupplots}
\pgfplotsset{compat=1.18}
\usepackage{microtype}
\usepackage[hidelinks]{hyperref}
\usepackage[capitalize,noabbrev]{cleveref}
\usepackage{multicol}
\usepackage{xurl}
\AtBeginDocument{%
  \setlength{\abovedisplayskip}{6pt plus 2pt minus 2pt}%
  \setlength{\belowdisplayskip}{6pt plus 2pt minus 2pt}}

\usepackage{graphicx}
\usepackage{enumitem}
\usepackage{algorithm}
\usepackage{algorithmic}
\usepackage{xcolor}
\usepackage{fancyvrb}
\DefineVerbatimEnvironment{smtcode}{Verbatim}{fontsize=\small,
  frame=leftline, framerule=1.2pt, rulecolor=\color{diagramblue},
  framesep=8pt, xleftmargin=1em}
\usepackage{tikz}
\usetikzlibrary{positioning,fit,arrows.meta,shapes.geometric}
\usetikzlibrary{calc,backgrounds,decorations.pathreplacing}

\definecolor{diagramblue}{HTML}{486A86}
\definecolor{diagramvm}{HTML}{A6533F}
\definecolor{diagramgreen}{HTML}{507765}
\tikzset{
    zkarc diagram/.style={>=Latex, x=1cm, y=1cm,
        every node/.style={font=\small, text=black,
            execute at begin node={\hyphenpenalty=10000\exhyphenpenalty=10000}},
        line width=0.55pt},
    zkarc box/.style={draw=black!65, rounded corners=2pt,
        align=center, inner sep=5pt, minimum height=0.95cm},
    zkarc stage/.style={zkarc box, fill=diagramblue!7},
    zkarc artifact/.style={zkarc box, fill=black!3},
    zkarc note/.style={align=left, font=\footnotesize, inner sep=2pt},
    zkarc heading/.style={anchor=west, font=\small\bfseries, inner sep=0pt},
    zkarc arrow/.style={->, draw=black!80, line width=0.65pt}
}

\theoremstyle{plain}
\newtheorem{theorem}{Theorem}[section]
\newtheorem{lemma}[theorem]{Lemma}
\newtheorem{corollary}[theorem]{Corollary}

\theoremstyle{definition}
\newtheorem{definition}[theorem]{Definition}

\newif\ifoldintro
\oldintrofalse
\newif\ifoldabstract
\oldabstractfalse

\def\thetitle{zkARc: Succinctly Verifiable Proofs for Neurosymbolic Natural Language
Formalization and Verification}
\title{\thetitle}
\hypersetup{
  pdftitle={\thetitle},
  pdfauthor={Wyatt Benno, Alberto Centelles, Aman Goel, Benjamin Kiesl-Reiter},
  pdfsubject={Zero-knowledge receipts for agent policy enforcement},
  pdfkeywords={zero knowledge, verifiable computation, automated reasoning, agent guardrails}
}

\usepackage{authblk}

\author[1]{Wyatt Benno}
\author[{}\protect\reviewchange{3}]{Alberto Centelles\thanks{\protect\reviewchange{Most of this research was done while the author was at ICME Labs.}}}
\author[2]{Aman Goel}
\author[2]{Benjamin Kiesl-Reiter}
\affil[1]{ICME Labs}
\affil[2]{Amazon Web Services}
\affil[3]{\protect\reviewchange{Imperial College London}}
\date{}

\newcommand{\reviewchange}[1]{{\color{black}#1}}
\providecommand{\reviewcolor}{\color{black}}

\begin{document}

\maketitle

\begin{abstract}
\noindent Autonomous agents increasingly execute consequential actions on behalf of users, but ordinary middleware guardrails do not provide cryptographic proof that a policy was actually enforced. Their verdicts leave counterparties dependent on the operator unless they repeat the checks themselves, potentially exposing sensitive policies and inputs. We present zkARc, a modular proof-carrying framework that wraps each stage of Automated Reasoning checks (ARc) in an independent, succinctly verifiable zero-knowledge proof. ARc is a neurosymbolic system that translates natural language policies into formal logic and verifies AI-generated content against them using SMT solvers. zkARc certifies the planner's output, the policy's formalization, the action's formalization and both solver queries. The receipt binds these computations to the exact action received and, under the stated assumptions, reveals no additional information about private policies and intermediate model values beyond the declared public information and verdict. This lets a third party check the claimed policy enforcement without re-running the pipeline or trusting the operator. We prove completeness, knowledge soundness and zero knowledge for the composed receipt under explicit assumptions. An implementation on Qwen2-0.5B enforces a spending policy end to end, approving two purchases and denying a third. A full receipt, which also proves model inference, takes about an hour to issue with CPU-only proving and verifies in under a minute, and the solver proof alone is 88\,KiB and verifies in 0.3 seconds.
\end{abstract}

\section{Introduction}
\label{sec:intro}

When autonomous agents act on behalf of their principals, the involved parties want assurance that the agents obey the rules imposed on them. Consider a consumer who deploys an AI shopping agent with a simple policy: \emph{spend at most \$500 in total, no single purchase over \$200, only approved vendors, never share payment credentials, and never commit to recurring payments.} Whenever the agent performs an action, the consumer and the seller want independently verifiable assurance that the action complies with these rules, without the consumer having to reveal the policy itself.

Today, such assurance rests on reputation systems, audit trails, and contractual liability. As agentic commerce scales, reputation becomes unreliable and audit trails become infeasible to review. What is needed is cryptographic assurance, meaning a proof that the policy was enforced, bound to the specific action, \reviewchange{and verifiable by anyone. Under the stated assumptions, the receipt reveals no additional information about the policy beyond the declared public information and verdict.}

Three classes of prior work each fall short. First, middleware guardrail frameworks such as LangChain~\cite{langchain2023}, NeMo Guardrails~\cite{nemoguardrails2023}, and the OpenAI Agents SDK intercept tool calls before execution, but they cannot \emph{prove} they did, so the operator's word is the only guarantee. Second, TEE-based approaches such as Proof-of-Guardrail~\cite{proofofguardrail2026} provide hardware-signed attestations that code ran, but require trusting the hardware manufacturer and its attestation chain. Third, Automated Reasoning checks (ARc)~\cite{arc2026} provides formal verification by translating natural language policies into SMT-LIB logic and checking AI-generated content against them with 99\%+ soundness, but ARc's results are only as trustworthy as the server that produced them. Auditability is not verifiability. A third party receiving an ARc verdict must either re-run the entire pipeline or take the result on faith. Contemporaneous proof-carrying guardrails~\cite{niyamai2026,gabbay2026certificates} prove different statements, which we compare in \S\ref{sec:related-work}.

We present \emph{zkARc}, a framework that extends ARc with succinctly verifiable zero-knowledge proofs. In ARc, LLMs translate both policy rules and agent actions into symbolic logic. For example, the rule ``no single purchase over \$200'' becomes $\mathit{purchaseAmount} > 200 \Rightarrow \neg\,\mathit{isPurchaseAllowed}$. An action such as ``buy running shoes for \$150 from an approved vendor'' becomes logical assertions capturing its attributes ($\mathit{purchaseAmount} = 150$, $\mathit{isApprovedVendor}$) and the compliance question to be checked ($\mathit{isPurchaseAllowed}$). A symbolic reasoning engine (an SMT solver) then checks whether these assertions, together with the policy, logically entail the compliance question. If so, the action is permitted, and otherwise it is blocked.

In zkARc, each step of this pipeline, from the planner's output and policy formalization through action formalization to solver execution, produces an independent zero-knowledge proof. The proofs compose into a single receipt as a tuple of per-component proofs with cryptographic bindings on every value that adjacent stages share, so an operator cannot combine valid proofs from unrelated executions. As a result, any third party can cryptographically verify that a policy was enforced correctly on the exact action it receives, without re-running anything, without trusting the server, and without seeing the policy. The asymmetry between proving cost (borne once by the operator) and verification in under a minute (borne by many) makes this practical for high-value, low-frequency actions such as large purchases or contract commitments. A receipt does not establish that a natural language translation is faithful, that an uncertified solver is correct, or that a downstream service executes the action (\S\ref{sec:deployment-acceptance}).

\paragraph{Contributions.}
This paper makes \reviewchange{four} contributions.
\begin{itemize}[leftmargin=1.5em, itemsep=3pt, topsep=4pt, parsep=0pt]
    \item \emph{A proof-carrying guardrail receipt and its security.} \reviewchange{We specify the required computations, the values they share, and the checks that link the proofs to the received action and authorize it,} and prove completeness, knowledge soundness and zero knowledge for the composed relation under explicit assumptions (\S\ref{sec:composed-security}).
    \item \emph{Proofs connecting model inference to the action.} We prove bounded Qwen generation with Jolt Atlas~\cite{joltatlas2026}, \reviewchange{specified conversions between text and tokens} and deterministic action extraction with Jolt~\cite{jolt2023}, and bind the extracted facts to both solver queries (\S\ref{sec:proof-carrying-guardrails}).
    \item \reviewchange{\emph{A HyperKZG construction for recursive verification.} We extend Dao et al.'s verifier-splitting method~\cite{joltrecursion2026} from Dory to HyperKZG, using auxiliary proofs to certify elliptic-curve computations, and establish conditions under which this preserves soundness and zero knowledge (\S\ref{sec:verifier-split}).}
    \item \emph{A stateful end-to-end implementation.} \reviewchange{A receiver verifies receipts, checks the spending total and reserves funds without allowing another request to change that total between the check and reservation. It approves two purchases and denies a third.} A full receipt takes about an hour to issue with CPU-only proving and about 33 seconds to verify. The solver proof alone takes about 30 seconds to produce and about 0.3 seconds to verify, with an 88\,KiB proof (\S\ref{sec:bench-action-receipt}).
\end{itemize}

\section{Related Work}
\label{sec:related-work}

\subsubsection*{Middleware guardrail frameworks}

\reviewchange{Middleware guardrails intercept tool calls before execution. LangChain~\cite{langchain2023}, LangGraph~\cite{langgraph2024}, NeMo Guardrails~\cite{nemoguardrails2023}, Guardrails AI~\cite{guardrailsai2023}, OpenAI Agents, AWS Strands Agents, Llama Guard~\cite{llamaguard2024} and Semantic Integrity Constraints~\cite{sic2025} require the verifier to trust the operator's deployment. They do not provide cryptographic evidence that the guardrail checked the actual input and produced the reported verdict, or prove compliance with a hidden policy.\par}

\subsubsection*{Pre-execution firewalls}

\leavevmode\reviewchange{AEGIS~\cite{aegis2026} intercepts tool calls across 14 frameworks, applies deep string extraction, content-first risk scanning and composable policy validation, and records an audit trail with Ed25519 signatures and SHA-256 hashes. Its signatures authenticate the service's claims, but do not prove which policy and input it checked or that the check was correct. zkARc instead binds proofs to the specified programs, inputs and verdict under the assumptions of Theorem~\ref{thm:zkarc-composition}.\par}

\subsubsection*{TEE-based attestation}

\leavevmode\reviewchange{Proof-of-Guardrail~\cite{proofofguardrail2026} runs the guardrail and agent inside a Trusted Execution Environment (TEE). Hardware-signed attestation can be verified offline at a cost independent of the workload, but requires trust in the provider and attestation chain and remains exposed to hardware and systems attacks. It keeps the developer's agent private while exposing the approved wrapper and guardrail. zkARc can hide the policy and replaces hardware trust with proof-system and commitment assumptions, at substantially higher per-action proving cost (\S\ref{sec:bench-action-receipt}).\par}

\reviewchange{\subsubsection*{Zero-knowledge proofs for solvers, neural networks and guardrails}
ZKSMT~\cite{zksmt2024} checks SMT proof certificates in zero knowledge, zkPi~\cite{zkpi2024} checks Lean proof terms, and zkLean~\cite{zklean} specifies ZK statements in Lean. Solver certificates in formats such as Alethe~\cite{schurr2021alethe} can be checked by Carcara~\cite{carcara2023}. A checked certificate establishes logical validity. Proved solver execution instead binds the verdict to a committed program, whose correctness remains assumed (\S\ref{sec:zk-smt-jolt}). Certificates may be substantially cheaper for UNSAT verdicts; a quantitative comparison is future work.\par}

A second line proves neural network execution. Standalone zkML systems such as EZKL~\cite{ezkl2023} and DeepProve~\cite{deepprove2025} prove that a model was executed correctly, but they do not address guardrail pipelines, policy formalisation, solver verification or the composition of heterogeneous proofs. Work on systems-security foundations for agentic computing~\cite{systemssecurity2025} identifies the need for unforgeable provenance chains, and Zhou~\cite{cryptobinding2026} implements a Groth16 circuit for binding agentic tool use that relies on a trusted replay service.

\reviewchange{NiyamAI~\cite{niyamai2026} commits to an intent contract, checks tool invocations with a neural judge and proves the result without revealing model parameters. Gabbay~\cite{gabbay2026certificates} proposes certificates of formal policy compliance, leaving implementation open. zkARc connects policy compilation, planner provenance, the received action and both SMT queries. Its proof components build on ARc~\cite{arc2026}, Jolt and Lasso~\cite{jolt2023,lasso2024}, Jolt Atlas~\cite{joltatlas2026}, Nova and HyperNova~\cite{nova2022,hypernova2023}, Spartan~\cite{spartan2020}, Twist and Shout~\cite{twistandshout2025}, Vega~\cite{vega} and Dao et al.'s verifier split~\cite{joltrecursion2026}. Section~\ref{sec:composed-security} states the additional assumptions their composition requires.\par}

\section{Background: Automated Reasoning checks}
\label{sec:arc-background}

zkARc extends ARc~\cite{arc2026}, a neurosymbolic system that verifies natural
language claims against policies formalized in logic. ARc has two main
components. The \emph{Policy Model Creator} (PMC) autoformalizes a natural
language policy document into a logical policy model. The \emph{Answer
Verifier} (AV) translates natural language claims into logic and verifies them
against that model.

\paragraph{Policy model creation.}
ARc represents natural language policies in logic using
SMT-LIB~\cite{BarFT-SMTLIB}, a standardized logic language that uses prefix
notation (e.g., the implication $x \Rightarrow y$ is written
\texttt{(=> x y)}). A natural language rule such as ``no single purchase over
\$200'' is represented as the assertion
\[
\texttt{(=> (> purchaseAmount 200) (not isPurchaseAllowed))}.
\]
Formally, a policy model is a tuple
$\mathcal{M} = (\mathcal{D}, \mathcal{V}, \mathcal{R})$, where $\mathcal{D}$
is a set of enumerated datatypes, $\mathcal{V}$ is a set of typed variables
(over $\texttt{Bool}$, $\texttt{Int}$, $\texttt{Real}$, or a datatype in
$\mathcal{D}$), and $\mathcal{R}$ is a set of logical rules expressed as
quantifier-free assertions over $\mathcal{V}$.

\paragraph{Answer verification.}
Given a policy model $\mathcal{M}$, ARc translates natural language claims
into so-called \emph{validation units}, each consisting of a premise $P$ and a
conclusion $C$, expressed as formulas over $\mathcal{V}$. For example, the
claim ``purchasing an item for \$150 is permitted'' might be translated into
the premise \texttt{(= purchaseAmount 150)} and the conclusion
\texttt{isPurchaseAllowed}. ARc then determines correctness via two SMT
queries,
\[
\underbrace{\mathcal{M} \land P \land C}_{\text{can-be-true}}
\qquad\text{and}\qquad
\underbrace{\mathcal{M} \land P \land \lnot C}_{\text{can-be-false}},
\]
checking whether the conclusion (resp.\ its negation) is consistent with the
premise and policy model.

\reviewchange{zkARc uses both results to derive a verdict. Only \textsf{VALID} authorizes dispatch (\S\ref{sec:deployment-acceptance}).\par}

\section{Background: Zero-Knowledge Proofs}
\label{sec:zk-background}

The proof pipeline of \S\ref{sec:system-overview} wraps each stage of the ARc pipeline in a zero-knowledge proof. This section recalls the primitives those proofs are built from.

\paragraph{Notation.} We write $\lambda$ for the security parameter, $\mathbb{F}$ for a finite field, $\mathsf{negl}(\lambda)$ for a negligible function in $\lambda$, and PPT for probabilistic polynomial time. Thaler~\cite{thaler2024} gives a comprehensive treatment of the building blocks below.

\begin{definition}[Polynomial Commitment Scheme]
\label{def:multilinear-pcs}
A polynomial commitment scheme (PCS) over a field $\mathbb{F}$ is a tuple of the form
\(
\mathsf{PCS}=(\mathsf{Setup},\mathsf{Commit},\mathsf{Open},\mathsf{Verify}).
\)
Given a polynomial $f:\mathbb{F}^k \to \mathbb{F}$,
\begin{itemize}[leftmargin=1.5em, itemsep=1pt, topsep=3pt, parsep=0pt]
    \item $\mathsf{Setup}(1^\lambda, k) \to \mathsf{pp}_{\mathsf{pcs}}$ outputs public parameters.
    \item $\mathsf{Commit}(\mathsf{pp}_{\mathsf{pcs}}, f) \to C$ outputs a commitment to $f$.
    \item $\mathsf{Open}(\mathsf{pp}_{\mathsf{pcs}}, f, \xi) \to (f(\xi),\pi)$ produces the evaluation of $f$ at $\xi \in \mathbb{F}^k$ together with an opening proof $\pi$.
    \item $\mathsf{Verify}(\mathsf{pp}_{\mathsf{pcs}}, C, \xi, y, \pi) \to \{0,1\}$ checks the claimed evaluation.
\end{itemize}
We require correctness and binding. zkARc's privacy target also requires a \emph{hiding} variant, in which $\mathsf{Commit}$ takes a blinding factor, since a deterministic commitment to a low-entropy witness (a short prompt, a five-rule policy) can be tested against guesses.
\end{definition}

\begin{definition}[Sum-check protocol]
\label{def:sumcheck}
Let $g:\mathbb{F}^n\to\mathbb{F}$ be a polynomial of degree at most $d$ in each variable. The sum-check protocol allows a prover to convince a verifier that
\[
C = \sum_{x\in\{0,1\}^n} g(x).
\]
The protocol proceeds in $n$ rounds. In round~$i$, the prover sends a degree-$d$ univariate polynomial
\[
s_i(X) = \sum_{x_{i+1},\dots,x_n \in\{0,1\}} g(r_1,\dots,r_{i-1},X,x_{i+1},\dots,x_n).
\]
Set $C_1=C$. The verifier checks $s_i(0)+s_i(1)=C_i$, samples $r_i\gets\mathbb F$ and sets $C_{i+1}=s_i(r_i)$. It finally checks $C_{n+1}=g(r_1,\ldots,r_n)$ against an oracle or a PCS opening. When $g$ is a product of a fixed number of multilinear polynomials given as tables, the prover uses $O(2^n)$ field operations~\cite{thaler2024}.
\end{definition}

\paragraph{Polynomials and curves.} A polynomial is \emph{multilinear} if its degree in each variable is at most one, and the multilinear extension (MLE) of a table is the unique such polynomial agreeing with it on the Boolean hypercube. We write $\widetilde{\mathsf{eq}}(r,x)=\prod_{j}\bigl(r_jx_j+(1-r_j)(1-x_j)\bigr)$ for the MLE of the equality predicate. A polynomial interactive oracle proof (PIOP) gives the verifier evaluation access to prover-supplied polynomials, which polynomial commitments realise in practice. BN254 and Grumpkin are elliptic curves forming a cycle: BN254 has base field $\mathbb F_q$ and scalar field $\mathbb F_r$, and Grumpkin the reverse. For the BN254 pairing $e:G_1\times G_2\to G_T$, a multi-pairing checks a product of pairing values, and a multi-scalar multiplication (MSM) computes $\sum_j[a_j]P_j$. Arithmetic is \emph{native} when it uses the field supported directly by the proof system. Hyrax~\cite{hyrax2018} commits to a polynomial row by row over a grid of its values.

\begin{definition}[Zero-Knowledge Succinct Non-Interactive Arguments of Knowledge (zkSNARKs)]
\label{def:aok}
Consider a relation $R$ over structure $s$, instance $u$, and witness $w$ tuples. An argument of knowledge for $R$ is a tuple of PPT algorithms $(G, K, P, V)$:
\begin{itemize}[leftmargin=1.5em, itemsep=1pt, topsep=3pt, parsep=0pt]
    \item $G(1^\lambda) \to pp$: outputs public parameters.
    \item $K(pp, s) \to (pk, vk)$: outputs a prover key and verifier key.
    \item $P(pk, u, w) \to \pi$: outputs a proof $\pi$.
    \item $V(vk, u, \pi) \to \{0, 1\}$: verifies the proof.
\end{itemize}
\reviewchange{A witness is the data that establishes the claim. Completeness means honest proofs pass. Knowledge soundness requires an extractor that can recover a valid witness from a successful prover. Zero knowledge means the proof can be simulated using only public information. Formally, an argument of knowledge satisfies:}
\begin{itemize}[leftmargin=1.5em, itemsep=1pt, topsep=3pt, parsep=0pt]
    \item \emph{completeness} if for all $(s, u, w) \in R$, $\Pr[V(vk, u, P(pk, u, w)) = 1] = 1$;
    \item \emph{knowledge soundness} if for every PPT prover $P^*$ there is a PPT extractor $\mathcal{E}$ such that, whenever $P^*$ outputs $(u, \pi)$ with $V(vk, u, \pi) = 1$, $\mathcal{E}$ outputs $w$ with $(s, u, w) \in R$, except with negligible probability;
    \item \emph{zero-knowledge} if there exists a PPT simulator $S$ such that the distributions $\{P(pk, u, w)\}$ and $\{S(vk, u)\}$ are computationally indistinguishable;
    \item \emph{succinctness} if $|\pi| = \mathsf{poly}(\lambda, |u|, \log |w|)$ and $V$ runs in time $\mathsf{poly}(\lambda, |u|, \log |w|)$;
    \item \emph{non-interactivity} if $P$ sends a single message to $V$, typically achieved via the Fiat--Shamir heuristic~\cite{fiat-shamir1987}.
\end{itemize}
\end{definition}

The composition theorem of \S\ref{sec:composed-security} uses three refinements of these properties. \emph{Adaptive multi-theorem} knowledge soundness and zero knowledge let the adversary choose polynomially many statements after seeing the public parameters, with one extractor (resp.\ one simulator) serving all of them. An extractor is \emph{straight-line} if it recovers $w$ from a single accepting transcript without rewinding $P^*$, so that several component extractors can run against the same adversary. Non-interactive zero knowledge is stated in the \emph{programmable} random-oracle model, in which the simulator may fix the oracle's answers on the points it programs.

Two further techniques underlie the component proofs. A \emph{lookup argument} proves that every entry of a vector $a \in \mathbb{F}^Q$ appears in a table $T \in \mathbb{F}^N$. Lasso~\cite{lasso2024} decomposes a large table into $c$ sub-tables of size $N^{1/c}$, reducing the prover's commitment cost from $O(N)$ to $O(c \cdot N^{1/c} + Q)$, and Twist and Shout~\cite{twistandshout2025} supersede its memory-checking machinery with one-hot addressing, achieving logarithmic verifier cost and over $10\times$ cheaper proving. A \emph{folding scheme}~\cite{nova2022} reduces the task of checking two instances of a relation to checking a single instance of the same relation.

zkARc uses folding to achieve zero knowledge per component, via the NovaBlindFold transform of HyperNova~\cite{hypernova2023} in the succinct form introduced by Vega~\cite{vega}. \reviewchange{Component proofs are bundled together with checks that they use the same shared values} (\S\ref{sec:proof-composition}) \reviewchange{and may be replaced by one outer proof using recursion or wrapping}~\cite{joltrecursion2026}.

Models are supplied in the Open Neural Network Exchange (ONNX) format~\cite{onnx2017}, which represents a trained neural network as a directed acyclic graph of typed operators (matrix multiplication, convolution, softmax, layer normalisation, etc.). Jolt Atlas proves execution of that graph operator by operator, and the graph also defines the verifier's preprocessing: its operator sequence, tensor shapes and hyperparameters determine the structure of the sum-check DAG and the lookup tables.

\section{System Overview}
\label{sec:system-overview}

\reviewchange{Consider the shopping agent of \S\ref{sec:intro}, proposing a \$150 purchase from an approved vendor. The counterparty needs evidence that the principal's five-rule policy was checked, without seeing the policy.\par}

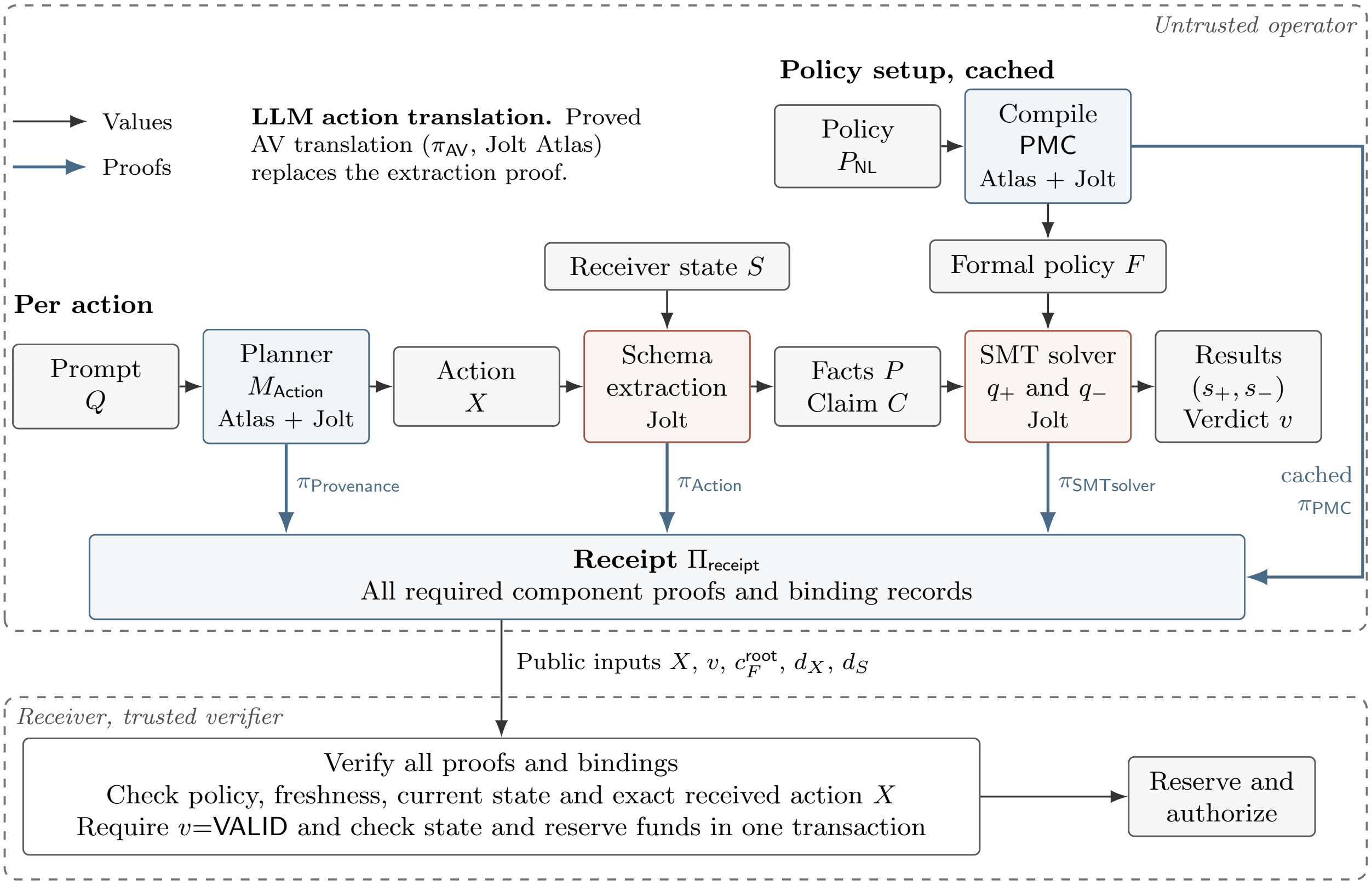
\begin{figure*}[!htbp]
\centering
\begin{tikzpicture}[zkarc diagram,
    proof arrow/.style={zkarc arrow, draw=diagramblue, line width=1pt},
    proof label/.style={font=\footnotesize, text=diagramblue, inner sep=3pt},
    boundary/.style={draw=black!55, dashed, rounded corners=4pt, line width=0.7pt},
    boundary label/.style={font=\footnotesize\itshape, text=black!70, inner sep=2pt}]
\node[zkarc heading] at (8.85,3.8) {Policy setup, cached};
\node[zkarc artifact, text width=1.65cm] (policy) at (9.8,2.9)
    {Policy\\$P_{\mathsf{NL}}$};
\node[zkarc stage, draw=diagramblue, text width=1.65cm] (compile) at (12.1,2.9)
    {Compile\\$\mathsf{PMC}$\\{\footnotesize Atlas + Jolt}};
\node[zkarc artifact, text width=2.5cm, minimum height=0.55cm]
    (formal) at (12.1,1.45) {Formal policy $F$};
\draw[zkarc arrow] (policy) -- (compile);
\draw[zkarc arrow] (compile) -- (formal);

\node[zkarc heading] at (-0.4,1.0) {Per action};
\node[zkarc artifact, text width=1.65cm] (prompt) at (0.6,0)
    {Prompt\\$Q$};
\node[zkarc stage, draw=diagramblue, text width=1.65cm] (planner) at (2.9,0)
    {Planner\\$M_{\mathsf{Action}}$\\{\footnotesize Atlas + Jolt}};
\node[zkarc artifact, text width=1.65cm] (action) at (5.2,0)
    {Action\\$X$};
\node[zkarc stage, draw=diagramvm, fill=diagramvm!6, text width=1.65cm]
    (extract) at (7.5,0) {Schema\\extraction\\{\footnotesize Jolt}};
\node[zkarc artifact, text width=1.65cm] (claims) at (9.8,0)
    {Facts $P$\\Claim $C$};
\node[zkarc stage, draw=diagramvm, fill=diagramvm!6, text width=1.65cm]
    (solver) at (12.1,0) {SMT solver\\$q_+$ and $q_-$\\{\footnotesize Jolt}};
\node[zkarc artifact, text width=1.65cm] (verdict) at (14.4,0)
    {Results\\$(s_+,s_-)$\\Verdict $v$};
\node[zkarc artifact, text width=2.6cm, minimum height=0.55cm]
    (state) at (7.5,1.45) {Receiver state $S$};
\draw[zkarc arrow] (prompt) -- (planner);
\draw[zkarc arrow] (planner) -- (action);
\draw[zkarc arrow] (action) -- (extract);
\draw[zkarc arrow] (state) -- (extract);
\draw[zkarc arrow] (extract) -- (claims);
\draw[zkarc arrow] (claims) -- (solver);
\draw[zkarc arrow] (formal) -- (solver);
\draw[zkarc arrow] (solver) -- (verdict);

\node[zkarc artifact, draw=diagramblue, fill=diagramblue!6,
      text width=13.6cm] (receipt) at (7.5,-2.3)
    {\textbf{Receipt $\Pi_{\mathsf{receipt}}$}\\
     All required component proofs and binding records};
\draw[proof arrow] (planner.south) -- (planner.south |- receipt.north)
    node[proof label, pos=0.45, right] {$\pi_{\mathsf{Provenance}}$};
\draw[proof arrow] (extract.south) -- (extract.south |- receipt.north)
    node[proof label, pos=0.45, right] {$\pi_{\mathsf{Action}}$};
\draw[proof arrow] (solver.south) -- (solver.south |- receipt.north)
    node[proof label, pos=0.45, right] {$\pi_{\mathsf{SMTsolver}}$};
\draw[proof arrow] (compile.east) -- (15.9,2.9) |- (receipt.east);
\node[proof label, anchor=east, align=right] at (15.9,-1.25)
    {cached\\$\pi_{\mathsf{PMC}}$};

\node[zkarc box, text width=11.2cm] (accept) at (5.5,-4.95)
    {Verify all proofs and bindings\\
     Check policy, freshness, current state and exact received action $X$\\
     \reviewchange{Require $v{=}\mathsf{VALID}$ and check state and reserve funds in one transaction}};
\node[zkarc artifact, text width=1.9cm] (dispatch) at (14.2,-4.95)
    {Reserve and\\authorize};
\draw[zkarc arrow] (receipt.south -| accept.north) -- (accept.north);
\node[zkarc note, anchor=west] at (5.6,-3.35)
    {Public inputs $X$, $v$, $c_F^{\mathsf{root}}$, $d_X$, $d_S$};
\draw[zkarc arrow] (accept) -- (dispatch);

\draw[boundary] (-0.5,-2.95) rectangle (15.95,4.6);
\node[boundary label, anchor=north east] at (15.9,4.55)
    {Untrusted operator};
\draw[boundary] (-0.5,-3.75) rectangle (15.95,-5.95);
\node[boundary label, anchor=north west] at (-0.45,-3.8)
    {Receiver, trusted verifier};

\draw[zkarc arrow] (-0.4,3.2) -- (0.5,3.2);
\node[zkarc note, anchor=west] at (0.6,3.2) {Values};
\draw[proof arrow] (-0.4,2.65) -- (0.5,2.65);
\node[zkarc note, anchor=west] at (0.6,2.65) {Proofs};
\node[zkarc note, text width=4.9cm, anchor=west] at (2.4,2.9)
    {\reviewchange{\textbf{LLM action translation.} Proved AV translation} ($\pi_{\mathsf{AV}}$,
     Jolt Atlas) replaces the extraction proof.};
\end{tikzpicture}
\caption{Required components, trust boundary and receiver checks. The upper dashed region runs on the untrusted operator, which holds every witness (prompt, policy, formal policy, model values and solver trace). Only the receipt and its public inputs cross to the receiver, and zero knowledge hides the witnesses from it. Black arrows carry values, and blue arrows carry proofs into the receipt. Each inference component includes its native Atlas proof and required Jolt conversion proofs. Extraction binds the received action $X$ and receiver state $S$ to facts $P$ and claim $C$, and the solver checks $q_+=F\land P\land C$ and $q_-=F\land P\land\neg C$. Adjacent relations bind their shared values, and the cached policy proof binds $F$ to \reviewchange{the authorized commitment} $c_F^{\mathsf{root}}$. The \reviewchange{implementation} reserves spending without executing a merchant transaction.}
\label{fig:zkarc-pipeline-compact}
\end{figure*}

\reviewchange{Figure~\ref{fig:zkarc-pipeline-compact} separates policy setup from the checks performed for each action.\par}

\begin{table}[tbp]
\centering
\small
\newcommand{\grp}[1]{\multicolumn{2}{@{}l}{\footnotesize\textbf{#1}}\\[1pt]}
\begin{tabular}{@{}l@{\hspace{14pt}}>{\raggedright\arraybackslash}p{0.66\columnwidth}@{}}
\toprule
\grp{Policy and action}
$P_{\mathsf{NL}}$, $P_{\mathsf{SMT}}$ & natural language policy, formal policy model ($F = P_{\mathsf{SMT}}$) \\
$Q$, $A_{\mathsf{NL}}$ & prompt, natural language action plan \\
$X$, $\Sigma$, $d_X$ & structured action, action schema, \reviewchange{action hash} \\
$S$, $d_S$ & \reviewchange{state supplied by the receiver}, its \reviewchange{authenticated state hash} \\
\addlinespace[3pt]
\grp{Solver}
$C_{\mathsf{SMT}} = (P, C)$ & \reviewchange{solver inputs: premise} $P$, conclusion $C$ \\
$q_+$, $q_-$ & $F \wedge P \wedge C$, \ $F \wedge P \wedge \neg C$ \\
$(s_+, s_-)$, $v$ & solver results, verdict with four possible outcomes \\
\addlinespace[3pt]
\grp{Models and component proofs}
$M_{\mathsf{PMC}}$, $M_{\mathsf{Action}}$, $M_{\mathsf{AV}}$ & \reviewchange{committed models for policy translation, planning, and optional action translation} \\
$\mathsf{Extract}_\Sigma$, $\pi_{\mathsf{Action}}$ & deterministic action extractor (main \reviewchange{configuration}), proof of action formalization \\
$\pi_{\mathsf{PMC}}$, $\pi_{\mathsf{AV}}$, $\pi_{\mathsf{SMTsolver}}$ & component proofs \\
$\pi_{\mathsf{Provenance}}$ & required proof of the planner execution \\
$\pi'_i$, $\widehat{V}_i$ & auxiliary recursion proof, extended verifier of $(\pi_i, \pi'_i)$ \\
\addlinespace[3pt]
\grp{Receipt and cryptography}
$\Pi_{\mathsf{receipt}}$, $\mathsf{bind}$, $\mathcal{I}$ & \reviewchange{proof bundle, records linking its proofs, required components} \\
$c_Q$, $c_F^{\mathsf{root}}$, $c_e$, $\mathsf{sid}$ & prompt commitment, authorized \reviewchange{policy commitment}, edge commitment, session identifier \\
$\mathbb{F}_r$, $\mathbb{F}_q$ & BN254 scalar field, base field (= Grumpkin scalar field) \\
\bottomrule
\end{tabular}
\caption{Notation.}
\label{tab:notation}
\end{table}

\subsubsection*{Offline stage: Policy formalization via $M_{\mathsf{PMC}}$}

At setup time, the principal's natural language policy $P_{\mathsf{NL}}$ is formalized by the policy model creator $M_{\mathsf{PMC}}$. For our example policy, $M_{\mathsf{PMC}}$ produces the permission predicate (full policy in Appendix~\ref{app:running-example-smtlib})
\begin{smtcode}
(and (<= total 500) (<= amount 200) approved
     (not credentials) (not recurring))
\end{smtcode}
\reviewchange{Each conjunct of this predicate $G$ represents one policy clause. An authenticated wrapper supplies declarations and sets $F=(\texttt{allowed}\leftrightarrow G)$, allowing the solver to establish permission rather than mere consistency. The cached proof $\pi_{\mathsf{PMC}}$ certifies the model's production of $G$ from $P_{\mathsf{NL}}$, its conversion to $F$ and agreement with the authorized policy commitment. It proves execution, not faithful translation. We cover a single model invocation, rather than ARc's richer formalization strategies.\par}

\subsubsection*{Online stage 1: Action provenance via $M_{\mathsf{Action}}$}

\reviewchange{The planner $M_{\mathsf{Action}}$ receives prompt $Q$ and produces a structured purchase $X$ of \$150 from \texttt{acme}, without credential sharing or recurring payments. The proof $\pi_{\mathsf{Provenance}}$ certifies $M_{\mathsf{Action}}(Q)=X$. The receiver compares the commitment to $Q$ with the prompt expected for that session.\par}

\subsubsection*{Online stage 2: Action formalization}

A deterministic extractor derives a \emph{validation unit} from $X$, consisting of a premise over the policy vocabulary and a conclusion to be checked. With a prior reserved total of \$200 and \texttt{acme} in the receiver's approved-vendor registry, it produces the premise
\[
P=(\texttt{amount}=150)\land(\texttt{total}=350)\land\texttt{approved}\land\neg\texttt{credentials}\land\neg\texttt{recurring}
\]
and conclusion $C=\texttt{allowed}$. We write $C_{\mathsf{SMT}}$ for this validation unit (the premise-conclusion pair). The zkARc proof $\pi_{\mathsf{Action}}$ certifies that the extractor produced exactly this $C_{\mathsf{SMT}}$ from the exact action $X$ the counterparty receives. The prior balance and vendor approval come from receiver state $S$, not from model execution. \reviewchange{An alternative uses an LLM to translate a natural language plan into logic, through the answer-verifier model} $M_{\mathsf{AV}}$ \reviewchange{with proof} $\pi_{\mathsf{AV}}$.

\subsubsection*{Online stage 3: SMT solving}

The SMT solver checks whether each of the following is satisfiable:
\[
q_+=F\land P\land C,
\qquad q_-=F\land P\land\neg C.
\]
\reviewchange{Here, $(\mathsf{SAT},\mathsf{UNSAT})$ entails permission and yields \textsf{VALID}. A purchase taking the total to \$650 instead yields $(\mathsf{UNSAT},\mathsf{SAT})$ and is blocked. The proof $\pi_{\mathsf{SMTsolver}}$ certifies the program's execution, results and derived verdict. The receiver verifies the receipt, checks current state and reserves the \$150 in one transaction. Section~\ref{sec:deployment-acceptance} gives the full verdict rule, and Appendix~\ref{app:running-example-smtlib} illustrates why prohibitions alone do not establish permission.\par}

\reviewchange{The proofs must agree on the planner output, policy $F$ and pair $(P,C)$. The operator holds the private witnesses, and the receiver checks the receipt against its authorized policy commitment and received action. The implementation reserves spending but does not execute a merchant transaction.\par}

\reviewchange{Section~\ref{sec:proof-carrying-guardrails} specifies the protocol, \S\ref{sec:composed-security} proves its conditional security, and \S\ref{sec:bench} evaluates the implementation.\par}

\subsection{Threat Model}
\label{sec:threat-model}

\reviewchange{\noindent\textbf{Parties and trust.} The policy owner authorizes the policy commitment and required computations. An untrusted operator runs the agent, guardrail and prover. The receiver obtains the action and receipt, and an enforcement point, possibly the receiver itself, checks them before dispatch. The honest verifier uses commitments, keys and requirements authenticated outside the operator's control. It learns the verdict and declared public inputs, without re-running inference or solving and without learning the policy or witnesses.\par}

\medskip
\noindent\textbf{Adversary model.} The operator controls every input and execution that is not proved, and may attempt to
\begin{enumerate}[leftmargin=*, itemsep=1pt, topsep=3pt, parsep=0pt]
    \item fabricate that a policy was checked when it was not, or substitute a different policy or action,
    \item reveal only a favorable subset of the computation, by omitting components or splicing proofs from different executions, or
    \item declare one action and dispatch another, present an action the planner did not produce, or replay a receipt across sessions or counterparties.
\end{enumerate}

\medskip
\noindent\textbf{Guarantees.} Cryptographic soundness rules out false receipt statements. The verifier separately checks authorization, freshness, single-use state, the exact received action and the verdict (\S\ref{sec:deployment-acceptance}). Together they close the gap between what an agent \emph{declares} and what the receiver \emph{gets}. The provenance proof $\pi_{\mathsf{Provenance}}$ binds the declared action to a specific prompt transcript and model execution, raising the bar from ``the agent lied about what it planned'' to ``the agent faithfully planned $X$ and then chose not to do $X$.'' The receiver rejects any action that arrives without a valid receipt (\S\ref{sec:section-six-deployment}), which binds the \emph{proven plan} to the \emph{action as received}. Neither proves that a downstream service executes that action. Closing that gap requires enforcement by the recipient, TEE-attested dispatch, or a proved dispatch program.

\medskip
\noindent\textbf{Scope and assumptions.} A receipt shows how supplied facts are used but does not authenticate them, so a deployment must authenticate prices, prior spending and vendor approval and enforce any cumulative budget against current state. It shows that the policy was \emph{checked}, not that it was \emph{good}. Whether a policy is correct, complete or well-intentioned is a question for reputation or audit, not cryptography. In \reviewchange{the configuration with LLM action translation}, $\pi_{\mathsf{AV}}$ shows that the committed model produced the premise--conclusion pair, not that the pair captures the whole action, and translator agreement controlled by the operator can be forged unless every execution is proved.

\reviewchange{Prompt injection is out of scope: a harmful but policy-compliant action can receive a \textsf{VALID} receipt. We do not evaluate prompt-injection defenses. A future component could prove a detector, such as an activation probe, with Jolt Atlas and bind it to the same planner execution.\par}

\reviewchange{Root access to the prover can expose witnesses or suppress service, but cannot make modified code verify as the authenticated program. Repeated chosen queries can reveal hidden thresholds through verdicts, so applications may need query limits. Inputs are bounded by the committed model configuration, and $d_X$ is public. Measured Dory proofs need no powers-of-tau ceremony~\cite{lee2021dory}; the alternative HyperKZG construction and a Groth16 wrapper require trusted setup.\par}

\subsection{Protocol Objects and \textcolor{black}{Action Authorization}}
\label{sec:deployment-acceptance}

The proof relation and the deployment decision are distinct. The proof establishes correct execution and consistency of the selected components. The verifier separately decides whether that receipt is authorized for a concrete action and context. \reviewchange{The protocol figures show the values shared by both checks.}

\reviewchange{Figures~\ref{fig:protocol-setup}--\ref{fig:protocol-action}
show how the policy, prompt, state and action are identified and checked.
Appendix~\ref{app:protocol-encodings} gives their exact hashes, commitments
and context fields.}

\paragraph{Receipt relation}
The public statement $u$ contains the receipt context, the exact component graph $\mathcal I$, and the identifiers of its relations and verification keys. It also contains the \reviewchange{policy commitment}, prompt commitment $c_Q$, \reviewchange{action hash} $d_X$, receiver-issued \reviewchange{state $S$ and its hash $d_S$}, the \reviewchange{shared-value identifiers}, the two certified solver results $(s_+,s_-)$, and the declared verdict $v$ with its four possible outcomes. Every supported \reviewchange{configuration} includes provenance, policy compilation, action formalization and solver execution. The relation $R_{\mathsf{receipt}}(u,w)$, defined formally in Definition~\ref{def:zkarc-relation}, requires every component relation to hold and every incident component to use the same canonical object or prove an opening to the same hiding commitment. Its action relation is fixed by the authenticated deployment \reviewchange{configuration}. It is either deterministic extraction $\mathsf{Extract}_{\Sigma}(X,S)=(P,C)$ or the explicitly identified proved-translation relation. In both cases, the relation binds the same $X$ named by $d_X$ to the exact $(P,C)$ consumed by the solver. The required provenance relation opens $c_Q$ to $Q$ and proves the planner execution on that same $Q$, and the action relation uses that exact output, directly as $X$ or through a proved plan-to-action relation.

The solver relation constructs, rather than accepts from the host,
\[
q_+=F\land P\land C,\qquad q_-=F\land P\land\mathsf{CanonicalNegate}(C),
\]
so $q_-$ contains the canonical negation of the exact $C$ in $q_+$ (\S\ref{sec:zk-smt-jolt}). Successful results $s_+,s_-\in\{\mathsf{SAT},\mathsf{UNSAT}\}$ determine
\[
\begin{array}{c|c}
(s_+,s_-) & v\\ \hline
(\mathsf{SAT},\mathsf{UNSAT}) & \mathsf{VALID}\\
(\mathsf{UNSAT},\mathsf{SAT}) & \mathsf{INVALID}\\
(\mathsf{SAT},\mathsf{SAT}) & \mathsf{SATISFIABLE}\\
(\mathsf{UNSAT},\mathsf{UNSAT}) & \mathsf{IMPOSSIBLE}.
\end{array}
\]
A separate query for $F\land P$ is diagnostic only, since $F\land P\equiv q_+\lor q_-$. Parse failure, panic, timeout, \textsf{unknown}, unsupported theory, and resource exhaustion do not satisfy the solver relation and never yield an authorizing verdict.

\paragraph{\reviewchange{Checks Before Accepting an Action}}
\reviewchange{A valid proof based on yesterday's spending total does not by itself
justify a purchase today. The receiver must check that the proved computation
applies to the intended request and current state.\par}

\reviewchange{\noindent\textbf{Approved policy and programs.}
The receiver chooses which policy and computations it accepts. Correct execution
of an unapproved policy or a substituted program does not satisfy this choice.\par}

\reviewchange{\noindent\textbf{Correct request and action.}
The receipt must concern the intended prompt, counterparty and received action.
Session-use and expiry checks prevent replay; the action hash detects a
substituted action.\par}

\reviewchange{\noindent\textbf{Valid proofs and approving verdict.}
Receipt verification checks the required computations and their shared values.
The result must also permit the action: a correct proof of denial cannot authorize it.\par}

\reviewchange{\noindent\textbf{Current external facts and state.}
The receiver authenticates the facts supplied to the proved computation and
checks that its state is still current. For spending limits, it checks the
available budget and reserves the funds in the same transaction, so another
request cannot spend that budget between the two operations.\par}

\reviewchange{Figure~\ref{fig:protocol-action}(b) gives the complete checklist.
We write $\mathsf{Accept}_{\mathsf{deploy}}(X_{\mathsf{recv}},u,\Pi)=1$ exactly when
all those checks pass. Acceptance concerns the received action; it does not prove
that a downstream service executes it (\S\ref{sec:threat-model}).\par}

\section{\reviewchange{The zkARc Protocol}}
\label{sec:proof-carrying-guardrails}

{\color{black}A receipt is a bundle of cryptographic proofs and the public inputs needed to verify them. It includes the information that links the proofs to the same action and policy.}

The number and placement of proofs determine exactly which claims a third party can verify without trusting the agent operator, the model host, or the guardrail service. zkARc applies two checks. First, $V_{\mathsf{receipt}}$ verifies that the required components describe one bound computation. It shows what the selected programs computed and how their inputs and outputs were connected. Second, $\mathsf{Accept}_{\mathsf{deploy}}$ decides whether that computation is sufficient for a particular action. It checks that the policy is authorized and the receipt is fresh. It also checks that the \reviewchange{configuration} is the required one, that the action matches what the counterparty received, and that the verdict is \textsf{VALID}. Only the second check authorizes the action.

\clearpage
\subsection{\reviewchange{What the Proofs Verify}}

A receipt \reviewchange{configuration} is a checklist and map of data flow approved by the verifier. Formally,
it is an authenticated graph
\[
G_{\mathsf{profile}}=
  (\mathcal I_{\mathsf{profile}},E_{\mathsf{profile}}).
\]
Its vertices identify the required relations, verifier keys, program or model
commitments, and versions. Its edges identify the values that two components
must share. The \reviewchange{configuration} also fixes the verdict rule. The verifier rejects a receipt with a missing component, a substituted implementation, or \reviewchange{fewer required computations or links between them}. The operator cannot choose a more convenient claim.

Table~\ref{tab:receipt-components} states the verified fact and the remaining
assumption for each component. A proof of execution shows that a committed
program ran on bound inputs. It does not show that a human approved those
inputs, that a learned translation captured every intended meaning, or that an
uncertified solver contains no bugs.

\begin{table}[H]
\reviewcolor
\centering
\footnotesize
\setlength{\tabcolsep}{5pt}
\newcommand{\comp}[1]{\textbf{#1}}
\begin{tabularx}{\textwidth}{@{}
    >{\raggedright\arraybackslash}p{0.155\textwidth}
    >{\raggedright\arraybackslash}X
    >{\raggedright\arraybackslash}X
    >{\raggedright\arraybackslash}p{0.085\textwidth}@{}}
\toprule
\textbf{Component} & \textbf{Verified fact} & \textbf{What remains assumed} & \textbf{Proved} \\
\midrule
\comp{Provenance} &
The committed planner used the prompt committed by $c_Q$ and produced the action object bound to the received action. &
The verifier authorizes the prompt and accepts that the planner output reflects the principal's intent. &
Per action \\
\addlinespace[2pt]
\comp{Policy compilation} &
The committed compiler produced $F$ from the authorized source policy. &
The policy owner accepts that $F$ faithfully represents the source policy. &
Per policy \\
\addlinespace[2pt]
\comp{Action formalization} &
The exact received action $X$ produced the bound solver instance $(P,C)$. &
Deterministic extraction is faithful to the declared schema. \reviewchange{A configuration with LLM action translation} also assumes that its declared translation rule is adequate. &
Per action \\
\addlinespace[2pt]
\comp{Solver} &
The committed solver processed $(F,P,C)$, returned both results, and derived the declared verdict. &
The solver program implements the intended SMT semantics correctly. &
Per action \\
\bottomrule
\end{tabularx}
\caption{\reviewchange{What the proofs verify and what remains assumed}. The policy proof is cached until the authorized policy changes.}
\label{tab:receipt-components}
\end{table}

Every zkARc \reviewchange{configuration} requires provenance, policy compilation, action
formalization and solver execution. The authenticated graph must connect the
planner output to the received action $X$. If the
planner emits the structured action directly, the provenance relation proves
$M_{\mathsf{Action}}(Q)=X$. If the planner emits a natural language plan
$A_{\mathsf{NL}}$, the \reviewchange{configuration} must include a proved relation from
$A_{\mathsf{NL}}$ to $X$. A detached proof of
$M_{\mathsf{Action}}(Q)=A_{\mathsf{NL}}$ does not satisfy the receipt relation without that link.

\subsection{\reviewchange{The Protocol, Step by Step}}
\label{sec:action-protocol}

\reviewchange{Figures~\ref{fig:protocol-setup}--\ref{fig:protocol-action}
show offline policy setup, online request preparation, and action proving and
authorization. Request preparation records the authorized prompt and current
state before the agent runs, so the later proofs must use those inputs.
Policy setup is repeated when the policy changes; the online steps run for each request.}

\reviewchange{In the figures, $H$ is a collision-resistant hash and
$\mathsf{Com}(m;r)$ is a binding, hiding commitment with fresh independent
randomness $r$. The brackets $\langle\cdot\rangle$ denote the distinct fixed
encodings in Appendix~\ref{app:protocol-encodings}. The hashes $d_S$ and $d_X$
identify state and action; $c_F^{\mathsf{root}}$, $c_Q$ and $c_e$ commit to
the policy, prompt and shared values, which may remain hidden. Each named
proof includes its required text and tensor conversions. The receiver obtains
approved models, programs, keys and schema independently of the operator.
In our spending implementation, $S$ is public and contains prior spending and
the approved-vendor registry. Solver correctness remains an assumption
(\S\ref{sec:threat-model}); authorization does not prove downstream execution
(\S\ref{sec:section-six-deployment}).}

\reviewchange{For example, a receipt for a purchase of \$150 must fail if the
receiver instead receives a purchase of \$250: recomputing $d_X$ detects the
change. If another request changes the spending total, the receiver rejects
the old state even if its proofs still verify. Appendix~\ref{app:protocol-encodings}
gives the exact encodings and context fields. The protocols retain all the
checks specified below.}

\clearpage
\begingroup
\setlength{\intextsep}{8pt}
\begin{figure}[H]
\centering
\reviewcolor
\setlength{\fboxsep}{7pt}
\fbox{\begin{minipage}{\dimexpr\linewidth-2\fboxsep-2\fboxrule\relax}
\small\raggedright
\setlength{\parskip}{4pt}
\setlength{\abovedisplayskip}{4pt}
\setlength{\belowdisplayskip}{4pt}
\textbf{Offline stage: policy setup}

\textbf{Inputs:} policy $P_{\mathsf{NL}}$; approved model for translating policies and verifier configuration.
\smallskip\hrule\smallskip
\begin{enumerate}[leftmargin=*,label=\arabic*.,itemsep=4pt,topsep=3pt]
\item \textbf{Prover:} translate the policy into logic, $P_{\mathsf{NL}}\longmapsto F$.
\item \textbf{Prover:} choose fresh $r_F$ and compute
\[
\begin{aligned}
\mathsf{ctx}_F&=(\text{owner, policy identifier, version, epoch}),\\
c_F^{\mathsf{root}}&=\mathsf{Com}
  (\langle\mathsf{ctx}_F,F\rangle_F;r_F)
  &&\text{(policy commitment).}
\end{aligned}
\]
\item \textbf{Prover:} generate $\pi_{\mathsf{PMC}}$ proving that the translation
computation produced $F$ from $P_{\mathsf{NL}}$ and that $c_F^{\mathsf{root}}$ commits to this $F$.
\item \textbf{Policy owner:} approve the policy commitment $c_F^{\mathsf{root}}$.
\item \textbf{Verifier during setup:} check and store the policy proof $\pi_{\mathsf{PMC}}$.
\end{enumerate}
\smallskip\hrule\smallskip
\textbf{Reuse:} $c_F^{\mathsf{root}}$, $\pi_{\mathsf{PMC}}$.
\quad\textbf{Prover retains:} $F,r_F$.
\end{minipage}}
\caption{\reviewchange{Offline policy setup, repeated when the authorized policy changes.}}
\label{fig:protocol-setup}
\end{figure}

\begin{figure}[H]
\centering
\reviewcolor
\setlength{\fboxsep}{7pt}
\fbox{\begin{minipage}{\dimexpr\linewidth-2\fboxsep-2\fboxrule\relax}
\small\raggedright
\setlength{\parskip}{4pt}
\setlength{\abovedisplayskip}{4pt}
\setlength{\belowdisplayskip}{4pt}
\textbf{Online stage: request preparation}

\textbf{Inputs:} authorized $c_F^{\mathsf{root}}$ and verifier configuration
(Figure~\ref{fig:protocol-setup}); private prompt $Q$.

\textbf{Requesting party:} authorizes $Q$ and is trusted by the receiver.
In our spending implementation, this is the receiver itself.
\smallskip\hrule\smallskip
\begin{enumerate}[leftmargin=*,label=\arabic*.,itemsep=4pt,topsep=3pt]
\item \textbf{Identify the request (receiver).}
Choose a fresh $\mathsf{sid}$, the counterparty and the validity interval.
\item \textbf{Record the state (receiver).}
Authenticate current state $S$ and compute
\[
d_S=H(\langle S\rangle_S)\qquad\text{(state hash).}
\]
Set $\mathsf{ctx}_X$ to the domain, counterparty, $\mathsf{sid}$, schema identifier,
policy context, $c_F^{\mathsf{root}}$ and $d_S$.
\item \textbf{Commit to the prompt (requesting party).}
Choose fresh $r_Q$ and compute
\[
c_Q^{\mathsf{exp}}=\mathsf{Com}(\langle\mathsf{ctx}_X,Q\rangle_Q;r_Q)
\qquad\text{(expected prompt commitment).}
\]
\item \textbf{Send the inputs (requesting party).}
Give $(Q,r_Q)$ privately to the prover and $c_Q^{\mathsf{exp}}$ to the receiver
through its authenticated request inputs.
\end{enumerate}
\smallskip\hrule\smallskip
\textbf{Prover:} compute $c_Q$ from $Q,r_Q,\mathsf{ctx}_X$.
\quad\textbf{Receiver:} keep $c_Q^{\mathsf{exp}}$.

\textbf{Later check:} the receipt must contain $c_Q=c_Q^{\mathsf{exp}}$
(Figure~\ref{fig:protocol-action}).
\end{minipage}}
\caption{Record the authorized prompt and current state before the agent runs.}
\label{fig:protocol-request}
\end{figure}

\endgroup
\begin{figure}[p]
\centering
\reviewcolor
\setlength{\fboxsep}{7pt}
\fbox{\begin{minipage}{\dimexpr\linewidth-2\fboxsep-2\fboxrule\relax}
\small\raggedright
\setlength{\parskip}{3pt}
\setlength{\abovedisplayskip}{4pt}
\setlength{\belowdisplayskip}{4pt}
\textbf{(a) Prover: generate the receipt}

\textbf{Inputs:} cached policy proof and $(F,r_F)$ (Figure~\ref{fig:protocol-setup});
prepared request (Figure~\ref{fig:protocol-request}).
\smallskip\hrule\smallskip
\textbf{1. Compute the action and solver results.}
\[
\begin{aligned}
X&=M_{\mathsf{Action}}(Q), & (P,C)&=\mathsf{Extract}_{\Sigma}(X,S),\\
d_X&=H(\langle\mathsf{ctx}_X,X\rangle_X)
&&\text{(action and context hash)},\\
q_+&=F\land P\land C, & q_-&=F\land P\land\neg C.
\end{aligned}
\]
Run both queries to obtain $(s_+,s_-)$ and verdict $v$.
Only $(\mathsf{SAT},\mathsf{UNSAT})$ gives $v=\mathsf{VALID}$.

\textbf{2. Before proving, fix the values shared by each pair of proofs.}
\[
\begin{array}{rcl}
\text{Planner proof} & \xleftrightarrow{\quad X\quad} & \text{Action proof}\\
\text{Action proof} & \xleftrightarrow{\quad(P,C)\quad} & \text{Solver proof}\\
\text{Policy proof} & \xleftrightarrow{\quad F\quad} & \text{Solver proof}
\end{array}
\]
Each pair uses the same public value or proves that both computations use
one hidden value $z_e$ committed as
\[
c_e=\mathsf{Com}_e(\langle\mathsf{ctx}_e,z_e\rangle_e;r_e)
\qquad\text{(shared-value commitment).}
\]
Reuse the cached policy commitment. If the solver uses a fresh commitment,
its proof also opens $c_F^{\mathsf{root}}$ to the same $F$.

\textbf{3. Generate proofs using these fixed values and commitments.}
\begin{enumerate}[leftmargin=*,label=(\alph*),itemsep=2pt,topsep=2pt]
\item $\pi_{\mathsf{Provenance}}$: planner execution and the opening of $c_Q$ to $Q$.
\item $\pi_{\mathsf{Action}}$: extraction bound to $d_X$ and $d_S$.
\item $\pi_{\mathsf{SMTsolver}}$: query construction, both results and verdict $v$.
\end{enumerate}
Include the required proofs of text and tensor conversions.

\textbf{4. Operator $\longrightarrow$ receiver:}
$(X,u,\Pi_{\mathsf{receipt}})$; the delivered action is denoted $X_{\mathsf{recv}}$.

\textbf{Receipt:} cached $\pi_{\mathsf{PMC}}$, the fresh proofs above,
their public inputs and records linking shared values.
\end{minipage}}

\medskip
\fbox{\begin{minipage}{\dimexpr\linewidth-2\fboxsep-2\fboxrule\relax}
\small\raggedright
\setlength{\parskip}{3pt}
\textbf{(b) Receiver: check the action and receipt}
\smallskip\hrule\smallskip
\begin{enumerate}[leftmargin=*,label=\arabic*.,itemsep=4pt,topsep=2pt]
\item \textbf{Approved policy and programs.} Require the authorized owner,
$c_F^{\mathsf{root}}$, version and epoch. The required components $\mathcal I$,
verifier keys, relation identifiers and binding mode must match the receiver's
approved configuration.
\item \textbf{Correct request and action.} Require the intended counterparty,
a fresh $\mathsf{sid}$, an unexpired validity interval and unused single-use state.
In the authenticated context, including $d_S$, require
$c_Q=c_Q^{\mathsf{exp}}$ and
$H(\langle\mathsf{ctx}_X,X_{\mathsf{recv}}\rangle_X)=d_X$.
\item \textbf{Valid proofs and approving verdict.} Require
$V_{\mathsf{receipt}}(u,\Pi_{\mathsf{receipt}})=1$ and $v=\mathsf{VALID}$.
\item \textbf{Current external facts and state.} Authenticate external facts
and bind them to this action and context. Require the proved state to match
current state. For cumulative limits, check state and reserve or update resources
in one transaction, also marking the session used for spending.
\end{enumerate}
\textbf{Authorize iff all checks pass} (\S\ref{sec:deployment-acceptance}).
\end{minipage}}
\caption{\reviewchange{Online proving and verification for structured actions.
Shared commitments are fixed before the proofs that use them are generated.}}
\label{fig:protocol-action}
\end{figure}

\clearpage

\subsection{Policy and Action Binding}

The core receipt binds four transitions. In the main structured \reviewchange{configuration},
\[
\begin{aligned}
Q
  &\xrightarrow{M_{\mathsf{Action}}} X,\\
P_{\mathsf{NL}}
  &\xrightarrow{\mathsf{PMC}} F,\\
(X,S)
  &\xrightarrow{\mathsf{ActionRel}_{\mathsf{profile}}} (P,C),\\
(F,P,C)
  &\xrightarrow{\mathsf{Solve2}} (s_+,s_-,v).
\end{aligned}
\]
\reviewchange{Section~\ref{sec:composition-protocol} specifies the equality checks and commitment openings that make the proofs agree on their shared values.\par}

\reviewchange{The cached PMC proof binds $F$, its source policy and compiler execution to the authorized commitment $c_F^{\mathsf{root}}$. An online proof may use a fresh commitment to the same $F$, but the public root or cached proof still links receipts to their registration. Opening the root requires $(F,r_F)$, which \reviewchange{must be supplied to the operator} as part of authorization. Any holder of $r_F$ can create fresh commitments to $F$.\par}

\reviewchange{Policy hiding protects thresholds, negotiation strategies and internal rules. The receiver still learns metadata, the verdict, the authorized commitment and information from repeated queries (\S\ref{sec:threat-model}). Proofs of selected policy properties could give further assurance, but are not implemented. A counterparty instead relies on an auditor who inspects and signs the policy, or on the principal's reputation.\par}

\reviewchange{For structured actions, schema} $\Sigma$ fixes a deterministic extractor
\[
\mathsf{Extract}_{\Sigma}(X,S)=(P,C).
\]
\reviewchange{The extractor rejects missing, duplicate, ill-typed, unsupported and noncanonical fields, checks $d_S$, and binds $d_X$ to the derived $(P,C)$. The receiver recomputes $d_X$ from the tool call. This construction requires no LLM translation of actions.\par}

\reviewchange{For natural language plans, ARc~\cite{arc2026} uses $k$ independent LLM translations and accepts a pair $\langle P,C\rangle$ when enough agree. If only the smallest translator is proved, the other $k-1$ votes remain a trust assumption: an erroneous translation and forged agreement can authorize a violating action. Alternatively, the counterparty can run the smallest translator on the received action, requiring disclosure of the policy vocabulary. A wrong translation then requires both translators to err identically. These options remain unevaluated.\par}

\subsection{\reviewchange{How the Solver Results Determine the Verdict}}

The proved solver program receives $F$, $P$, and $C$. It constructs both
queries itself:
\[
q_+=F\land P\land C,
\qquad
q_-=F\land P\land\mathsf{CanonicalNegate}(C).
\]
The second query contains the canonical negation of the exact $C$ used in the
first. The host never supplies either query. The pair $(s_+,s_-)$ determines the
verdict with four possible outcomes defined in \S\ref{sec:deployment-acceptance}. Only
$(\mathsf{SAT},\mathsf{UNSAT})$ yields \textsf{VALID}. Parse failure, panic,
timeout, \textsf{unknown}, unsupported theory, and resource exhaustion fail
closed.

zkARc runs the committed SMT solver inside Jolt. The proof binds the solver
program, both inputs, both results, and the verdict. It therefore removes trust
in the operator's report of what the solver did. It does not prove that the
uncertified solver is mathematically correct. Interpreting the proved verdict as an SMT entailment result additionally
assumes the solver implements the intended semantics correctly. This is an
application assumption beyond the theorem about execution of the specified
programs. Section~\ref{sec:full-solver-jolt} discusses this boundary and the
independent checks that can reduce the remaining risk.

\subsection{Receipt Composition}

Individually valid proofs are not enough. Without shared bindings, an operator could combine a policy proof from one run, an action proof from another, and a favorable solver result from a third. The receipt therefore binds every value that adjacent components share, either as an equal public value or as an opening of the same domain-separated commitment, and its context prevents replay for another session or counterparty. Section~\ref{sec:proof-composition} defines the composed relation, proves its security under explicit hypotheses, and describes an optional verifier split~\cite{joltrecursion2026} for heterogeneous verification.

\subsection{Deployment Modes and Enforcement}
\label{sec:section-six-deployment}

\reviewchange{Receipts can be generated by a standalone prover or at a framework's tool-call hook. Both use the receiver checks in \S\ref{sec:deployment-acceptance}.\par}

\reviewchange{\medskip
\noindent\textbf{Standalone proof generation.} The operator attaches a receipt to the outbound action. For observable transactions or API calls, the counterparty can compare the proved action with the outcome. Otherwise, dispatch remains trusted.\par}

\reviewchange{\medskip
\noindent\textbf{Proof generation at a tool-call hook.} Hooks such as AWS Strands \texttt{BeforeToolCallEvent}, OpenAI Agents tool guardrails, LangChain \texttt{wrap\_tool\_call} and AEGIS's \texttt{tool\_use} interception across 14 frameworks~\cite{aegis2026} can capture exact outbound arguments. zkARc generates the receipt there, permits dispatch after the receiver's checks pass with \textsf{VALID}, and delivers the receipt alongside the action. The receiver independently recomputes $d_X$, so mismatched arguments reject in either deployment mode.\par}

\medskip
\noindent\reviewchange{\textbf{Request construction and execution.} Acceptance does not establish that the receiving service executed the action. The following mechanisms provide different guarantees:}
\begin{enumerate}[leftmargin=*, itemsep=1pt, topsep=3pt, parsep=0pt]
    \item \reviewchange{\textit{TEE-attested execution.} A receiving-side Trusted Execution Environment can attest that specified code ran on the received action. This adds a hardware trust assumption and does not by itself establish external effects beyond the TEE.}
    \item \reviewchange{\textit{Proved request construction in a zkVM.} A zkVM can prove that a specified program constructed the request bytes from the accepted action. The proof alone establishes neither transmission nor execution by an external service.}
    \item \textit{Verified code generation.} When the agent emits a program rather than a tool call, the natural language~$\to$~specification~$\to$~verified-code pipeline (sometimes termed \textit{vericoding}~\cite{bursuc2025vericoding}) adds a compile-time correctness layer; we leave its integration to future work.
\end{enumerate}
\reviewchange{Actual transmission and external execution require separate enforcement or authenticated evidence tied to the action and session. External facts and state still require authentication.\par}

\section{Proving Model Inference with Jolt Atlas}
\label{sec:zk-jolt-atlas}

\reviewchange{Jolt Atlas~\cite{joltatlas2026} proves that a registered model produced the declared output from bound inputs. The measured implementation verifies \reviewchange{Atlas proofs built with Dory and BlindFold} directly. Section~\ref{sec:verifier-split} describes an alternative HyperKZG construction for an optional outer proof.\par}

For model $M$, input $x$, output $y$, and execution trace $\tau$, the proved
relation is
\[
R_{\mathsf{Atlas}}(M,x,y,\tau)=1
\quad\Longrightarrow\quad
y=M(x).
\]
\reviewchange{The configuration declares public values and hidden witnesses. Provenance and policy compilation combine \reviewchange{Atlas inference} with proved text/token conversions and parsing to obtain $X$ or $A_{\mathsf{NL}}$ and $\mathsf{PMC}(P_{\mathsf{NL}})=F$. An LLM action translator requires its own proofs; structured actions use deterministic extraction.\par}

\subsection{From ONNX Execution to a Succinct Proof}

\reviewchange{\reviewchange{Atlas proves the model's tensor operations directly}}~\cite{joltatlas2026}. \reviewchange{The implementation converts the fixed integer ONNX model into a registered graph of tensor operations whose execution is proved,} which fixes the shapes, parameters, lookup tables and generation rules. The prover evaluates this graph, commits to its private tensors with hiding Dory commitments, proves the operator relations with sum-checks and lookup arguments, and reduces their evaluation claims to a shared opening argument with BlindFold. The verifier checks the resulting transcript instead of replaying the model, and required Jolt proofs bind canonical token encoding and decoding to the original hidden tensors.

\reviewchange{The prover constructs the model trace and witness polynomials before proving and retains them for the batched opening; it does not stream them.\par}

\begin{figure*}[!htbp]
\centering
\begin{tikzpicture}[zkarc diagram,
    zkarc hidden/.style={zkarc box, dashed, fill=black!10}]
\node[zkarc artifact, text width=4.2cm] (model) at (4.1,0)
    {Registered model\\{\footnotesize graph, shapes, parameters}};
\node[zkarc box, text width=2.3cm] (pprep) at (9.5,0)
    {Prover\\preprocessing};
\node[zkarc box, text width=2.3cm] (vprep) at (14.9,0)
    {Verifier\\preprocessing};
\draw[zkarc arrow] (model) -- (pprep);
\draw[zkarc arrow] (model.north) -- ++(0,0.45) -| (vprep.north);

\node[zkarc hidden, text width=1.5cm] (inputs) at (1.0,-2.0)
    {Input\\tokens $x$};
\node[zkarc stage, draw=diagramgreen, fill=diagramgreen!8,
      text width=1.9cm] (eval) at (4.1,-2.0)
    {Model\\evaluation};
\node[zkarc hidden, text width=1.9cm] (witness) at (6.9,-2.0)
    {Output $y$\\and trace};
\node[zkarc stage, draw=diagramblue, text width=1.5cm] (prover) at (9.5,-2.0)
    {Atlas\\prover};
\node[zkarc artifact, draw=diagramblue, fill=diagramblue!10,
      text width=2.1cm] (proof) at (12.2,-2.0)
    {Proof $\pi$\\{\footnotesize + commitments}};
\node[zkarc stage, draw=diagramvm, fill=diagramvm!6,
      text width=1.5cm] (verifier) at (14.9,-2.0)
    {Atlas\\verifier};
\draw[zkarc arrow] (inputs) -- (eval);
\draw[zkarc arrow] (model) -- (eval);
\draw[zkarc arrow] (eval) -- (witness);
\draw[zkarc arrow] (witness) -- (prover);
\draw[zkarc arrow] (pprep) -- (prover);
\draw[zkarc arrow] (prover) -- (proof);
\draw[zkarc arrow] (proof) -- (verifier);
\draw[zkarc arrow] (vprep) -- (verifier);

\node[zkarc stage, text width=3.4cm, font=\footnotesize] (commit) at (2.0,-4.5)
    {\textbf{1. Commit}\\Hiding commitments\\per graph node};
\node[zkarc stage, text width=3.5cm, font=\footnotesize] (checks) at (6.3,-4.5)
    {\textbf{2. Prove relations}\\Operator sum-checks\\Lookup and range checks};
\node[zkarc stage, text width=3.4cm, font=\footnotesize] (opening) at (10.6,-4.5)
    {\textbf{3. Open}\\Shared opening reduction\\Dory hiding openings};
\draw[zkarc arrow] (commit) -- (checks);
\draw[zkarc arrow] (checks) -- (opening);
\node[zkarc box, text width=7.6cm, font=\footnotesize,
      minimum height=0.7cm] (hiding) at (8.45,-5.85)
    {\textbf{NovaBlindFold}: hides the sum-check rounds
     and proves the deferred verifier checks};
\begin{scope}[on background layer]
    \node[draw=diagramblue, dashed, rounded corners=3pt,
          inner sep=7pt, fit=(commit)(checks)(opening)(hiding)]
          (details) {};
\end{scope}
\node[zkarc note, text=diagramblue, anchor=north west]
    at ([xshift=4pt,yshift=-4pt]details.south west |- hiding.north)
    {Inside the\\Atlas prover};
\draw[dashed, draw=diagramblue] (prover.south west) -- (details.north west);
\draw[dashed, draw=diagramblue] (prover.south east) -- (details.north east);

\node[zkarc hidden, minimum height=0.4cm, minimum width=0.7cm,
      inner sep=0pt] (lh) at (13.7,-4.2) {};
\node[zkarc note, anchor=west] at (lh.east) {Hidden witness};
\node[zkarc artifact, minimum height=0.4cm, minimum width=0.7cm,
      inner sep=0pt] (lp) at (13.7,-4.85) {};
\node[zkarc note, anchor=west] at (lp.east) {Public};
\end{tikzpicture}
\caption{\reviewchange{Atlas model proving and verification}. Registration fixes the graph,
tensor shapes and model parameters, which are public in the measured
configuration. Tokens and intermediate values remain hidden: the prover
commits to them before deriving challenges, and the verifier receives the
proof body and the fresh tensor commitments. Required canonical conversion
proofs (not shown) bind text to the original token tensors. The alternative
HyperKZG \reviewchange{configuration} is used only in the separate verifier-split construction.}
\label{fig:jolt-atlas-pipeline-wrapper-sumchecks}
\end{figure*}
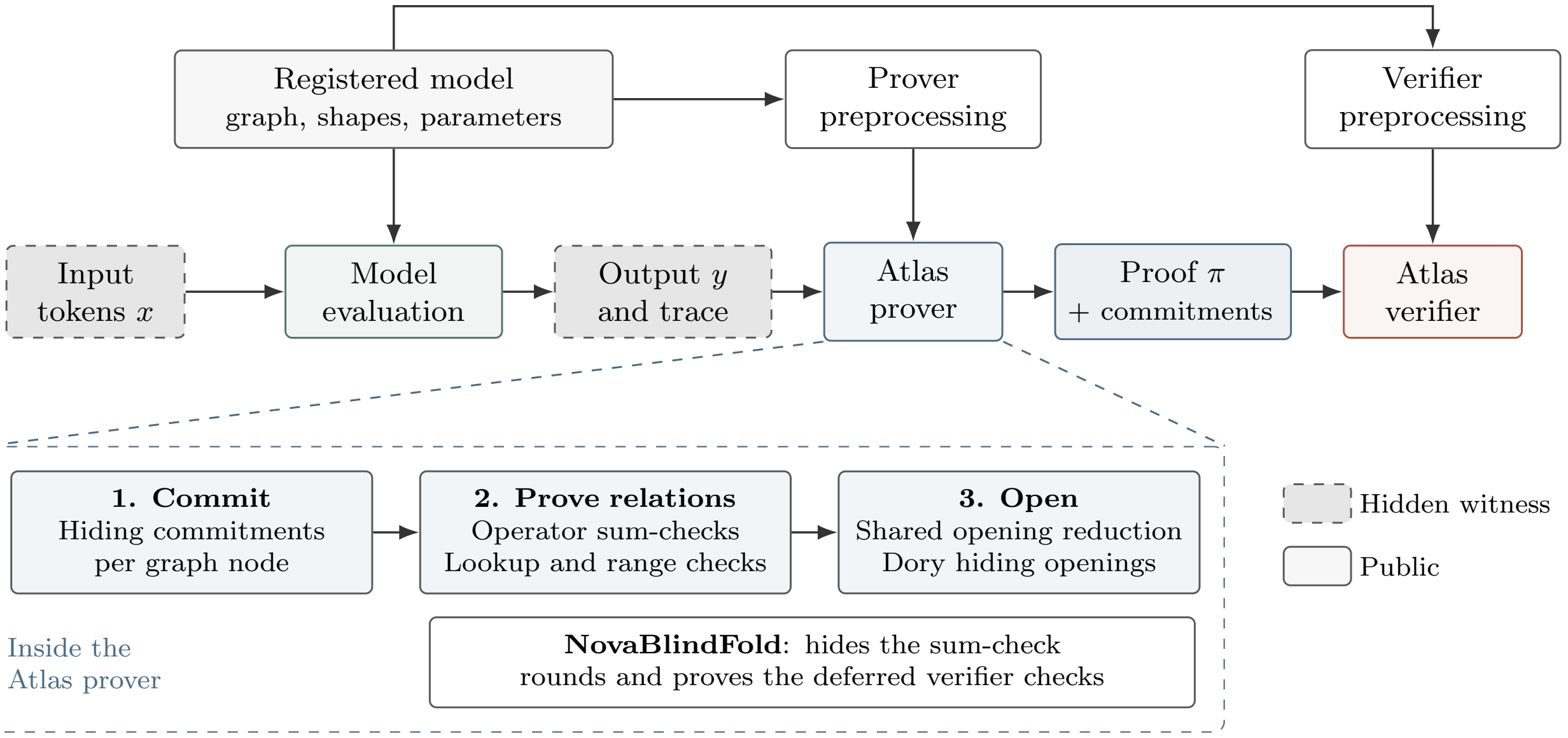

\subsection{Hiding the Inference Witness}

\reviewchange{Plain sum-check exposes witness-dependent round polynomials and evaluations. Atlas combines hiding commitments and openings with Succinct NovaBlindFold~\cite{hypernova2023,vega}: it commits to the round coefficients, encodes their consistency checks as a rank one constraint system (R1CS), folds this verifier instance with a random satisfying instance, and proves the result with Spartan. Appendix~\ref{app:blindfold-jolt-atlas} gives the construction.\par}

\reviewchange{For fixed graph and tensor shapes, randomizing the verifier R1CS costs the same regardless of which witness entries are private. Hiding tensor commitments and openings still scale with the hidden data. We do not measure hiding overhead or compare it with commit-and-prove or masking polynomials~\cite{thaler2024}.\par}

\reviewchange{The Qwen measurements use public parameters, private tensors and separate conversion proofs. Their implementation checks (Appendix~\ref{app:native-privacy}) do not establish all composition-theorem assumptions or the cost of hiding model parameters.\par}

Verifier preprocessing still reveals the operator graph and tensor shapes.
Hiding the model architecture would require additional machinery, for example
proving the verifier inside an outer proof, and is unimplemented.

\subsection{Guarantees and Assumptions}

\reviewchange{Knowledge soundness requires secure operator and lookup arguments and binding commitments. Native Dory and alternative BN254 HyperKZG are not post-quantum; HyperKZG also requires a universal, updatable powers-of-tau setup. Zero knowledge requires hiding commitments and openings and secure Fiat--Shamir compilation of BlindFold. Theorem~\ref{thm:zkarc-composition} states the simulation and extraction assumptions, which remain unproved for the combined implementation. Application assumptions are stated in \S\ref{sec:threat-model}.\par}

\section{Proving SMT Solver Execution with Jolt}
\label{sec:zk-smt-jolt}

The specified solver component proves one exact execution claim:
\[
\mathsf{Solve2}(F,P,C)=(s_+,s_-)
\quad\land\quad
v=\mathsf{Verdict}(s_+,s_-).
\]
The proof binds the formal policy, the action instance, both solver results, the derived verdict, and the solver program. It removes trust in the operator's report of what ran. It does not prove that an uncertified solver implements SMT semantics without bugs.

\subsection{The Full-Solver Construction}
\label{sec:full-solver-jolt}

The following program specifies the target relation for arbitrary supported
$(F,P,C)$. The measured \reviewchange{spending example} fixes the five policy rules and
queried claim, accepts bounded typed spending inputs, constructs both queries,
and returns both solver results. It does not parse arbitrary $(F,P,C)$ or
derive the four-way receipt verdict. The separate bounded public receipt
example also derives that verdict (\S\ref{sec:bench-public-receipt}).

The most direct approach is to compile an SMT solver to RISC-V and execute it
inside Jolt. The specified program receives canonical $F$, $P$, and $C$. It then performs the following steps:
\begin{enumerate}
    \item parse and validate the three inputs
    \item construct
    $q_+=F\land P\land C$
    and
    $q_-=F\land P\land\mathsf{CanonicalNegate}(C)$
    \item run the solver on both queries
    \item reject parse failure, panic, timeout, \textsf{unknown}, unsupported
    theory, and resource exhaustion
    \item derive $v$ from the two successful results
\end{enumerate}
Jolt proves the entire execution trace, producing $\pi_{\mathsf{SMTsolver}}$
that attests both results, the verdict and the identity of the solver program
(via commitment to the RISC-V bytecode). The public statement binds $(F,P,C)$ through the receipt edges
and exposes $(s_+,s_-,v)$ as declared outputs.

\reviewchange{The proved program constructs both queries and the verdict, ensuring that $q_-$ negates the same $C$ as $q_+$ and that the verdict follows from both results.\par}

\subsection{Full Solver or Certificate Checker}

\reviewchange{Table~\ref{tab:smt-instantiations} contrasts proved solver execution with certificate checking: the former establishes which program ran, the latter that its result has a valid certificate.\par}

\reviewchange{\reviewchange{A solver running outside the zkVM can produce certificates} for a smaller checker to verify inside the zkVM. Keeping certificates private preserves policy hiding. Checking is near-linear in certificate size, but depends on changing SMT certificate formats and does not identify the solver that produced them.\par}

\begin{table*}[t]
\centering
\footnotesize
\setlength{\tabcolsep}{6pt}
\newcommand{\dimn}[1]{\textbf{#1}}
\begin{tabularx}{\textwidth}{@{}
    >{\raggedright\arraybackslash}p{0.19\textwidth}
    >{\raggedright\arraybackslash}X
    >{\raggedright\arraybackslash}X@{}}
\toprule
& \textbf{Full solver in Jolt} & \textbf{Certificate checker in Jolt} \\
\midrule
\dimn{Proved claim} &
The committed solver constructs both queries and derives the verdict. &
The committed checker accepts evidence for both results and derives the verdict. \\
\addlinespace[2pt]
\dimn{Private witness} &
The solver execution traces. &
A SAT model or UNSAT certificate per query. \\
\addlinespace[2pt]
\dimn{Primary guarantee} &
Solver-execution binding. The receipt identifies the solver that ran. &
Certificate validity. The producing solver need not be identified or trusted. \\
\addlinespace[2pt]
\dimn{\reviewchange{What remains assumed}} &
The committed solver implements SMT semantics correctly. &
The checker is sound and the format covers every relevant theory step. \\
\addlinespace[2pt]
\dimn{Cost driver} &
Solver cycles for both queries. &
Certificate size and checker cycles, lower when compact certificates exist. \\
\addlinespace[2pt]
\dimn{Theory support} &
Any theory the committed solver implements. &
Only theories the format and checker cover. \\
\addlinespace[2pt]
\dimn{Failure behavior} &
Every non-success outcome fails closed. &
Missing, malformed or unsupported certificates fail closed. \\
\bottomrule
\end{tabularx}
\caption{Two ways to certify the solver stage with both queries.}
\label{tab:smt-instantiations}
\end{table*}
zkARc uses the full-solver construction because solver identity and execution
provenance are part of the receipt claim. A checker \reviewchange{configuration} may provide stronger
evidence of logical validity when a sound checker and complete theory
certificates exist. It does not provide solver-execution binding unless a
separate proved or authenticated relation identifies the certificate producer.

\subsection{\reviewchange{Cost and Remaining Risks}}

SMT solving is worst-case exponential, so the proving cost depends on the solver's runtime on both queries, not just the formula size. A trace longer than one Jolt bounded-cycle segment would require continuation across segments, which is future work and distinct from the folding used for zero knowledge (\S\ref{sec:zk-background}).

\reviewchange{An independent solver reduces the risk of implementation errors only if its execution and the agreement rule are proved or attested by a trusted party. Untrusted claims of agreement add no soundness, and even proved agreement cannot exclude shared bugs.\par}

\section{Composing the Guardrail Receipt}
\label{sec:proof-composition}

\reviewchange{Receipt composition prevents an operator from combining proofs that used different policies, actions or solver inputs.\par}

This section defines a receipt that combines Jolt proofs over Dory with native
Atlas proofs, gives both proof families a common verifier interface that also
supports an Atlas \reviewchange{configuration} using HyperKZG, and proves conditional completeness,
knowledge soundness and zero knowledge for the bound receipt
(\S\ref{sec:composed-security}). The construction has three layers:
\[
\begin{aligned}
\pi_i
  &\xrightarrow{\text{optional verifier extension}}
    \widehat\pi_i=(\pi_i,\pi'_i),\\
(\widehat\pi_i,\mathsf{pub}_i)_{i\in\mathcal I}
  &\xrightarrow{\text{\reviewchange{form the receipt}}}
    \Pi_{\mathsf{receipt}},\\
\Pi_{\mathsf{receipt}}
  &\xrightarrow{\text{\reviewchange{an optional outer proof}}}
    \Omega_{\mathsf{final}}.
\end{aligned}
\]
\reviewchange{The second line collects the proofs, their public inputs and outputs, and the previously fixed records linking their shared values into one receipt.} The third changes its
delivery format and verification cost, not the claim.

\subsection{Constructing the Receipt}
\label{sec:composition-protocol}

Let $\mathcal I\supseteq\{\mathsf{Provenance},\mathsf{PMC},\mathsf{Action},\mathsf{SMTsolver}\}$ be the components required by the \reviewchange{\reviewchange{configuration approved by the verifier}}. The four listed components are mandatory. A \reviewchange{configuration} may add translators or solver checks, but cannot omit provenance or any of the other required stages. For each $i\in\mathcal I$, let $\pi_i$ prove relation $R_i$, let $\pi'_i$ be its auxiliary verifier proof when the \reviewchange{configuration} uses a verifier split, and set $\widehat\pi_i=(\pi_i,\pi'_i)$. For direct verification, take $\pi'_i=\varnothing$ and $\widehat V_i=V_i^{\mathsf{base}}$. Let $\mathsf{pub}_i$ be the public component statement. The receipt is
\[
\Pi_{\mathsf{receipt}}=\bigl((\widehat\pi_i,\mathsf{pub}_i)_{i\in\mathcal I},\mathsf{bind}\bigr).
\]
The value $\mathsf{bind}$ contains one authenticated record $(e,m_e,b_e)$ for every selected graph edge, naming the edge, stating whether it is public or hidden, and carrying either a canonical public value or a commitment to a hidden value. It is data checked by the verifier, not an additional proof. The construction has three steps.

\paragraph{\reviewchange{1. Fix the configuration}}
\leavevmode\reviewchange{Before proving, the \reviewchange{configuration approved by the verifier} fixes each relation, proving system, code or model commitment, verifier key, and direct or extended verification method. Direct verification uses the base verifier; a split additionally requires $\pi'_i$ for its selected verifier equations. Both use $\widehat V_i(\mathsf{pub}_i,\pi_i,\pi'_i)$ under the refinement conditions of \S\ref{sec:verifier-split}.\par}

\paragraph{\reviewchange{2. Fix shared values and commitments}}
\leavevmode\reviewchange{As in Figure~\ref{fig:protocol-action}, compute the action and solver results, then fix the shared values and commitments before generating the online proofs.\par}
Let $Z_{\mathsf{Action}}$ be the action object fixed by the \reviewchange{configuration}. In the main structured \reviewchange{configuration}, $Z_{\mathsf{Action}}=X$. In \reviewchange{a configuration with LLM action translation}, it may be a plan in natural language, but the action relation must also bind that plan to $X$. The selected edges enforce
\[
\begin{aligned}
Z_{\mathsf{Action}}^{(\mathsf{Provenance})} &=Z_{\mathsf{Action}}^{(\mathsf{Action})},\\
(P,C)^{(\mathsf{Action})} &=(P,C)^{(\mathsf{SMTsolver})},\\
F^{(\mathsf{PMC})} &=F^{(\mathsf{SMTsolver})}.
\end{aligned}
\]
All three edges are required. A detached provenance proof or a receipt without the provenance component fails receipt verification.

For a public edge, both component statements contain the same canonical value. For a hidden edge other than the cached policy, both component relations open the same public, domain-separated commitment $c_e$ to the value they use. Binding prevents two different hidden values from satisfying one edge, and hiding conceals the shared value. The \reviewchange{\reviewchange{configuration approved by the verifier}} fixes the commitment scheme, parameters, type tags, contexts, and encodings. A change of representation, such as tokens to text, adds a proved conversion relation. The measured Qwen chain keeps its original tensor commitments, which need not be identical, and uses a joint conversion proof to require equal decoded and encoded text.

\reviewchange{Online commitments include the receipt session. The cached policy proof retains $c_F^{\mathsf{root}}$ and its stable context. The SMT proof opens this root to its policy $F$ and, if it uses a fresh commitment, proves that it contains the same $F$ (Definition~\ref{def:zkarc-relation}).\par}

\paragraph{\reviewchange{3. Generate proofs and form the receipt}}
\leavevmode\reviewchange{Generate each online $\pi_i$ using the fixed statements and commitments, and reuse the registered cached proofs. Native Atlas proves inference (\S\ref{sec:zk-jolt-atlas}); Jolt proves extraction and solver execution (\S\ref{sec:zk-smt-jolt}). An LLM translation configuration also requires its translator proofs and \reviewchange{the link between the plan and the action}. For online components using a split, generate $\pi'_i$ from the completed base proof and transcript. \reviewchange{Collect the proofs, public statements and binding records into} $\Pi_{\mathsf{receipt}}$.\par}

\paragraph{Receipt verification}
The verifier first evaluates $\mathsf{check\_public}(u,\Pi_{\mathsf{receipt}})$. This check requires the receipt's component public statements and binding records to equal those in $u$. It also requires every repeated public field to agree with the corresponding component statements, namely contexts, graph and verifier identifiers, \reviewchange{identifiers for the policy and shared values}, $c_Q$, $d_X$, $S$, $d_S$, solver results, and verdict. In particular, the top-level $(s_+,s_-,v)$ must equal the solver component's certified output. The \reviewchange{\reviewchange{configuration approved by the verifier}} fixes these field projections and their canonical encodings. The verifier then
\begin{enumerate}
    \item runs $\widehat V_i(\mathsf{pub}_i,\pi_i,\pi'_i)$ for every selected component,
    \item requires exactly the binding records named by the authenticated graph and checks them against the component statements according to Definition~\ref{def:zkarc-relation}, and
    \item accepts only if $\mathsf{check\_public}$, every component and every edge accept.
\end{enumerate}
We call this algorithm $V_{\mathsf{receipt}}$. Under the hypotheses of Theorem~\ref{thm:zkarc-composition}, its acceptance establishes the composed relation of Definition~\ref{def:zkarc-relation}. It does not authorize dispatch, which the enforcement point decides separately with $\mathsf{Accept}_{\mathsf{deploy}}$ (\S\ref{sec:deployment-acceptance}).

\begingroup\reviewcolor
\subsection{\reviewchange{Verifier Splitting for Recursive Proofs}}
\label{sec:verifier-split}
\label{sec:atlas-hyperkzg}

\reviewchange{Following Dao et al.~\cite{joltrecursion2026}, we reduce the cost of recursive verification by proving selected expensive verifier computations separately. Sum-check round consistency, evaluation claim chaining and Fiat--Shamir consistency checks use the native field $\mathbb F_r$ of Jolt and Jolt Atlas, with \mbox{$O(n_s d_s)$} field operations per sum-check of $n_s$ rounds and degree $d_s$, apart from transcript hashing. Dory's commitment-opening checks instead use scalar multiplications in $G_1$ and $G_2$, exponentiations in $G_T$ and a size-$3$ multi-pairing. Their coordinates lie in $\mathbb F_q$, $\mathbb F_{q^2}$ or $\mathbb F_{q^{12}}$, whose arithmetic is costly inside an outer proof over $\mathbb F_r$.}

\reviewchange{An auxiliary proof $\pi'_i$ over Grumpkin certifies selected curve equations. Its two-cycle with BN254 makes the relevant base-field arithmetic native. For an $n$-coefficient polynomial arranged as a $2^\nu\times2^\sigma$ matrix, Dory~\cite{lee2021dory} contributes $\sigma$ rounds of $G_T$, $G_1$ and $G_2$ accumulator updates. In the partition used here, the final pairing remains a direct verifier check.}

\paragraph{\reviewchange{Extension to HyperKZG.}}
\leavevmode\reviewchange{zkARc extends this method to an alternative Atlas configuration using HyperKZG. The measured Atlas configuration uses hiding Dory commitments and direct verification; the extension supports an optional outer proof. We write}
\[
\reviewchange{V_{\mathsf{Atlas}}=V_{\mathsf{IT}}\land V_{\mathsf{HyperKZG}}.}
\]
\reviewchange{$V_{\mathsf{IT}}$ contains the transcript, sum-check, lookup, consistency and BlindFold checks. $V_{\mathsf{HyperKZG}}$ checks the commitment opening: a $G_1$ joint-commitment fold and two batched linear combinations, $L$ and $R$, followed by}
\[
\reviewchange{e(L,[1]_2)\,e(-R,[\tau]_2)=1.}
\]
\reviewchange{The auxiliary proof certifies the selected $G_1$ equations, while the verifier checks $V_{\mathsf{IT}}$ and the final pairing directly. Unlike Dory, this requires no $G_T$ accumulator updates or $G_2$ scalar-multiplication PIOP. The batching scalars come from the base Fiat--Shamir transcript. The statement, transcript and base-proof format are unchanged.}

\reviewchange{Figure~\ref{fig:verifier-split} and Table~\ref{tab:split} show how the two verifiers divide these checks. Appendices~\ref{app:verifier-ops-dory} and~\ref{app:verifier-ops-hyperkzg} give the equations and operation counts; Appendix~\ref{app:efficient-recursion} describes packing, wiring and Hyrax openings. Appendix~\ref{sec:bench-atlas-pi-prime} reports the HyperKZG auxiliary implementation.}

\begin{figure*}[!htbp]
\centering
\begin{tikzpicture}[zkarc diagram]
\node[zkarc artifact, text width=4.2cm] (dory) at (4.6,0)
    {\textbf{Dory verifier}\\{\footnotesize split of Dao et al.}};
\node[zkarc artifact, text width=4.2cm] (hyperkzg) at (11.3,0)
    {\textbf{HyperKZG verifier}\\{\footnotesize split extended in zkARc}};

\node[zkarc stage, text width=10.6cm] (transcript) at (7.95,-1.6)
    {Bind the public statement and replay the base transcript\\
     Derive the operands and challenges used by all checks};
\draw[zkarc arrow] (dory.south) -- (dory.south |- transcript.north);
\draw[zkarc arrow] (hyperkzg.south) -- (hyperkzg.south |- transcript.north);

\node[zkarc box, text width=3.7cm] (native) at (2.3,-3.55)
    {\textbf{Native checks}\\{\footnotesize transcript, field equations}};
\node[zkarc stage, draw=diagramblue, text width=3.7cm] (aux) at (7.95,-3.55)
    {\textbf{Auxiliary proof $\pi'_i$}\\{\footnotesize selected curve equations}};
\node[zkarc box, text width=3.7cm] (pairing) at (13.6,-3.55)
    {\textbf{Direct check}\\{\footnotesize final pairing}};
\draw[draw=black!80, line width=0.65pt] (transcript.south) -- (7.95,-2.6);
\draw[zkarc arrow] (7.95,-2.6) -| (native.north);
\draw[zkarc arrow] (7.95,-2.6) -- (aux.north);
\draw[zkarc arrow] (7.95,-2.6) -| (pairing.north);
\draw[zkarc arrow, draw=diagramblue] (aux.east) -- (pairing.west)
    node[midway, below, font=\footnotesize, text=diagramblue, align=center,
         inner sep=3pt] {certified\\outputs};

\node[zkarc artifact, text width=6.0cm] (result) at (7.95,-5.75)
    {Accept only if all checks pass\\
     $\widehat V_i(\mathsf{pub}_i,\pi_i,\pi'_i){=}1$};
\draw[draw=black!80, line width=0.65pt] (native.south) |- (7.95,-4.8);
\draw[draw=black!80, line width=0.65pt] (pairing.south) |- (7.95,-4.8);
\draw[zkarc arrow] (aux.south) -- (result.north);
\end{tikzpicture}
\caption{The target common verifier interface under the partition in
Table~\ref{tab:split}. The verifier split for Dory follows
Dao et al.~\cite{joltrecursion2026}, and zkARc extends it to HyperKZG. The
auxiliary statement binds the operands derived from the base transcript, and
the direct pairing consumes its certified outputs. The current artifact does
not fully implement the binding hypotheses of the refinement lemma. Receipt
binding and deployment acceptance are separate checks.}
\label{fig:verifier-split}
\end{figure*}
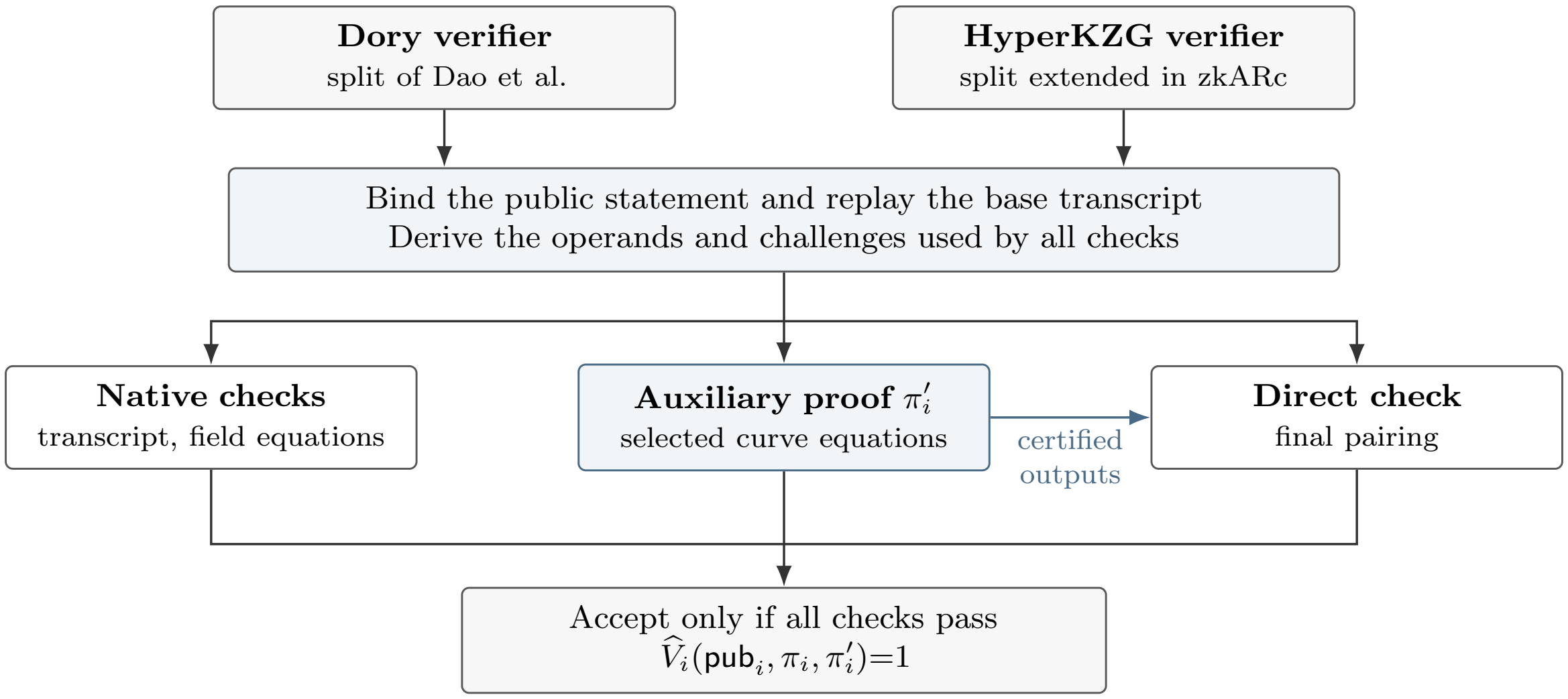

\begin{table*}[t]
\centering
\footnotesize
\setlength{\tabcolsep}{5pt}
\begin{tabularx}{\textwidth}{@{}p{0.10\textwidth}X p{0.32\textwidth}p{0.18\textwidth}@{}}
\toprule
\textbf{System} & \textbf{Checked natively} & \textbf{Certified by $\pi'_i$} & \textbf{Checked directly} \\
\midrule
Dory
& Sum-checks, evaluation claims, and Fiat--Shamir challenges.
& Packed $G_1$ and $G_2$ group operations, packed $G_T$ operations, wiring, and one Hyrax opening.
& Final size-$3$ multi-pairing. \\
HyperKZG
& Joint-evaluation reduction, folding equations, and Fiat--Shamir challenges.
& The $G_1$ joint-commitment fold and the final linear combinations $L$ and $R$, with one Hyrax opening.
& $e(L,[1]_2)e(-R,[\tau]_2)=1$. \\
\bottomrule
\end{tabularx}
\caption{Verifier partition used by the extended Dory and HyperKZG proofs.}
\label{tab:split}
\end{table*}

\paragraph{\reviewchange{Security conditions.}}
\leavevmode The auxiliary proof extends the base Fiat--Shamir transcript. Its prefix binds the extension tag, component and relation versions, verifier-key digest, public statement, and digest of the complete base transcript, and it absorbs every auxiliary commitment before deriving a challenge. A separate transcript domain for each component prevents an auxiliary proof for one statement or version from being replayed as another.

\begin{lemma}[Verifier-split refinement]
\label{lem:verifier-split-refinement}
For component $i$, partition the original verifier equations into $\mathcal C_i^{\mathsf{IT}}$, checked directly as native-field equations, $\mathcal C_i^{\mathsf{dir}}$, checked directly as cryptographic equations, and $\mathcal C_i^{\mathsf{aux}}$, certified by $\pi'_i$. Assume that these sets cover every base-verifier equation exactly once. Assume that the base and extended verifiers derive the same challenges from the same public statement and base transcript. Assume further that the operands of every equation in $\mathcal C_i^{\mathsf{aux}}$ (proof group elements and transcript-derived scalars) are public inputs of the auxiliary statement computed by $\widehat V_i$ from $(\mathsf{pub}_i,\pi_i)$ and the shared transcript, and that the certified outputs of $\mathcal C_i^{\mathsf{aux}}$ are the exact values $\widehat V_i$ consumes in $\mathcal C_i^{\mathsf{dir}}$. If the auxiliary proof soundly certifies its packing, wiring, EC-PIOP, and Hyrax-opening checks with error $\varepsilon_i$, then
\[
\Pr\bigl[\widehat V_i(\mathsf{pub}_i,\pi_i,\pi'_i)=1\land V_i^{\mathsf{base}}(\mathsf{pub}_i,\pi_i)=0\bigr]\leq\varepsilon_i.
\]
If the base proof has knowledge error $\kappa_i$, then $\widehat\kappa_i\leq\kappa_i+\varepsilon_i$. If the base and auxiliary honest provers have completeness errors $\delta_i$ and $\delta'_i$, then $\widehat\delta_i\leq\delta_i+\delta'_i$. Zero knowledge requires a joint simulator for $(\pi_i,\pi'_i)$. The next lemma supplies one for the public verifier extensions used here.
\end{lemma}

\begin{proof}
The extended verifier evaluates $\mathcal C_i^{\mathsf{IT}}\cup\mathcal C_i^{\mathsf{dir}}$ itself. Auxiliary soundness reduces the packed claim to every equation in $\mathcal C_i^{\mathsf{aux}}$. Wiring fixes their internal inputs and outputs, and the boundary hypothesis fixes their operands and the values consumed by $\mathcal C_i^{\mathsf{dir}}$, so the certified equations are the base verifier's equations on the base verifier's values. Transcript identity fixes their challenges and public statement. Since the three sets cover the base verifier exactly once, an extended proof that accepts while the base verifier rejects breaks the auxiliary argument except with probability $\varepsilon_i$. The remaining bounds follow from the base extractor and union bounds.
\end{proof}

\begin{lemma}[Zero knowledge under public verifier extension]
\label{lem:public-extension-zk}
Suppose the base proof $\pi_i$ is adaptive multi-theorem zero knowledge with advantage $\zeta_i$. Suppose further that $\mathsf{AuxProve}_i(\mathsf{pub}_i,\pi_i,\rho_i)$ generates $\pi'_i$ from the public statement, base proof and transcript, public verifier state, and fresh randomness $\rho_i$, without using the witness for $R_i$. Then $(\pi_i,\pi'_i)$ is jointly adaptive multi-theorem zero knowledge with advantage at most $\zeta_i$, in the same programmable-random-oracle model.
\end{lemma}

\begin{proof}
Simulate $\pi_i$, reconstruct the public verifier state from it, and run $\mathsf{AuxProve}_i$ honestly with fresh $\rho_i$ in the simulated transcript. Conditioned on the simulated base proof, the distribution of auxiliary proofs is identical to the real conditional distribution because auxiliary proving uses no relation witness. The distinguishing gap is therefore only that of the base simulator.
\end{proof}

The verifier split of Dao et al.~\cite{joltrecursion2026} meets the no-witness condition of Lemma~\ref{lem:public-extension-zk}, because the EC-operation trace is reconstructed from the public proof and verifier transcript. Verification of the resulting $\pi'_i$ reduces to two Grumpkin multi-scalar multiplications, and the extended verifier also evaluates the one Dory or HyperKZG pairing in $\mathcal C_i^{\mathsf{dir}}$.
\par\endgroup

\subsection{Security of the Composed Receipt}
\label{sec:composed-security}

\reviewchange{\paragraph{Main result}
We prove completeness, knowledge soundness and zero knowledge for the composed relation under the assumptions below. Application assumptions remain as stated in \S\ref{sec:threat-model}.\par}

For each $i\in\mathcal I$, let
$\widehat\pi_i=(\pi_i,\pi'_i)$ be its extended proof, let $\widehat V_i$ be its
extended verifier, and let $R_i(\mathsf{pub}_i,w_i)$ be its relation. A logical
component may contain more than one proof: the inference components include
their required canonical input and output conversions, and the PMC component
includes the registered interpretation of the generated predicate as the
formal policy $F$. The required graph contains
\[
\begin{aligned}
R_{\mathsf{Prov}} &: M_{\mathsf{Action}}(Q)=Z_{\mathsf{Action}}\\
&\quad\land \mathsf{Open}_Q(c_Q,Q,r_Q,\mathsf{ctx}_X)=1,\\
R_{\mathsf{PMC}} &: \mathsf{PMC}(P_{\mathsf{NL}})=F,\\
R_{\mathsf{Action}} &: \mathsf{ActionRel}_{\mathsf{profile}}
  (Z_{\mathsf{Action}},X,S)=(P,C),\\
R_{\mathsf{SMT}} &: \mathsf{Solve2}(F,P,C)=(s_+,s_-)\\
&\quad\land v=\mathsf{Verdict}(s_+,s_-).
\end{aligned}
\]
Here $\mathsf{Open}_Q$ checks the prompt-commitment equation from
\reviewchange{Appendix~\ref{app:protocol-encodings}}, and $Q$ and $r_Q$ are part of the provenance
witness. Checking that $c_Q$ equals the receiver's authorized expectation is
the separate deployment check.
In the main structured \reviewchange{configuration}, $Z_{\mathsf{Action}}=X$ and
$\mathsf{ActionRel}$ applies deterministic schema extraction. In \reviewchange{a configuration with LLM action translation}, $Z_{\mathsf{Action}}=A_{\mathsf{NL}}$ and the relation proves the
declared translation and its binding to $X$. In every \reviewchange{configuration}, the action
relation binds canonical $X$ to $d_X$ and produces the exact $(P,C)$ consumed
by the solver.
Let $E_{\mathcal I}$ be the selected edges of the computation graph. Each edge
$e=(i,j,z_e)$ has an injective canonical encoding $\mathsf{enc}_e$, a type tag
$\mathsf{tag}_e$, and a context $\mathsf{ctx}_e$. The \reviewchange{\reviewchange{configuration approved by the verifier}}
fixes a concrete randomized commitment $\mathsf{Com}_e$ and its public
parameters for each hidden edge. Per-action contexts contain the fresh session
identifier $\mathsf{sid}$. The cached-policy context contains the owner, policy
identifier, version, and epoch. In public mode, adjacent statements contain the
same encoded value. For hidden edges other than the cached policy, the binding
record and both statements contain
\[
c_e=\mathsf{Com}_e\!\left(
 \mathsf{EncodeTuple}(\mathsf{tag}_e,\mathsf{ctx}_e,
 \mathsf{enc}_e(z_e)),r_e\right),
\]
and both relations prove an opening of $c_e$ to the value they use. The cached
policy proof and its statement retain the authorized $c_F^{\mathsf{root}}$.
The SMT relation opens this root in the cached-policy context to the $F$ it uses.
If it also uses a fresh $c_e$, only the SMT statement contains that commitment,
and the SMT relation opens it in the receipt context to the same $F$.
Otherwise, $c_e=c_F^{\mathsf{root}}$. The cached policy proof is reused unchanged.

\begin{definition}[Composed receipt relation]
\label{def:zkarc-relation}
The public statement $u$ contains the session identifier, \reviewchange{graph of required computations}, and the exact binding records $\mathsf{bind}$. It also contains the
prompt commitment $c_Q$, the \reviewchange{action hash} $d_X$, the \reviewchange{state $S$ supplied by the receiver and its hash $d_S$}, \reviewchange{the contexts and identifiers for shared values}, the component and
verification-key identifiers, the component public statements, the public
parameters, the solver results $(s_+,s_-)$, and the verdict $v$. The witness $w$ contains the component
witnesses, including $Q$ and $r_Q$, together with all hidden edge values and openings.

The pair $(u,w)$ belongs to $R_{\mathsf{receipt}}$ exactly when
\begin{enumerate}
    \item every repeated public field in $u$ agrees with its corresponding component statement under the \reviewchange{field projections specified by the configuration approved by the verifier},
    \item the graph contains all four required components and all three shared-value edges, and every relation $R_i(\mathsf{pub}_i,w_i)$ holds,
    \item the provenance relation opens $c_Q$ in the receipt context to the exact $Q$ used by the planner,
    \item every public edge has equal canonical encodings, and every hidden edge other than the cached policy is opened inside both incident relations to the same public commitment recorded in $\mathsf{bind}$,
    \item for the hidden policy edge, the PMC and SMT relations open the authorized $c_F^{\mathsf{root}}$ in the cached-policy context to the $F$ each uses. If $\mathsf{bind}$ records a fresh $c_e$, the SMT statement also contains $c_e$ and its relation opens it in the receipt context to the same $F$,
    \item the action relation proves a canonical $X$ \reviewchange{whose hash is $d_X$}, checks that $S$ has the \reviewchange{state hash} $d_S$ in that action context, and binds its output $(P,C)$ to the exact solver input according to the selected \reviewchange{configuration}, and
    \item the SMT outputs are $(s_+,s_-)$ and the public verdict is $v=\mathsf{Verdict}(s_+,s_-)$.
\end{enumerate}
\end{definition}

\begin{lemma}[Exact-action binding]
\label{lem:action-handle-binding}
Suppose the encoding of structured actions and typed tuple encoding are injective,
and $H$ is collision resistant. If a receipt witness contains $X$ and the
receiver accepts $X_{\mathsf{recv}}$ under the same action context because both
recompute to $d_X$, then
\[
X=X_{\mathsf{recv}}
\]
except with probability at most $\mathsf{Adv}^{\mathsf{cr}}_H$.
\end{lemma}

\begin{proof}
If the actions differ, injectivity gives two distinct typed hash inputs. \reviewchange{Equality of their hashes} is then a collision in $H$.
\end{proof}

\noindent This lemma binds the receipt to the action received by the verifier. It does not prove that the same action later executes downstream.

Partition the components as
$\mathcal I=\mathcal I_{\mathsf{on}}\mathbin{\dot\cup}
\mathcal I_{\mathsf{cache}}$. Online proofs are created for the current
receipt. Cached proofs are accepted once during registration. Normally
$\mathcal I_{\mathsf{cache}}=\{\mathsf{PMC}\}$.

A cached component is \emph{registration-extractable} with error $\chi_i$ if
an extractor with access to registration obtains a witness $w_i$ satisfying
$R_i(\mathsf{pub}_i,w_i)$ whenever registration accepts
$(\mathsf{pub}_i,\widehat\pi_i)$, except with probability $\chi_i$. This is
stronger than ordinary soundness for a proof that another party may later
reuse.

\paragraph{Compatible simulation}
The component simulators receive the fixed public parameters and statements
of Definition~\ref{def:aok}. They share one record of random oracle queries
and answers. Each simulator may program only its
own proof domain and must preserve answers already returned. Its error bound
includes any failure caused by a prior query. The component zero-knowledge
bounds must hold with the other component transcripts and shared public data
as auxiliary input. Domain labels alone do not establish these properties.

The specified zkARc Fiat--Shamir transcripts start with an injectively encoded prefix.
The prefix contains the protocol tag, component and relation versions,
verification-key digest, public-statement digest, and either $\mathsf{sid}$ or
the stable cached context. The transcript absorbs every commitment before it
derives the challenge that binds that commitment. The auxiliary transcript
also absorbs its extension tag and the digest of the complete base transcript before
deriving its challenges.

\begin{theorem}[Security of the composed receipt]
\label{thm:zkarc-composition}
Assume the following conditions.
\begin{enumerate}[label=\textup{(H\arabic*)}, leftmargin=*, itemsep=2pt, topsep=4pt]
    \item \textup{\textbf{Base proofs.}} Each base proof is complete and
    adaptive multi-theorem zero knowledge under the common setup and random
    oracle, with compatible simulators as specified above, and any auxiliary
    argument used by the \reviewchange{configuration} is complete and satisfies the
    public-extension condition of Lemma~\ref{lem:public-extension-zk}.
    \item \textup{\textbf{Extraction.}} Online base proofs are adaptive
    multi-theorem knowledge sound with straight-line extractors in the stated
    security model, while cached extended proofs are registration-extractable.
    \item \textup{\textbf{Verifier splits.}} Every verifier split used by the
    \reviewchange{configuration} satisfies Lemma~\ref{lem:verifier-split-refinement}.
    \item \textup{\textbf{Transcripts.}} Every transcript is domain separated
    and binds its relation version, verification key, public statement,
    context, and prior commitments.
    \item \textup{\textbf{Commitments.}} Each commitment on a hidden edge and
    the prompt commitment $c_Q$ is adaptively binding and hiding over its
    encoded-message domain.
    \item \textup{\textbf{Encodings.}} All canonical encodings are injective
    and enforced inside the incident relations.
    \item \textup{\textbf{Freshness.}} Honest receipt generation uses fresh
    session identifiers. Each new edge commitment samples its randomness after
    fixing its message, independently of setup and proof randomness and the
    randomness of all other commitments.
    \item \textup{\textbf{Authentication.}} Verifier policy authenticates the
    required graph, verdict predicate, verifier keys, model and program
    commitments, encodings, preprocessing, and cached \reviewchange{policy commitment}.
\end{enumerate}

\smallskip\noindent
For direct verification, the extended bounds below equal the base bounds and
the auxiliary error is zero. For a component with a split, let
$\widehat\delta_i$ and $\widehat\kappa_i$ be the bounds
obtained from Lemma~\ref{lem:verifier-split-refinement}, and let
$\widehat\zeta_i$ be the joint zero-knowledge bound from
Lemma~\ref{lem:public-extension-zk}. Let
$E_{\mathcal I}^{\mathsf{hid}}\subseteq E_{\mathcal I}$ be the hidden edges and
let
$E_{\mathcal I}^{\mathsf{fresh,hid}}
 \subseteq E_{\mathcal I}^{\mathsf{hid}}$
be those whose commitments are sampled for this receipt. Then:
\begin{enumerate}[label=\textup{(\roman*)}, leftmargin=*, itemsep=2pt, topsep=4pt]
    \item \textup{\textbf{Completeness.}} The receipt is complete for
    $R_{\mathsf{receipt}}$ with error at most
    \[
    \sum_{i\in\mathcal I}\widehat\delta_i .
    \]
    \item \textup{\textbf{Knowledge soundness.}} It is knowledge sound for
    $R_{\mathsf{receipt}}$ with error at most
    \[
    \sum_{i\in\mathcal I_{\mathsf{on}}}\widehat\kappa_i
    +\sum_{i\in\mathcal I_{\mathsf{cache}}}\chi_i
    +\sum_{e\in E_{\mathcal I}^{\mathsf{hid}}}
      \mathsf{Adv}^{\mathsf{bind}}_e .
    \]
    \item \textup{\textbf{Zero knowledge.}} Conditioned on the public
    statement and registered cache, it is zero knowledge with advantage at
    most
    \[
    \sum_{i\in\mathcal I_{\mathsf{on}}}\widehat\zeta_i .
    \]
    If registration is part of the simulated view, add
    $\sum_{i\in\mathcal I_{\mathsf{cache}}}\widehat\zeta_i$ and simulate each
    cached proof once for all later reuse.
    \item \textup{\textbf{Matched-leakage privacy.}} For two valid executions
    with identical declared leakage, a fixed registered cache and the same
    authenticated expected prompt commitment, their receipt views are
    indistinguishable with advantage at most
    \[
    2\sum_{i\in\mathcal I_{\mathsf{on}}}\widehat\zeta_i
    +\sum_{e\in E_{\mathcal I}^{\mathsf{fresh,hid}}}
      \mathsf{Adv}^{\mathsf{hide}}_e .
    \]
    If registration views are also compared, add
    $2\sum_{i\in\mathcal I_{\mathsf{cache}}}\widehat\zeta_i$.
\end{enumerate}

\smallskip\noindent
Declared leakage includes the selected graph, public parameters, code and model
identifiers, public action $X$ and receiver state $S$, other intentionally
public values, verdict, prompt commitment $c_Q$, stable \reviewchange{policy commitments}, cached
proofs, persistent identifiers, and unpadded length and proof-size metadata.
The matched-leakage comparison holds the expected prompt commitment fixed.
It does not compare different prompt registrations. Zero knowledge hides its
opening conditional on this public commitment.
\end{theorem}

\begin{proof}
\smallskip\noindent\textit{Completeness.}
Run every honest component prover on a global satisfying witness. Each extended
verifier accepts except with probability $\widehat\delta_i$. Public bindings
are equal by construction. Hidden bindings open correctly inside both incident
relations, using the stable root for the policy edge. Any fresh policy commitment
opens to the same $F$ inside the SMT relation. The deterministic binding and verdict checks therefore accept. A
union bound gives the stated error.

\smallskip\noindent\textit{Knowledge soundness.}
Run the adaptive extractors against every online component and the registration
extractors against every cached component.
Except with the stated extraction errors, this produces one witness for every
component. Because the extractors are straight-line, all of them run on the
single adversarial transcript without nested rewinding, so their failure events
combine by a union bound rather than multiplicatively. The transcript and extraction hypotheses bind each recovered witness to its
authenticated component statement and context.

The public-statement check fixes the extracted component statements to $u$
and equates their repeated fields with its top-level fields. On a public edge,
verifier equality and injective encoding make the extracted
values equal. On a hidden edge, both extracted witnesses open the same
commitment under the same tag and context: for the policy edge, this is
$c_F^{\mathsf{root}}$ in the cached-policy context. Different extracted values
would break binding. If the SMT statement also contains a fresh policy commitment,
its extracted witness opens that commitment to the same $F$ as the root. The extracted
provenance witness opens $c_Q$ to the planner's exact input. The extracted action recomputes to $d_X$. The action relation uses the public
\reviewchange{\reviewchange{state supplied by the receiver} $S$ with hash $d_S$} and yields the exact $(P,C)$ in the
extracted solver witness. The component
witnesses therefore form one witness for $R_{\mathsf{receipt}}$.

\smallskip\noindent\textit{Zero knowledge.}
Keep the public registered cache fixed. Replace each online extended proof in
turn with its joint simulator output. Each simulator receives its component's
public statement, including the values and commitments specified above. The compatible simulators share one setup and
oracle table. Each replacement preserves the other proof domains and all
previously returned answers, with failures charged to its component bound.
A hybrid argument gives the first zero-knowledge bound. If
registration is hidden, simulate each cached proof once and reuse it.

\smallskip\noindent\textit{Matched-leakage privacy.}
First simulate the online proofs for the first execution. Replace each fresh
hidden commitment once across the whole graph with the corresponding fresh
commitment for the second execution. Regenerate all incident simulated proofs
after each replacement. Independent fresh randomness lets adaptive
commitment hiding bound each change even when edge messages are correlated. Finally replace the simulated proofs
with real proofs for the second execution. The forward and reverse simulation
steps contribute the factor of two.
\end{proof}

\paragraph{Cached proofs}
An online operator may copy a registered PMC proof without knowing its witness. Full receipt knowledge soundness therefore requires extraction during registration, a retained registration witness, or another extraction property that applies to the registered proof. Without it, the remaining hypotheses still support completeness, zero knowledge and extraction of the online component witnesses, but not extraction of a global witness that includes the cached policy witness, because ordinary soundness of a cached proof establishes existence of its witness without supplying it.

\paragraph{\reviewchange{What the theorem establishes}}
The theorem proves execution of the specified component relations and
consistency of their inputs and outputs. It does not establish semantic
faithfulness of translation from natural language, mathematical correctness of
an uncertified solver, or downstream execution. Authorization, freshness,
required-component selection, verdict gating and comparison with the received
action are enforced by the deployment checks in
\S\ref{sec:deployment-acceptance}.

The result is conditional. \reviewchange{The implementation produces and verifies the complete Qwen receipt for the tested configuration with public model parameters. It verifies both native Atlas proofs and every required Jolt proof directly. Its receiver checks spending and reserves funds in one transaction. Acceptance therefore does not depend on the optional auxiliary split.} \reviewchange{We have not proved that the implementation satisfies all the theorem's assumptions about component proofs, commitments, the shared random oracle and registration.} \reviewchange{Successful verification and tests that substitute different proofs or inputs show how the implementation behaves on those inputs. They do not establish the required simulation or extraction properties}. Private model parameters and a single outer proof of the complete receipt remain open.

At each \reviewchange{connection between an Atlas proof and a Jolt proof}, the conversion witness is committed before
the evaluation point is derived, and \reviewchange{the Atlas opening} and the conversion
relation check the same hidden tensor evaluation. A concrete security
reduction must account for equality testing at that point, its Fiat--Shamir
challenges, and the hiding of the salted hash commitments; collision
resistance alone supplies no hiding guarantee. \reviewchange{The theorem assumes these properties of the components and commitments. The implementation tests do not prove them.}

\subsection{\reviewchange{Proving Receipt Verification with One Outer Proof}}
\label{sec:wrap}

\reviewchange{The receipt is already complete as a bundle of proofs. An optional outer proof certifies that the receipt verifier accepted, replacing the bundle with one proof}
$\Omega_{\mathsf{final}}$. It changes the delivery format and verifier cost while preserving the intended
receipt relation. Transferring the full knowledge guarantee requires the
nested-extraction condition stated below. The outer relation is
\[
R_{\mathsf{final}}(u,\Pi_{\mathsf{receipt}})
 := [V_{\mathsf{receipt}}(u,\Pi_{\mathsf{receipt}})=1].
\]
\reviewchange{The outer proof preserves the public statement $u$ of the receipt, including all shared values and contexts. The receiver must still perform the authorization checks in \S\ref{sec:deployment-acceptance}.\par}

\begin{corollary}[\reviewchange{Security of the outer proof}]
\label{cor:zkarc-finalisation}
\reviewchange{A complete and sound outer proof system} proves that an accepting inner receipt
exists. \reviewchange{If the outer proof system is knowledge sound, its extractor recovers that receipt.} It is a proof of
knowledge for $R_{\mathsf{receipt}}$ if the component and registration
extractors from Theorem~\ref{thm:zkarc-composition} remain applicable to the
inner proofs recovered by the outer extractor. If the outer proof hides the
inner receipt, \reviewchange{the outer proof system} must also be zero knowledge.
\end{corollary}

\begin{proof}
Completeness follows by using an honestly generated receipt as the outer
witness. The outer extractor recovers an accepting
$\Pi_{\mathsf{receipt}}$. Under the nested extraction condition, the component
and registration extractors then recover a witness for
$R_{\mathsf{receipt}}$. The total knowledge error is at most the outer
extraction error plus the receipt extraction error. Without the nested
condition, extraction stops at the accepting proof tuple. Zero knowledge follows
from \reviewchange{the outer proof simulator} when the tuple is hidden. If the tuple is also
delivered, its privacy follows from
Theorem~\ref{thm:zkarc-composition}.
\end{proof}

\reviewchange{Following the terminology of~\cite{joltrecursion2026}, zkARc would support two ways to produce an outer proof:} \emph{recursion}, \reviewchange{which proves all extended verifier computations and binding checks inside Jolt itself}, and \emph{wrapping}, \reviewchange{which proves the verifier computation with another SNARK} for deployments whose end verifier cannot tolerate Jolt's proof size (smart-contract calldata, on-chain pairing verifiers).\footnote{\reviewchange{A wrapping implementation} would prove the same $R_{\mathsf{final}}$ with a downstream SNARK. Spartan with HyperKZG avoids a circuit-specific ceremony but is not transparent, and Groth16~\cite{groth16} gives a smaller proof but requires a circuit-specific setup. Either must satisfy the conditions of Corollary~\ref{cor:zkarc-finalisation}.}

\reviewchange{A recursive implementation} would compile
\[
\begin{aligned}
\mathsf{verify}(u,\Pi_{\mathsf{receipt}})
 &= \mathsf{check\_public}(u,\Pi_{\mathsf{receipt}})\\
 &\quad\land \bigwedge_{i\in\mathcal I}
 \widehat V_i(\mathsf{pub}_i,\pi_i,\pi'_i)\\
 &\quad\land \mathsf{check\_records}(u,\mathsf{bind})
\end{aligned}
\]
to RISC-V and prove with Jolt that it returns $\mathsf{true}$. \reviewchange{The outer
proof} would be a single Jolt RV64 proof. Verifying it requires one Jolt
verification call independent of the number of components. The recursion is
enabled by the subprotocols of~\cite{joltrecursion2026}: the inner verifiers are
dominated by elliptic-curve operations (Dory openings and pairings in the
measured \reviewchange{configuration}, HyperKZG MSMs in the alternative Atlas \reviewchange{configuration}), which admit
efficient PIOPs inside Jolt (Appendix~\ref{app:efficient-recursion}).
Native-field copy-constraint sum-checks would handle the bindings. A \reviewchange{configuration} may retain complete
direct verification or use a split whose refinement hypotheses have been
established. In either case, every component proof and binding check remains
required. The complete-receipt implementation currently has accepted host
verification and bounded guest diagnostics, but no completed outer proof.

\section{Evaluation}
\label{sec:bench}

We evaluate a \reviewchange{complete receipt for a model with fixed input and generation limits} with public Qwen2-0.5B parameters, hidden model inputs and intermediate tensors, deterministic action extraction, and \reviewchange{a receiver that checks spending and reserves funds in one transaction}. Two independently audited runs each complete three \reviewchange{requests initiated by the receiver}, two approvals followed by a proved denial. An additional \reviewchange{comparison using the same executable} measures \reviewchange{reuse of authenticated preparation data in the \reviewchange{Atlas prover}} with independently verified receipts in both modes \reviewchange{(Appendix~\ref{sec:bench-request-cache})}. Separate component experiments measure \reviewchange{Atlas model generation}, solver recursion and auxiliary arithmetic, with detailed results in Appendix~\ref{app:evaluation-support}. The single outer proof of the complete receipt remains unfinished.

\subsection{Stateful Qwen Receipt}
\label{sec:bench-action-receipt}

\reviewchange{The experiment uses the five-rule permission policy and two queries of \S\ref{sec:system-overview} (complete formula in Appendix~\ref{app:stateful-running-example}). Exhaustive comparison over the restricted grammar's order regions checks this fixed policy, but does not establish reliable translation of arbitrary natural language.\par}

Policy generation uses 128 positions, a 100-token prompt and a 27-token
response. The planner uses 128 positions, a 63-token prompt and a 25-token
response that parses as a \$150 purchase from \texttt{acme}, without
credential sharing or a recurring commitment. The policy proof is cached,
while each request has fresh planner, prompt, extraction and solver proofs.
Every receipt \reviewchange{includes two \reviewchange{Atlas proofs}, all four original hidden
tensor boundaries, and five Jolt proofs: source-policy binding, policy
compilation, planner-prompt authentication, exact action extraction and both
solver queries.}

After verification the receiver rechecks authorization, expiry and current
state inside one database transaction, reserves the amount only for
\textsf{VALID}, and consumes a successful session. Starting at \$200, the two
approved requests reserve \$150 each, reaching \$350 and then \$500. The third
proposes \$650, proves \textsf{INVALID}, and leaves the total at \$500. Replay
of successful sessions and an unused session against stale state reject. No
merchant purchase occurs.

\begin{table*}[t]
\centering
\small
\setlength{\tabcolsep}{7pt}
\begin{tabular}{@{}c rrr l rrr r@{}}
\toprule
& \multicolumn{3}{c}{\textbf{Spending total (\$)}} &
& \multicolumn{3}{c}{\textbf{Time (s)}} & \\
\cmidrule(lr){2-4}\cmidrule(lr){6-8}
\textbf{Request} & Prior & Proposed & Final & \textbf{Verdict} &
\reviewchange{\shortstack{Request to\\decision}} & Verify & Verify now & \textbf{Proof bytes} \\
\midrule
0 & 200 & 350 & 350 & \textsf{VALID}   & 3,903.81 & 33.03 & 8.35 & 23,695,245 \\
1 & 350 & 500 & 500 & \textsf{VALID}   & 3,529.27 & 33.27 & 8.26 & 23,695,242 \\
2 & 500 & 650 & 500 & \textsf{INVALID} & 3,401.80 & 32.71 & 8.19 & 23,692,263 \\
\bottomrule
\end{tabular}
\caption{\reviewchange{Three spending requests on the Xeon E7-8890 v4.} Spending totals are whole
US dollars. \reviewchange{Request-to-decision time runs from session creation to the authorization}
outcome. Verify is the complete verification command on the \reviewchange{server generating the proofs}, and
\reviewchange{Verify now checks the same receipts again} with the current default verifier on an AMD EPYC 9755
(\S\ref{sec:bench-verifier-build}). Proof bytes count complete proof messages.}
\label{tab:stateful-receipts}
\end{table*}

\subsection{Costs and Measurement Scope}

\leavevmode\reviewchange{The stateful experiment uses a Xeon E7-8890 v4 at 2.20\,GHz, 32 \reviewchange{Atlas proving workers}, 16 Jolt workers and a recorded quota of 30.72 CPU cores, with no GPU acceleration. Each request is a single sequential observation with different state, session and verdict, not a matched optimization comparison. Sizes are serialized byte lengths; KiB, MiB, GiB and TiB use powers of two.\par}

Issuing and authorizing a request takes 56.70 to 65.06 minutes (Table~\ref{tab:stateful-receipts}). The \reviewchange{Atlas planner command} accounts for 2,377.58 to 2,866.49 seconds of this, of which witness construction and commitments take 620.61 to 686.60 seconds and graph proving 792.89 to 835.67 seconds. Numerical generation, cache validation, boundary proofs and other work make up the rest, and Table~\ref{tab:stateful-native-phases} gives each request's phases. Honest complete verification takes 32.71 to 33.27 seconds, including 6.75 to 7.23 seconds of producer graph verification, and \reviewchange{additional validation tests} run separately. On an AMD EPYC 9755, the current default verifier re-checks all six saved receipts in 8.04 to 8.35 seconds of internal time each and returns the original verdict and byte count. The September 28 build takes 9.41 to 9.56 seconds on that host, so the current build is 12.6 to 15.5\% faster.

\paragraph{Setup and reuse.}
Registration-time preprocessing is excluded from request latency, but first-request setup and the loading of retained data are included. Administrative action preprocessing, shared by the prompt and extraction program, takes 79.51 seconds including serialization for a maximum padded trace of $2^{26}$ rows, and solver preprocessing takes 11.75 seconds for $2^{21}$ rows. Both precede session issuance and keep their registered online keys fixed across the sequence. \reviewchange{The first request also includes} 201.63 seconds of registered public-model preprocessing and a 122.66-second Dory setup call, which count toward that request and are reused by later ones. Loading and validating retained Jolt preprocessing remains online work, taking 80.80 to 83.30 seconds per prompt or extraction proof and 6.81 to 7.64 seconds per solver proof. Private commitments remain fresh. These observations do not measure regeneration of all policy proofs or initial \reviewchange{model proof parameters}.

\paragraph{Memory and communication.}
The \reviewchange{Atlas proving service} peaks at 129.15 to 140.26\,GiB of \reviewchange{sampled RAM usage per request}, and each receipt carries about 22.6\,MiB of proof messages. Sampling is once per second, so brief peaks are unmeasured, and these values do not cover simultaneous memory across the whole workflow (Appendix~\ref{app:evaluation-support}). The 22.6\,MiB counts every fresh \reviewchange{Atlas commitment} and required boundary and the proofs reused from the registered policy. Reusable verification keys, receiver trust inputs and the received action, whose canonical encoding is 86 bytes, are separate objects.

\subsection{Fresh Reproduction and Rejection Controls}

A separate Linux run rebuilt five \reviewchange{programs that run outside the zkVM} from the authenticated source archive and created both online Jolt setups before registering fresh keys and issuing new sessions. It reused authenticated archived action and solver \reviewchange{programs compiled for the zkVM}, \reviewchange{model proof parameters}, model inputs and cached policy proofs. It therefore establishes a \reviewchange{rebuild of those programs from source} and fresh complete proofs with \reviewchange{recorded zkVM executables}, but not a \reviewchange{recompilation of those zkVM programs}. Separately built local verifiers accept all three new receipts, and the independent database audit again finds exactly two reservations and a final spending total of \$500. The run reproduces the original verdicts, with \reviewchange{times from request to decision} of 3,401.56 to 3,890.07 seconds and complete verification of 32.73 to 34.87 seconds on the \reviewchange{server generating the proofs}, or 8.04 to 8.23 seconds with the current verifier on the AMD EPYC 9755. This is a \reviewchange{repeat experiment using the same source code} with one observation per request, not a measured software speedup.

\reviewchange{Both runs reject 36 receipts with substituted proofs or inputs and pass five groups of tests that alter Jolt's internal data, two that alter the model proof system's internal data, and 14 tests with malformed model-proof messages.} \reviewchange{The tests cover} missing proofs and altered actions, state, contexts, roots, boundaries, keys and \reviewchange{configurations}, including substitution of a separately valid extraction proof for a different state. They test the implemented checks and do not prove the composition theorem's cryptographic hypotheses. The fresh audit authenticates 4,352 source and dependency files and all 4,519 public inventory files, checks the private inputs and \reviewchange{order of setup steps}, and requires the independent verifier executables to match accepted build hashes. \reviewchange{Both the public and private archives pass the authenticity checks.}

\subsection{\reviewchange{Verification Cost}}
\label{sec:bench-verifier-build}

\leavevmode\reviewchange{Checking a complete receipt takes about 8 seconds on the tested AMD EPYC 9755 server, using 32 workers. Software optimizations reduce this time by approximately 13--16\% on the same machine, while retaining the same proofs and verification checks.\par}

\reviewchange{For six saved receipts from the original run and its reproduction, verification takes 8.04--8.35 seconds per receipt, compared with 9.41--9.56 seconds for the reference version. These measurements exclude input preparation and program startup. The roughly 33-second times in Table~\ref{tab:stateful-receipts} use a different Xeon server and include additional overheads.\par}

\reviewchange{The optimized verifier continues to reject altered proofs and inputs. Appendix~\ref{app:verifier-optimizations} gives the optimization and comparison details.\par}

\begingroup\reviewcolor
\subsection{\reviewchange{Proof recursion}}
\label{sec:bench-outer-limits}

\reviewchange{Complete solver-verifier recursion produces a 104,222-byte proof verified in 0.223 seconds (\S\ref{sec:bench-solver-recursion}). It covers the spending example, not the complete Qwen receipt. Appendix~\ref{app:evaluation-support} reports this and the other component experiments separately.\par}

\reviewchange{One 105,078-byte M2 proof covers the five Jolt proofs and their shared public
input, including the four conversion statements. The planner and policy
\reviewchange{model-inference proofs} remain outside it and must be verified separately.}

\reviewchange{The complete} verifier accepts all three original stateful receipts on the
\reviewchange{server, but running it inside the zkVM exceeds} the selected $2^{32}$-cycle proving capacity on
an actual receipt (Appendix~\ref{app:complete-guest-limits}). \reviewchange{Three child
proofs divide the Jolt checks into groups that fit within this capacity}.
Child P checks \reviewchange{planner-prompt authentication} and action extraction, child Q
\reviewchange{checks source-policy binding and policy compilation}, and child D \reviewchange{checks} the
decision proof. \reviewchange{M1 verifies P and Q, and M2 verifies M1 and D. The merge
programs check that their children carry the same public input and the
required output digests.}

\reviewchange{Inside the zkVM program that verifies child proofs}, verifying the P outer proof takes 1,945,795,727 cycles
after certificate-backed target-group membership checks\reviewchange{. The certificate is}
bound to the exact proof bytes before any challenge. \reviewchange{A two-proof execution}
takes 3,783,700,162 cycles, \reviewchange{or} 0.88 times the capacity\reviewchange{.} M1 \reviewchange{uses} one joint
certificate \reviewchange{for P and Q}. \reviewchange{Each completed outer proof} \reviewchange{accepts} in a separate process on its \reviewchange{server generating the proofs} and
independently on \reviewchange{a Mac}, and each rejects ten altered inputs, proofs and keys.

\reviewchange{Verification of the two} \reviewchange{model-inference proofs} \reviewchange{inside the zkVM still exceeds}
the capacity by about eight times. \reviewchange{No} outer proof of the complete receipt is
reported.

\reviewchange{The result for this configuration} uses public model parameters and deterministic structured extraction. Private-model execution, \reviewchange{action translation with an LLM} and \reviewchange{ensuring that the receiving service executes the proved action} are separate targets. \reviewchange{Section~\ref{sec:composed-security} states the outstanding security assumptions}. Plain ARc's reported 5--15-second validation latency~\cite{arc2026} is measured on a different workload, and a quantitative comparison requires matched tasks, model calls and assurance requirements.
\par\endgroup

\section{Conclusion}
\label{sec:conclusion}

\reviewchange{zkARc connects proved policy formalization, planner execution, action extraction and SMT checks in a receipt bound to the received action. Its security depends on the stated cryptographic and application assumptions. A receiver used these receipts to enforce a budget, approving two requests and denying a third, with an independently verified reproduction.\par}

\reviewchange{The solver proof is 88\,KiB and verifies in 0.3 seconds. One 105\,KB proof covers the receipt's five Jolt proofs, while its two \reviewchange{model-inference proofs} are verified separately. Issuing a complete receipt takes about an hour, dominated by model inference proving on CPUs. Future work includes accelerating proof generation on GPUs through CUDA and Metal, completing a single outer proof, and transitioning to lattice-based proofs with Lattice Jolt~\cite{latticejolt2026} and Lattice Jolt Atlas.\par}

\clearpage
\begingroup
\small
\setlength{\baselineskip}{10.5pt}
\let\oldthebibliography\thebibliography
\renewcommand{\thebibliography}[1]{%
  \oldthebibliography{#1}%
  \setlength{\itemsep}{1pt plus 0.3ex}%
  \setlength{\parskip}{0pt}}
\let\reviewbibitem\bibitem
\renewcommand{\bibitem}[1]{%
  \color{black}%
  \def\reviewbibkey{#1}%
  \def\reviewnewbibkey{latticejolt2026}%
  \ifx\reviewbibkey\reviewnewbibkey\reviewcolor\fi
  \reviewbibitem{#1}}
\begin{multicols}{2}
\bibliographystyle{plain}
\bibliography{bib}
\end{multicols}
\endgroup
\clearpage

\appendix
\small

\section{Additional Evaluation and Component Measurements}
\label{app:evaluation-support}

\reviewchange{\reviewchange{These experiments use different workloads from the requests in \S\ref{sec:bench} and time different parts of the computation.} Here, ``full validation'' includes rejection tests. Table~\ref{tab:stateful-receipts} times honest verification only.\par}

\subsection{\reviewchange{Time Spent on Each Request}}
\label{app:stateful-phases}

\reviewchange{Tables~\ref{tab:stateful-native-phases} and \ref{tab:stateful-consumer-phases} report request phases. \reviewchange{Public setup for the Atlas prover} is reused after request 0. Jolt preprocessing is registered before issuance and loaded and validated for each proof. Complete commands also include capacity checks, serialization and \reviewchange{coordination between programs}, so proof phases do not sum to \reviewchange{time from request to decision}.\par}

\begin{table}[H]
\centering
\small
\setlength{\tabcolsep}{8pt}
\newcommand{\grp}[1]{\multicolumn{4}{@{}l}{\footnotesize\textbf{#1}}\\[1pt]}
\begin{tabular}{@{}l rrr@{}}
\toprule
& \multicolumn{3}{c}{\textbf{Request}} \\
\cmidrule(l){2-4}
& 0 & 1 & 2 \\
\midrule
\grp{\reviewchange{Model proving stages (s)}}
\quad Public model preprocessing & 201.63 & 0.00 & 0.00 \\
\quad Dory setup call & 122.66 & 0.00 & 0.00 \\
\quad Public cache identity check & 31.89 & 30.60 & 24.63 \\
\quad Witness and commitments & 686.60 & 660.49 & 620.61 \\
\quad Hidden boundary proofs & 2.45 & 2.37 & 2.04 \\
\quad Graph proof after commitments & 835.67 & 824.33 & 792.89 \\
\quad Producer graph verification & 6.75 & 7.12 & 7.23 \\
\addlinespace[2pt]
\quad \reviewchange{Complete model-proving command} & 2,866.49 & 2,494.79 & 2,377.58 \\
\addlinespace[3pt]
\grp{Process memory (GiB)}
\quad \reviewchange{Peak RAM during request (RSS)} & 129.15 & 140.26 & 135.65 \\
\quad \reviewchange{Last recorded process peak (VmHWM)} & 129.01 & 140.56 & 140.56 \\
\bottomrule
\end{tabular}
\caption{\reviewchange{Model-proving measurements for the spending requests.} \reviewchange{Resident set size (RSS), the memory held in RAM, is sampled once per second.} VmHWM is the last sampled lifetime high-water counter, which can
carry an earlier request. Request 0's reported VmHWM is lower than its largest
sampled RSS. Both raw observations are retained. No total-workflow memory peak
or speedup comparison is inferred.}
\label{tab:stateful-native-phases}
\end{table}

\begin{table}[H]
\centering
\small
\setlength{\tabcolsep}{8pt}
\newcommand{\grp}[1]{\multicolumn{4}{@{}l}{\footnotesize\textbf{#1}}\\[1pt]}
\begin{tabular}{@{}l rrr@{}}
\toprule
& \multicolumn{3}{c}{\textbf{Time (s)}} \\
\cmidrule(l){2-4}
& Cache load & Prove & Command \\
\midrule
\grp{Planner prompt}
\quad Request 0 & 83.30 & 352.92 & 440.46 \\
\quad Request 1 & 82.49 & 350.53 & 437.07 \\
\quad Request 2 & 81.89 & 354.68 & 441.17 \\
\addlinespace[3pt]
\grp{Exact action extraction}
\quad Request 0 & 82.91 & 319.52 & 405.39 \\
\quad Request 1 & 83.19 & 315.13 & 401.25 \\
\quad Request 2 & 80.80 & 314.05 & 397.89 \\
\addlinespace[3pt]
\grp{Both solver queries}
\quad Request 0 & 7.00 & 34.15 & 41.70 \\
\quad Request 1 & 7.64 & 33.95 & 42.34 \\
\quad Request 2 & 6.81 & 31.04 & 38.34 \\
\bottomrule
\end{tabular}
\caption{Fresh online \reviewchange{Jolt proofs} in the original stateful sequence, by
Jolt relation. Complete commands include cache loading and validation,
capacity execution and serialization. Administrative setup and serialization
takes 79.51 seconds for the action program ($2^{26}$ maximum padded rows) and
11.75 seconds for the solver ($2^{21}$). It precedes session issuance.}
\label{tab:stateful-consumer-phases}
\end{table}

\begingroup
\reviewcolor
\subsection{\reviewchange{Reusing Prover Preparation}}
\label{sec:bench-request-cache}

\leavevmode The prover can save checked preparation data and reuse it across requests. In this experiment, reuse reduces the mean time for later requests from 53.8 to 38.4 minutes (28.72\%; Table~\ref{tab:request-cache}). Model proof generation itself takes about the same time: the saving comes from avoiding repeated preparation.

Both modes use the \reviewchange{same Linux server}, program, public model and registered keys, with 32 workers for model proving and 16 for Jolt. Both reuse public cryptographic setup data and verify proofs sequentially. Every request creates fresh private commitments, randomness and proofs.

\begin{table}[H]
\reviewcolor
\centering
\small
\setlength{\tabcolsep}{8pt}
\begin{tabularx}{\linewidth}{@{}X rr@{}}
\toprule
\textbf{Time (s)} & \textbf{Without reuse} & \textbf{With reuse} \\
\midrule
\reviewchange{First request: time to decision} & 3,627.89 & 3,602.39 \\
\reviewchange{Later requests: time to decision (mean)} & 3,229.12 & 2,301.84 \\
Model prover including preparation, later requests (mean) & 2,251.07 & 1,327.69 \\
Model proof generation in later requests (mean) & 749.25 & 760.55 \\
\bottomrule
\end{tabularx}
\caption{Effect of preparation reuse. Request time includes proving and
verification. Each later-request mean combines one approval and one denial.
The last row measures model proof generation after commitments.}
\label{tab:request-cache}
\end{table}

Each mode processes three requests: two approvals followed by a denial. Independent checks accept all six receipts and both spending sequences, confirm fresh proof data, and reject altered inputs and proofs. These runs are separate from the main evaluation. Each mode was measured once, with reuse tested first, so the observed reduction applies to this workload and run order.

\endgroup

\subsection{\reviewchange{Limits of Verifying a Receipt Inside the zkVM}}
\label{app:complete-guest-limits}

\reviewchange{On a stateful receipt with 328 public-input bytes and 23,999,549 \reviewchange{bytes of private input to the zkVM}, \reviewchange{the full receipt verifier running inside the zkVM stops} after 4,294,967,297 cycles at its $2^{32}$-cycle bound. A matched comparison of the first 512 planner commitments in \reviewchange{one compiled zkVM program} uses one accepted execution per mode. Canonical-field checks and exact batched inversion reduce direct-check cycles from 1,585,042,236 to 319,501,951 (79.84\%), \reviewchange{while private input grows} from 98,356 to 240,295 bytes. This excludes \reviewchange{Atlas model-proof verification}, tensor boundaries, Jolt proofs and outer proving, so it is not a complete-receipt speedup.\par}

\reviewchange{Certificates for all 117,878 \reviewchange{Atlas commitments} \reviewchange{increase the private input from} 23,999,549 to 35,859,465 bytes. \reviewchange{Construction and checking outside the zkVM} take 1,913.90 seconds in one observation \reviewchange{while other jobs were running}. The batched candidate accepts the honest receipt and rejects all 44 substitutions \reviewchange{outside the zkVM, but exceeds the zkVM's} $2^{32}$-cycle bound. With allocator reuse, it fails a 24,616,704-byte allocation after 33,267,659,727 cycles in a 1\,GiB heap with a 768\,MiB Rust arena. These resource failures do not imply invalid proofs or establish proving capacity at a larger execution bound. Outer setup, proving, verification and receiver commands exist, but \reviewchange{tests of the receiver's request handling use simulated cryptographic checks}. No complete outer proof exists.\par}

\reviewchange{A child program retains every receipt check except the two \reviewchange{Atlas model-proof verification}s, takes their four conversion statements privately and returns a digest for merging with native children. It accepts the saved receipt in 38,439,566,072 cycles, 8.95 times capacity. Each of the five Jolt checks costs 7.38--7.78 billion cycles, too large even for separate proofs. Changing one conversion byte rejects at the first statement comparison after 3.77 billion cycles, suggesting decoding consumes about half each check. \reviewchange{The verifier running outside the zkVM accepts the honest input} and rejects all 22 substitutions.\par}

\reviewchange{Target-group subgroup checks dominate decoding. Decoding each distinct key once, since two proofs share a key, lowers the child to 36,151,197,492 cycles. Checking cyclotomic membership before the subgroup test lowers it to 25,093,031,824 cycles. This accepts exactly nonzero $x$ with $x^r=1$ and agrees with the original on all 15,520 decoded elements. Embedding the four keys, validated at registration and authenticated by the program digest, lowers it to 23,316,243,270 cycles (5.43 times capacity). All variants accept the honest input and \reviewchange{reject 26 altered inputs in tests run outside the zkVM}. Jolt verification then occupies 86\% of the child, about 4 billion cycles per proof.\par}

\reviewchange{Per proof, the opening reduction and Dory check cost about 2.64 billion cycles and BlindFold 1.13 billion. General target-group exponentiation accounts for 57\%. Cyclotomic scaling, shared squarings across Dory sums, delayed BlindFold commitment folds into one MSM per opening, and lazy claim-source lookup preserve every check and transcript byte. Together they reduce the child to 9,857,771,913 cycles (2.30 times capacity), with each proof check at 1.94--1.98 billion, below $2^{31}$. Three children take 3.93, 3.94 and 2.01 billion cycles. Every variant accepts the honest input and rejects 26 altered receipts plus 52 Dory and BlindFold controls. Exactness rests on documented reasoning and differential tests, not a formal proof; \S\ref{sec:bench} reports the first child's outer proof.\par}

\subsection{\reviewchange{Hardware and Size Units}}

\leavevmode\reviewchange{\reviewchange{The machines used for the component experiments are} an AMD EPYC 9534 with 256\,GB RAM (solver), a Threadripper 3970X with 32 Rayon workers (\reviewchange{Qwen without BlindFold}), an EPYC 9684X (short-prompt Qwen), and a Xeon Platinum 8272L (policy Qwen). \reviewchange{Both Qwen cases proved by Atlas} use 48 workers and 1,024 scalars per BlindFold commitment row. \reviewchange{The policy proving server} runs source and compilation proofs with 16 workers. Its solver cases use an E5-2699 v3 with 16 workers. Canonical conversions use a Platinum 8272L with 52 workers. Solver recursion uses an EPYC 7662 with 64 threads. \reviewchange{Other jobs run on these machines during the measurements.} Unless stated otherwise, timings are single samples, not controlled speedup comparisons.\par}

\reviewchange{The \reviewchange{Qwen comparison without BlindFold} uses Atlas \texttt{cfcffb0} without native zero knowledge. The base solver uses Jolt \texttt{36f549f7c}. Solver recursion uses modular prover \texttt{7b364bfb3} and the complete Jolt verifier with BlindFold. \reviewchange{Atlas programs} use Rust 1.97.1 and \reviewchange{the zkVM program uses Rust} 1.95.0.\par}

\reviewchange{KiB, MiB, GiB and TiB denote $2^{10}$, $2^{20}$, $2^{30}$ and $2^{40}$ bytes. KB and TB denote $10^3$ and $10^{12}$. Sizes count serialized proofs, transmitted public statements, fresh tensor commitments and required boundary messages. Reusable registration is separate. Model graphs and public parameter commitments are reconstructed from the verifier key.\par}

\begin{table*}[t]
\centering
\small
\setlength{\tabcolsep}{8pt}
\newcommand{\grp}[1]{\multicolumn{3}{@{}l}{\footnotesize\textbf{#1}}\\[1pt]}
\begin{tabularx}{\textwidth}{@{}l r >{\raggedright\arraybackslash}X@{}}
\toprule
\textbf{Measured artifact} & \textbf{Proof size} & \textbf{Verified computation} \\
\midrule
\grp{Qwen generation and receipts}
\quad Short prompt and conversions & 22.8\,MiB & Bounded \reviewchange{Atlas model generation} with canonical prompt and response \\
\quad Two linked generations & 45.3\,MiB & Both \reviewchange{Atlas model stages}, canonical conversions and equal hidden text \\
\quad Policy generation, bindings and SMT & 10.4\,MiB & Native policy generation, exact \reviewchange{source and response conversion checks}, and both solver queries \\
\quad Static structured-action receipt & 22.6\,MiB & Policy and planner generation, \reviewchange{all required Jolt checks}, exact action extraction and both queries \\
\addlinespace[3pt]
\grp{Solver proofs in Jolt}
\quad Generated policy SMT & 87.6\,KiB & Both queries on the formula bound to the native Qwen response \\
\quad \reviewchange{SMT proofs for the spending example} & 88.7\,KiB & Both queries for the \reviewchange{spending example} \\
\quad SMT verifier & 101.9\,KiB & Complete verification of the structured SMT proof \\
\bottomrule
\end{tabularx}
\caption{Complete proof messages of the measured computations. The Qwen
rows include \reviewchange{Atlas commitments}, tensor boundaries and every required
conversion proof, and the linked pair also requires the joint proof of equal
hidden text. The static structured-action row combines the cached policy proof
with a fresh planner proof using shared range commitments, and the policy-only
row retains its separate prover measurement. The inner SMT proof is about
91\,KB. \reviewchange{Complete receipt verification is implemented; a single outer proof covering the entire receipt remains unimplemented.} The bounded public example
is reported in Table~\ref{tab:public-spending-receipts}, and no numerical
projection for recursion or a downstream wrapper is included.}
\label{tab:end-to-end-receipt}
\end{table*}

\subsection{\reviewchange{Qwen Execution Without BlindFold}}
\label{app:ordinary-qwen-control}

\reviewchange{\reviewchange{The comparison without BlindFold proves} one Qwen2-0.5B forward pass on the PMC input at sequence length 32 and scale $2^{14}$. Its final logits match the floating-point reference's next token and top ten candidates, with cosine similarity $0.999874$ and mean absolute error $0.05177$. Twelve Threadripper 3970X proofs average 60.58 seconds to prove, 3.85 seconds to verify and 30.89 GiB peak memory. Each proof is 5,188,295 bytes. Full verifier input including weights is 1,996,946,516 bytes. Independent verification accepts every proof and rejects altered inputs and outputs. This tests numerical execution and cost, not translation or generation.\par}

\subsection{\reviewchange{Privacy of Atlas Model Proofs}}\label{app:native-privacy}

\reviewchange{Native Atlas combines BlindFold and hiding Dory openings at the same hidden evaluation. Equality proofs link original tensor commitments, including hidden-index lookups and row gathers. Supported graphs and layout changes share a grouped opening and BlindFold proof, reusing public commitments and refreshing private ones. The frozen Qwen graph preserves 2,461 node shapes and matches Atlas on 2,040 computed outputs. \reviewchange{Verification in a new process} rejects altered registration, public parameters, private commitments and missing claims.\par}

\reviewchange{Table~\ref{tab:native-qwen} measures greedy causal generation: 32 context positions and two new tokens for the short prompt, and 128 positions and 27 new tokens for the policy, whose introductory text alone takes 35 tokens. Proofs check stopping, trailing zero padding and causal masks in all twenty-four attention blocks. Sequences match independent execution. Model constants, positions and masks are public, with all 416 constants preserved at the larger context. Hiding tensor commitments bind prompt and response tokens to canonical Jolt conversions using \reviewchange{a 32 MiB heap for the zkVM program}.\par}

\reviewchange{Short-prompt conversion takes 141.62 seconds to prove and 0.310 seconds to verify. Response conversion takes 126.76 and 0.325 seconds. Each proof body is 91.8 KiB. Conversion files total 491.7 KiB. Native messages and conversions total 22.8 MiB, with a 244.38-second validation command including decoding and rejection checks. A second generation consumes the first's hidden response. A joint Jolt proof checks canonical decoding, re-encoding and equal hidden text, rejecting a valid second generation on different input. It takes 209.64 seconds to prove and 0.334 seconds to verify, with a 94.7 KiB body and 251.9 KiB transmitted. The pair totals 45.3 MiB and validates in 489.40 seconds. This establishes bounded generation and linkage, not policy compilation, action extraction or solver execution.\par}

\begin{table}[ht]
\centering
\small
\setlength{\tabcolsep}{8pt}
\newcommand{\grp}[1]{\multicolumn{3}{@{}l}{\footnotesize\textbf{#1}}\\[1pt]}
\begin{tabular}{@{}l rr@{}}
\toprule
\textbf{Native Qwen generation} & \textbf{Short prompt} & \textbf{Policy} \\
\midrule
Context positions / generated tokens & 32 / 2 & 128 / 27 \\
\addlinespace[3pt]
\grp{Setup and proving (s)}
\quad Dory setup call & 86.00 & 102.82 \\
\quad Public parameter preprocessing & 161.76 & 164.01 \\
\quad Witness construction and commitments & 583.46 & 673.49 \\
\quad Proving after commitments & 810.16 & 613.49 \\
\quad Commitments and proving & 1,393.62 & 1,286.98 \\
\addlinespace[3pt]
\grp{Verification (s)}
\quad Decode saved verifier objects & 42.86 & 3.63 \\
\quad Fresh graph verification & 65.12 & 6.94 \\
\quad Original tensor boundary checks & 0.09 & 0.30 \\
\addlinespace[3pt]
\grp{Size and memory}
\quad Proof size & 22.3 MiB & 10.2 MiB \\
\quad\quad Proof body, included above & 237.7 KiB & 244.4 KiB \\
\quad\quad Fresh commitments, included above & 22.1 MiB & 9.9 MiB \\
\quad Reusable verifier key & 49.9 MiB & 49.9 MiB \\
\quad Peak proving process memory & 126.1 GiB & 140.9 GiB \\
\bottomrule
\end{tabular}
\caption{Complete native Atlas BlindFold proofs with 48 workers. The short
prompt uses an EPYC 9684X and the policy uses a Xeon Platinum 8272L, so
\reviewchange{different workloads and machines} prevent a software speedup comparison. Whole
native invocations take 1,855.76 and 2,495.76 seconds, and policy message
encoding takes 84.50 seconds separately. Policy verification times average
two checks and exclude decoding and \reviewchange{tests that invalid inputs or proofs are rejected}. Sizes count fresh
commitments (384 bytes each for the short prompt, 192 for the policy), the
proof body and tensor boundary messages. Jolt conversion and application
proofs are separate.}
\label{tab:native-qwen}
\end{table}

\reviewchange{The registered prompt combines the spending policy with an example using different limits: 100 prompt tokens and 27 response tokens fit in 128 positions. The response parses as $G$ over \texttt{amount}, \texttt{total}, \texttt{approved}, \texttt{credentials} and \texttt{recurring}, with whole-dollar amounts and permission exactly when $G$ holds. The parser accepts comparisons and Boolean operations over these fields and rejects commands or extra text. All 27 selections match sequential execution, and all 2,040 operator blocks match numerical execution. Exhaustive comparison over the grammar's order regions checks the five restrictions, without establishing faithful translation of arbitrary policies.\par}

\reviewchange{Source policy and formula use separate SHA-256 commitments with distinct tags, policy context and independent 32-byte private nonces. The source proof opens the policy commitment and checks exact prompt encoding at the original tensor boundary. The compilation proof decodes and parses the same hidden response bytes and commits to the formula. The solver opens it and constructs both queries. Registration independently supplies expected commitments, context, keys, session and action facts.\par}

\begin{table}[tbp]
\centering
\small
\setlength{\tabcolsep}{7pt}
\newcommand{\grp}[1]{\multicolumn{6}{@{}l}{\footnotesize\textbf{#1}}\\[1pt]}
\begin{tabular}{@{}l rrr rr@{}}
\toprule
& \multicolumn{3}{c}{\textbf{Time (s)}} & \textbf{Proof} & \textbf{Peak} \\
\cmidrule(lr){2-4}
\textbf{\reviewchange{Policy check}} & Setup & Prove & Verify & \textbf{(KiB)} & \textbf{(GiB)} \\
\midrule
\grp{Conversions, Xeon Platinum 8272L}
\quad Source policy and prompt & 68.82 & 276.42 & 0.298 & 92.4 & 50.7 \\
\quad Response and compilation & reused & 237.56 & 0.300 & 92.3 & 50.7 \\
\addlinespace[3pt]
\grp{Solver, Xeon E5-2699 v3}
\quad \textsf{VALID} & 10.38 & 30.17 & 0.292 & 87.6 & 10.6 \\
\quad \textsf{INVALID} & 10.50 & 29.52 & 0.292 & 87.6 & 10.5 \\
\quad \textsf{SATISFIABLE} & 10.43 & 30.01 & 0.303 & 87.6 & 10.5 \\
\bottomrule
\end{tabular}
\caption{Complete Jolt proofs for the native policy instance with 16 workers.
Source and compilation share one program setup, charged once in the source
row, and their peak is measured over the combined process. Sizes include one
proof, public IO, framing and the registered program digest. Setup uses the
measured padded trace: $2^{26}$ rows for the conversions (32 MiB heap) and
$2^{21}$ for the solver (1 MiB stack and heap). The separate capacity
measurement takes at most 1.85 seconds. Peak memory covers setup and proving.
These phases do not measure full application latency.}
\label{tab:policy-consumers}
\end{table}

All three solver cases use the same generated formula. The valid purchase has
total spending of \$350, an amount of \$150, an approved vendor, no credential
sharing and no recurring payment. Raising the amount to \$250 gives
\textsf{INVALID}. Leaving vendor approval unspecified at \$150 gives
\textsf{SATISFIABLE}. Their query results are respectively
$(\mathsf{SAT},\mathsf{UNSAT})$, $(\mathsf{UNSAT},\mathsf{SAT})$ and
$(\mathsf{SAT},\mathsf{SAT})$.

\reviewchange{The 10.4 MiB policy branch contains the \reviewchange{Atlas proof}, both Jolt conversions and the valid solver proof. Its validation command averages 20.97 seconds over two checks, including decoding and rejection tests. Thirteen controls cover missing proofs, altered commitments, context, session, facts, \reviewchange{Atlas proofs} and boundaries, unsupported native \reviewchange{configurations} and a valid solver proof for another formula. Seven more reject malformed \reviewchange{Atlas proof messages}. Dispatch remains disabled. Section~\ref{sec:bench-action-static} adds planner provenance and extraction, and \S\ref{sec:bench-action-receipt} issues sessions before proving.\par}

\subsection{\reviewchange{Qwen Spending Example}}
\label{sec:bench-action-static}

\reviewchange{This static receipt uses cached policy compilation and fresh planner, prompt, extraction and solver proofs. The planner uses 128 positions, 63 prompt tokens and 25 response tokens, producing a \$150 purchase from \texttt{acme} without credential sharing or recurring payments. Extraction binds the original hidden response to the received action and authenticated state, taking prior spending from \$200 to \$350. The receiver independently supplies the vendor registry, authorized context and expected prompt commitment, as required by \S\ref{sec:deployment-acceptance}.\par}

\reviewchange{Verification accepts the \textsf{VALID} receipt with two \reviewchange{Atlas proofs}, four original tensor boundaries and five Jolt proofs. Thirty-six rejection tests cover missing or altered components and public inputs, including a valid extraction proof for different state whose key is allowed by the control. Each \reviewchange{Atlas proof branch} also passes seven malformed-message tests. The receiver re-verifies, checks authorization, expiry and state, reserves \$150 and consumes the session in one transaction. Replay and fresh sessions against old state reject; a relocated public copy verifies. No purchase executes. The session is installed after proving, so this test does not measure \reviewchange{time from a live request to sending the action}.\par}

\reviewchange{The static case uses public parameters, a Xeon E7-8890 v4, 32 native and 16 Jolt proving workers, and 32 verification workers. The planner bounds commitment and lookup batches, shares range commitments, reclaims opening buffers and compacts indices. Arithmetic copies release compact coefficients and unused field tails after binding. Original committed polynomials remain intact. Full 256-entry lookup tables use one-byte addresses. Each sparse commitment row sums the public generators for one address. Caching the all-column sum lets a fully populated block recover its largest row by subtracting the others. Model context, registered key, proving parameters and cached policy bytes are unchanged. Compact transport preserves every distinct fresh commitment and restores the exact statements and proofs.\par}

\reviewchange{The verifier uses target-group MSMs for large commitment sums and a hash set to deduplicate openings without changing order or constraints. With BN254 base-field characteristic $p$ and scalar order $r$, decoding rejects zero and checks $x^{p^4}x=x^{p^2}$ for cyclotomic membership. Every target-group element satisfies this because $r$ divides $p^4-p^2+1$. It then checks $x^p=x^{p-r}$ using Frobenius and a 127-bit cyclotomic exponent, retaining $x^r=1$ and all canonical field and message checks.\par}

\reviewchange{Planner witness construction and commitments take 667.44 seconds and proving 862.90 seconds. Its full command takes 2,953.94 seconds and peaks at 124.60\,GiB, including numerical and parity checks, registration and other work. Prompt and extraction proofs each use $2^{26}$ rows: setup takes 81.72/82.27 seconds, proving 339.27/291.57 seconds and peak memory 33.79/33.76\,GiB. Both SMT queries use $2^{21}$ rows, 9.98 seconds of setup, 33.18 seconds of proving and 10.56\,GiB. The alternate extraction rejection test is excluded. This single sequential run includes process counters and commitment tracing and does not establish repeatability. Messages including cached policy proofs total 23,695,235 bytes (22.6\,MiB), plus an 86-byte action. Reusable keys and trust inputs are separate. Full validation, including rejection tests, takes 55.21 seconds.\par}

\subsection{Measured Solver Recursion}
\label{sec:bench-solver-recursion}

\reviewchange{\reviewchange{The program running in Jolt verifies} the structured SMT proof with BlindFold, including all openings and pairings, in 4,170,582,955 cycles (Table~\ref{tab:bench-components}). For base modulus $p$, group order $r$ and BN254 parameter $x$, it uses $p-r=6x^2$ for target-group membership and processes cyclotomic powers four exponent bits at a time after validation, retaining zero and subgroup checks.\par}

\reviewchange{The outer proof has a 104,222-byte body, 135 \reviewchange{bytes of public inputs and outputs} and 69,267 bytes of separate preprocessing. Its $2^{32}$-row trace proves in 5,328.76 seconds (88.8 minutes), verifies in 0.223 seconds and sets up in 967.60 seconds, peaking at 1.082\,TiB. Program commitments use 64 chunks. Measurements include Perfetto capture. Separate processes on the \reviewchange{server generating the proofs} and another machine verify the saved proof. Controls check inner-statement binding and reject altered public statements, results and proof commitments. This proves solver-verifier recursion, not a complete receipt.\par}

\reviewchange{Repeated jobs copy retained prover and verifier data for the \reviewchange{registered zkVM program and configuration}, with fresh program-commitment randomness and opening hints. On an EPYC 7B13 with 32 threads, fresh setup takes 242.19 seconds versus 0.60 seconds for \reviewchange{a call that reuses the prepared data}, averaged over three calls. Initial setup remains required. These measurements cover setup alone.\par}

\subsection{Per-Component Cost}

\reviewchange{Table~\ref{tab:bench-components} reports component costs without auxiliary verifier-split proofs. Complete receipt verification is implemented; a single outer proof covering the entire receipt remains unimplemented. \S\ref{sec:bench-public-receipt} measures the separate bounded public example.\par}

\begin{table}[tbp]
\centering
\small
\setlength{\tabcolsep}{7pt}
\newcommand{\grp}[1]{\multicolumn{5}{@{}l}{\footnotesize\textbf{#1}}\\[1pt]}
\begin{tabular}{@{}l rr rr@{}}
\toprule
& \multicolumn{2}{c}{\textbf{Time (s)}} & \textbf{Proof} & \textbf{Peak} \\
\cmidrule(lr){2-3}
\textbf{Component} & Prove & Verify & \textbf{size} & \textbf{memory} \\
\midrule
\grp{Short-prompt generation}
\quad Qwen short prompt, first & 1,394.24 & 65.21 & 22.3\,MiB & 126.1\,GiB \\
\quad Qwen short prompt, second & 1,394.51 & 66.93 & 22.3\,MiB & 123.4\,GiB \\
\quad Conversions, one generation & 268.38 & 0.63 & 491.7\,KiB & 38.7\,GiB \\
\quad Conversions, linked pair & 476.62 & 0.96 & 743.6\,KiB & 53.2\,GiB \\
\addlinespace[3pt]
\grp{Policy generation}
\quad Qwen policy generation & 1,289.11 & 7.24 & 10.2\,MiB & 140.9\,GiB \\
\quad Policy prompt and compilation & 513.98 & 0.60 & 184.7\,KiB & 50.7\,GiB \\
\quad Generated policy SMT & 30.17 & 0.29 & 87.6\,KiB & 10.6\,GiB \\
\addlinespace[3pt]
\grp{Solver proofs in Jolt}
\quad \reviewchange{SMT proofs for the spending example} & 69.15 & 0.21 & 88.7\,KiB & n/a \\
\quad SMT verifier & 5,328.76 & 0.22 & 101.9\,KiB & 1.082\,TiB \\
\bottomrule
\end{tabular}
\caption{Component measurements from separate runs. Sizes are rounded
serialized byte lengths of every transmitted file, and peak memory covers the
measured worker including setup; n/a marks a value not reported. Qwen rows
include witness commitments, proving and tensor boundary checks, and exclude
setup and public parameter preprocessing (Table~\ref{tab:native-qwen}).
Conversion rows sum separately measured proofs and are not application
latency. The short prompt and linked pair retain duplicated bundle data,
while policy messages contain each proof once. The generated policy SMT row
is the valid case in Table~\ref{tab:policy-consumers}. The \reviewchange{row for the spending example}
proves both solver queries with BlindFold on an EPYC 9534 with 16 threads.
The verifier row measures complete SMT verification inside Jolt on an
EPYC 7662 with 64 threads, includes Perfetto capture, and excludes setup and
the inner proof.}
\label{tab:bench-components}
\end{table}

\subsection{\reviewchange{What the Measurements Include}}
\label{sec:bench-endtoend}

\leavevmode\reviewchange{The fixed case installs sessions after proving and validates with \reviewchange{additional validation tests}; \S\ref{sec:bench-action-receipt} measures \reviewchange{requests initiated by the receiver} through authorization. Both reuse policy generation and compilation. Ordinary Qwen and \reviewchange{Atlas model generation} measurements alone cover neither receipt.\par}

\subsection{\reviewchange{What the Reported Sizes Include}}
\reviewchange{Table~\ref{tab:end-to-end-receipt} counts canonical message encodings checked against saved bytes. \reviewchange{In the comparison without BlindFold, the} 1,996,946,516-byte verifier input includes weights, inputs, outputs, parameters and proof. It is not the native BlindFold configuration. The first \reviewchange{Atlas model generation} sends 23,401,006 bytes including fresh commitments, boundaries and a 243,411-byte proof body. Its reusable key is 52,286,327 bytes. Conversions add 503,516 bytes, yielding 23,904,522 bytes per generation. Registration reconstructs the graph and public commitments. All fresh commitments are transmitted.\par}

\reviewchange{The SMT message is 90,847 bytes: a 90,692-byte proof body plus 155 \reviewchange{bytes of public inputs and outputs} and framing. The verifier proof is 104,357 bytes: a 104,222-byte body plus 135 \reviewchange{bytes of public inputs and outputs}. Program commitments reduce solver preprocessing from 90,922,838 to 53,899 bytes with an approximately 91\,KB proof (Table~\ref{tab:bench-smt-pi-prime}).\par}

\reviewchange{Policy messages encode a norm-one target-group element $a+bu$ in a quadratic extension with basis $u$ as $b/(a+1)$ in 192 bytes. Decoding checks subgroup membership and canonical encoding. Identical range-indicator polynomials share one freshly blinded commitment. Distinct polynomials remain separate. The registered graph fixes their order and references. Parallel weighted commitment sums preserve every term and coefficient.\par}

\reviewchange{The policy \reviewchange{Atlas proof} and boundaries total 10,670,088 bytes, including a 250,283-byte proof body. Its reusable key is 52,314,178 bytes. Source and compilation messages add 94,617 and 94,552 bytes, the valid solver 89,688 and the decision record 1,634, totaling 10,950,579 bytes. Source and compilation share 61,583 bytes of preprocessing. The solver needs 53,903 bytes. Each Jolt message includes its registered key digest. The static receipt (\S\ref{sec:bench-action-static}) adds planner and action checks to compact policy messages and has a separate size total.\par}

\subsection{HyperKZG Auxiliary Implementation}
\label{sec:bench-atlas-pi-prime}

\reviewchange{The HyperKZG auxiliary bundle certifies the final two $G_1$ linear combinations of \S\ref{sec:atlas-hyperkzg}. Native field checks and the final pairing remain direct. Unlike Dory, it has no $G_2$ or $G_T$ accumulator chain or cross-group wiring, but each MSM must link its scalar multiplications and additions.\par}

Each joint opening reduces to two MSMs that the $\pi'$ bundle certifies
separately:
\begin{itemize}
    \item $L = \sum_i q_i \cdot \mathsf{com}_i$, an $(\ell + 4)$-term MSM
    where $\ell$ is the joint polynomial dimension and the coefficients $q_i$
    are computed from transcript challenges and claimed evaluations, and
    \item $R = W_0 + d_0 \cdot W_1 + d_0^2 \cdot W_2$, a size-$3$ linear
    combination over the quotient commitments $W_0, W_1, W_2$.
\end{itemize}
\reviewchange{Atlas directly computes the preceding joint-commitment fold, although Table~\ref{tab:split}'s target assigns it to the auxiliary relation. Each $N$-term MSM uses $N$ scalar-multiplication and $N-1$ addition PIOPs, wiring, claim merging and a committed-witness opening. The implementation also recomputes the curve operations. The extended verifier directly checks $e(L,[1]_2)e(-R,[\tau]_2)=1$. An all-native split would additionally need Miller-loop and final-exponentiation PIOPs. Qwen measurements use the base verifier without this bundle. Synthetic auxiliary measurements appear in \S\ref{sec:auxiliary-kernels}.\par}

\subsection{SMT Component over Dory}
\label{sec:bench-smt-pi-prime}

\reviewchange{The SMT program fixes} the five spending rules and constructs
$q_+=F\land P\land C$ and $q_-=F\land P\land\neg C$ from typed inputs.
It bounds the amounts, validates the Boolean flags, recomputes both public
input digests and requires exactly one solver response for each query.
The \reviewchange{compliant spending example} yields (SAT, SAT).

\reviewchange{Jolt proves both queries with Dory and BlindFold. Recursion retains all opening and pairing checks. Table~\ref{tab:bench-smt-pi-prime} separates proof and preprocessing sizes for full program tables and program commitments. Table~\ref{tab:auxiliary-dory} separately measures synthetic auxiliary proofs. Neither establishes faithful LLM translation.\par}

\begin{table}[tbp]
\centering
\small
\setlength{\tabcolsep}{8pt}
\newcommand{\grp}[1]{\multicolumn{3}{@{}l}{\footnotesize\textbf{#1}}\\[1pt]}
\begin{tabular}{@{}l rr@{}}
\toprule
& \textbf{Full program} & \textbf{Program commitment} \\
\midrule
\grp{Size}
\quad Proof body (KiB) & 88.4 & 88.6 \\
\quad Public IO and framing (bytes) & 155 & 155 \\
\quad Proof size (KiB) & 88.6 & 88.7 \\
\quad Verifier preprocessing (bytes) & 90,922,838 & 53,899 \\
\quad Complete verifier input (bytes) & 91,013,553 & 144,746 \\
\addlinespace[3pt]
\grp{Time (s)}
\quad Prove & 16.76 & 69.15 \\
\quad Verify & 2.08 & 0.21 \\
\addlinespace[3pt]
\grp{Execution}
\quad Padded execution trace & $2^{21}$ & $2^{21}$ \\
\quad Solver results $(s_+,s_-)$ & (SAT, SAT) & (SAT, SAT) \\
\bottomrule
\end{tabular}
\caption{Accepted Jolt proofs of both queries for the \reviewchange{spending example}. Program commitments replace the full code tables in verifier
preprocessing. The complete verifier input contains the proof, preprocessing
and public IO. Runs use 16 Rayon threads on the EPYC 9534, with other jobs
present, so these single samples do not establish a controlled timing
comparison.}
\label{tab:bench-smt-pi-prime}
\end{table}

\subsection{\reviewchange{Spending Example with Public Inputs}}
\label{sec:bench-public-receipt}

\reviewchange{This public example uses two small registered ONNX graphs with fixed integer weights. One maps restricted policy features to limits and a permission mode; the other maps a request to an amount, resulting total and action flags. Ordinary Atlas with Dory proves both executions. \reviewchange{A program running in Jolt parses} the policy, replays both graphs, binds the received action, and derives both solver results and the verdict. All three proofs are checked against an independently expected public statement and registered models and parameters. Replaying the graphs establishes this bounded relation independently of the Atlas opening requirements.\par}

\begin{table}[tbp]
\centering
\small
\setlength{\tabcolsep}{8pt}
\begin{tabular}{@{}l c c rr@{}}
\toprule
& & & \multicolumn{2}{c}{\textbf{Time (s)}} \\
\cmidrule(l){4-5}
\textbf{Case} & $\boldsymbol{(s_+,s_-)}$ & \textbf{Authorize} & Jolt prove & Receipt verify \\
\midrule
Allowed & (SAT, UNSAT) & Yes & 29.644 & 1.438 \\
Blocked & (UNSAT, SAT) & No & 29.854 & 1.483 \\
Ambiguous & (SAT, SAT) & No & 29.949 & 1.468 \\
\bottomrule
\end{tabular}
\caption{Bounded public receipts on ARM64 macOS with four Rayon workers.
Each receipt is 172,130 bytes. These are \reviewchange{measurements from one local run per case}. Jolt proving
excludes the two Atlas proofs, and verification excludes setup.}
\label{tab:public-spending-receipts}
\end{table}

The allowed case uses a \$150 purchase and an explicit permission rule. The
blocked case is a \$250 purchase against a \$200 limit, and the ambiguous
case is the \$150 purchase under prohibitions alone. All three receipts pass
fresh verification and \reviewchange{reject nine deliberately altered statements or proofs each}.
Each receipt contains Atlas proofs of 40,933 and 40,741 bytes and a Jolt
proof of 90,173 bytes, plus framing and the public statement. Registered
parameters occupy a separate 3,158,480 bytes, and Jolt setup takes
11.92--12.07 seconds, separate from Table~\ref{tab:public-spending-receipts}.

\reviewchange{This fixed-grammar example exposes all policy, request and action data and treats prices, prior spending and vendor approval as declared facts. It excludes Qwen generation, hidden-edge openings, verifier splitting, an outer proof hiding the receipt and payment dispatch.\par}

\subsection{\reviewchange{Measurements of Auxiliary Proof Operations}}
\label{sec:auxiliary-kernels}

\reviewchange{These synthetic benchmarks exclude inference, solver execution, recursion and complete receipts. Let $\ell$ count polynomial variables, $d$ bound each variable's degree, $c$ count evaluation claims and $N$ count MSM terms. Claim merging brings claims to one point for a shared opening. Appendix~\ref{app:reproducibility} identifies configurations.\par}

\paragraph{Sumcheck and claim verification.}
The merger, curve operation checks and wiring checks evaluate equality factors in $O(\ell)$ field operations and constant auxiliary space using
\[
\sum_{x\in\{0,1\}^{\ell}}
 \widetilde{\mathsf{eq}}(r,x)\widetilde{\mathsf{eq}}(s,x)
=\prod_{j=1}^{\ell}\bigl(r_js_j+(1-r_j)(1-s_j)\bigr).
\]
The sumcheck prover folds tables in place, generates consecutive linear evaluations by addition, and recovers $p(1)$ from the round claim and $p(0)$. Interpolation reuses denominators for fixed nodes across rounds, and Hyrax row commitments use batch normalization to convert points to affine form.

\paragraph{Claim merging and witness construction.}
\reviewchange{Claim merging combines equality factors and folds one table per source polynomial using a specialized quadratic round. Scalar-multiplication witnesses batch-normalize blocks of 128 rows to bound temporary storage. Shifted coordinates reuse the next accumulator row with a zero boundary.\par}

\paragraph{Sparse wiring and parallel proving.}
\reviewchange{Wiring tracks each selector's single nonzero Boolean point and weight as variables are bound. MSM claim merging borrows tables from the committed witness. The \texttt{parallel} feature uses Rayon for independent Hyrax rows and MSM proofs, each with its own transcript. Results retain specified order, with commitments fixed before challenges.\par}

\paragraph{Measurements.}
Table~\ref{tab:auxiliary-kernels} reports medians of 21 samples in three batches with Rust~1.89.0 release builds on \reviewchange{two Linux machines}, an AMD Ryzen 7 2700X with one untimed warmup per batch and a Ryzen 5 7600 with one warmup per case and batch. Serial measurements use one pinned CPU core, and parallel measurements use six distinct physical cores. Input generation, generator setup, serialization and the initial correctness check are outside the timed region, and the witness benchmark measures coordinate generation alone. Merge cases use two evaluation claims per source polynomial. On the 2700X, the MSM bundles for $N=3$ and $N=14$ occupy 27,296 and 116,544 bytes, and the largest peak resident set size (RSS), the memory resident in RAM, across the three runs is 68,500\,KiB. This measures the whole process including setup, rather than prover memory alone. On the 7600, the separate Hyrax commitment benchmark for $N=14$ takes 38.459\,ms with six cores, peak RSS for the serial stage benchmark is 35,752\,KiB, and the complete kernel benchmark process uses about 51\,MiB.

\begin{table}[t]
\centering
\small
\setlength{\tabcolsep}{7pt}
\newcommand{\grp}[1]{\multicolumn{4}{@{}l}{\footnotesize\textbf{#1}}\\[1pt]}
\begin{tabular}{@{}l l c r@{}}
\toprule
\textbf{Operation} & \textbf{Host} & \textbf{Cores} & \textbf{Time (ms)} \\
\midrule
\grp{Sum-check}
\quad Prove, $\ell=16$, $d=2$ & 2700X & 1 & 18.127 \\
\addlinespace[3pt]
\grp{Claim merging}
\quad Verify, $\ell=12$, $c=32$, 16 polynomials & 2700X & 1 & 0.254 \\
\quad Verify, $\ell=16$, $c=16$, 8 polynomials & 2700X & 1 & 0.183 \\
\quad Prove, $\ell=12$, $c=32$, 16 polynomials & 7600 & 1 & 9.897 \\
\quad Prove, $\ell=16$, $c=16$, 8 polynomials & 7600 & 1 & 89.256 \\
\addlinespace[3pt]
\grp{Sparse wiring}
\quad $\ell=8$, 32 edges, 16 polynomials & 7600 & 1 & 0.196 \\
\quad $\ell=12$, 32 edges, 16 polynomials & 7600 & 1 & 2.262 \\
\quad $\ell=16$, 16 edges, 8 polynomials & 7600 & 1 & 18.308 \\
\addlinespace[3pt]
\grp{MSM}
\quad Prove, $N=3$ & 2700X & 1 & 126.491 \\
\quad Prove, $N=14$ & 2700X & 1 & 465.605 \\
\quad Verify, $N=14$ & 2700X & 1 & 17.141 \\
\quad Prove, $N=14$ & 7600 & 1 & 264.503 \\
\quad Prove, $N=3$ & 7600 & 6 & 19.092 \\
\quad Prove, $N=14$ & 7600 & 6 & 63.395 \\
\addlinespace[3pt]
\grp{Coordinate witness generation}
\quad $G_1$, 256 rows & 7600 & 1 & 0.131 \\
\quad $G_2$, 256 rows & 7600 & 1 & 0.437 \\
\bottomrule
\end{tabular}
\caption{Synthetic auxiliary kernels on an AMD Ryzen 7 2700X and a Ryzen 5 7600. MSM rows use BN254 scalars spanning the full field and include operation proofs, wiring, claim merging and the Hyrax opening. Witness rows exclude those steps.}
\label{tab:auxiliary-kernels}
\end{table}

\paragraph{Dory auxiliary layouts.}
The layout with separate instances proves each operation separately. The family layout packs operations of the same kind into a shared argument. Table~\ref{tab:auxiliary-dory} measures three operations in each family on \reviewchange{an ARM64 Mac} without CPU affinity, using Rust~1.89.0 release builds and one Rayon thread. Each median covers 15 samples in three batches, with one warmup per case and batch. Proving includes witness construction and generator setup. Times vary substantially between batches, limiting the timing comparison. Process peak RSS across the three runs ranges from 120.4 to 159.6\,MiB.

\begin{table}[t]
\centering
\small
\setlength{\tabcolsep}{8pt}
\begin{tabular}{@{}l rr r@{}}
\toprule
& \multicolumn{2}{c}{\textbf{Time (ms)}} & \textbf{Payload} \\
\cmidrule(lr){2-3}
\textbf{Layout} & Prove & Verify & \textbf{(bytes)} \\
\midrule
Per instance & 1,293.150 & 35.496 & 106,280 \\
By family & 368.192 & 10.295 & 39,640 \\
\bottomrule
\end{tabular}
\caption{Dory auxiliary bundles with three synthetic operations per family. Payload sizes exclude public instance metadata. These workloads do not execute a complete SMT receipt.}
\label{tab:auxiliary-dory}
\end{table}

\reviewchange{Direct curve checks remain in these verifier timings.\par}

\subsection{Implementation Checks and Limits}
\label{sec:auxiliary-audit}

\reviewchange{The aggregate transcript binds the Hyrax commitment, layout and source-polynomial count before claim-merging and packing challenges. MSM proving commits the witness before operation and wiring PIOPs and derives their domains from it. Substituting a constant witness with an honest sum-check transcript rejects. Both Dory layouts commit before family PIOPs and finally open the retained polynomial and masks. Tests confirm that a witness change affects all families and that each family rejects altered commitments.\par}

\reviewchange{The full relation still needs an audit of operand, scalar-bit and boundary binding to the base transcript to establish Lemma~\ref{lem:verifier-split-refinement}'s premises. The MSM verifier recomputes scalar multiplications and additions when constructing public instances, so its measured cost includes more than native field PIOPs and two Grumpkin MSMs. Implementation measurements do not prove the theorem's assumptions.\par}

\noindent\begin{minipage}{\linewidth}
\subsection{\reviewchange{Verification Optimizations}}
\label{app:verifier-optimizations}

\leavevmode\reviewchange{\textbf{Compiler comparison.} We hold the receipt, keys, statements and five Jolt proofs fixed on an AMD EPYC 9755 with 48 CPUs and a 46.08-CPU quota. Verification uses 32 Rayon workers, checking policy and planner branches in parallel. Only the native verifier executable changes; registration records its digest. Each version is measured twice, reversing the order. All accept the same statements, facts, verdict and 23,695,245 bytes. Thin and full link-time optimization with one code generation unit reduce mean verification time by 8.6--9.6\% relative to the standard decoder without link-time optimization. These observations do not distinguish the two optimized versions.\par}

\reviewchange{\textbf{Proof decoding.} The batched decoder combines exact torus inversions and removes redundant commitment re-encoding, retaining every subgroup check. With full link-time optimization, it reduces verification time by 16.7\% relative to the standard decoder without optimization. A separate comparison on another EPYC 9755 \reviewchange{uses the same machine} and isolates decoding: time falls from 8.970 to 8.391 seconds.\par}

\reviewchange{\textbf{Selected verifier.} The six-receipt comparison in \S\ref{sec:bench-verifier-build} uses the batched decoder with thin link-time optimization. Per-receipt reductions are 12.6--15.5\%, with the original verdicts and byte counts. It rejects all 36 altered receipts and a valid extraction proof for another state. Receiver reservation, replay and stale-state tests pass with serial and parallel schedules. Overflow checks remain enabled. Proofs and proving computations are unchanged.\par}

\reviewchange{\textbf{Timing.} The timer starts after preparation of the \reviewchange{authenticated copy of the inputs}. Full command time also includes that preparation, startup and \reviewchange{checking whether the process has finished} at one-second intervals. Appendix~\ref{app:reproducibility} identifies the executables and measurement records.\par}
\end{minipage}
\par

\section{\reviewchange{Experiment Records}}
\label{app:reproducibility}

\begingroup
\reviewcolor
\subsection{\reviewchange{Where to Find the Experiment Records}}

\reviewchange{All paths are relative to the implementation's source directory.} Dated directories in the index are under \path{bench-logs/}. The records identify source and executable hashes, models and keys, machine configuration, timings, proofs and verification results.

\medskip
\noindent
\begin{tabularx}{\linewidth}{@{}>{\raggedright\arraybackslash}p{0.23\linewidth}>{\raggedright\arraybackslash}X@{}}
\toprule
\textbf{Experiment} & \textbf{Records} \\
\midrule
Receipt sequence and reproduction
& \path{bounded_system/results.md} and \path{bounded_system/results.json}.\newline
  \path{2026-09-21-bounded-system/}: \path{linux-stateful/} contains the original run and \path{finalization-retry/} its audit.\newline
  \path{2026-09-21-bounded-reproduction-retry/} contains the audited reproduction. \\
\addlinespace[5pt]
Verifier comparisons (\S\ref{sec:bench-verifier-build})
& \path{2026-09-30-lto-epyc/}: build comparison.\newline
  \path{2026-09-28-recursive-verifier-target/}: decoder comparison.\newline
  \path{2026-09-30-default-verifier/}: selected-verifier validation.\newline
  \path{2026-10-02-reverify-stateful/}: six-receipt comparison. \\
\addlinespace[5pt]
Preparation reuse (Table~\ref{tab:request-cache})
& \path{docs/request-cache-full-results-2026-09-27.md} and matching JSON.\newline
  \path{2026-09-27-request-cache-recovery/} contains the collected records. \\
\addlinespace[5pt]
Recursive Jolt proofs (\S\ref{sec:bench-outer-limits})
& \path{2026-10-02-outer-p/}, \path{2026-10-02-outer-q/} and \path{2026-10-02-outer-d/}: child proofs.\newline
  \path{2026-10-04-merge-m1/} and \path{2026-10-04-merge-m2/}: merge proofs and \reviewchange{verification records from the server and a separate Mac}. \\
\addlinespace[5pt]
Verifier execution measurements
& \path{final_proof/README.md}, \path{final_proof/results.md} and \path{final_proof/results.json}.\newline
  \path{2026-09-21-final-proof/} \reviewchange{records hashes of the zkVM program, source files, inputs and execution script}. \\
\bottomrule
\end{tabularx}

\noindent\begin{minipage}{\linewidth}
\subsection{\reviewchange{Check a Previously Generated Receipt}}

\leavevmode Build and register the verifiers as described in \path{bounded_system/README.md}, then run:
\begin{verbatim}
python3 -m bounded_system verify-export \
  --artifact /path/to/public-receipt \
  --registration /path/to/trusted-registration.json \
  --expected /path/to/expected.json \
  --received-action /path/to/received-action.json
\end{verbatim}
Supply registration, expected context and received action independently of the prover. This command verifies an exported public receipt without a private witness or proving setup.
\end{minipage}
\par\medskip

\noindent\begin{minipage}{\linewidth}
\subsection{Reproduce the Experiment}

\leavevmode The Linux reproduction uses the private evidence archive and sufficient \reviewchange{memory for the model prover}:
\begin{verbatim}
python3 -m bounded_system.reproduce \
  --evidence /path/to/original-collected-run \
  --output /path/to/new-reproduction \
  --max-seconds 18000 --build-jobs 16 \
  --use-archived-orchestrator --remove-build-cache
\end{verbatim}
This rebuilds five \reviewchange{programs that run outside the zkVM}, authenticates the archived action and solver \reviewchange{programs compiled for the zkVM}, creates fresh online setup and issues three new sessions. Each run retains public receipts and a separate private reproduction archive.
\end{minipage}
\endgroup

\section{Running-Example Policy and Action}
\label{app:running-example-smtlib}

\subsection{\reviewchange{Spending Rules for Structured Actions}}
\label{app:stateful-running-example}

\reviewchange{The stateful Qwen model emits only $G$ (\S\ref{sec:system-overview}). A deterministic wrapper declares variables, asserts $F=(\texttt{allowed}\leftrightarrow G)$ and inserts $(P,C)$ from the received action and \reviewchange{state supplied by the receiver}. The first request expands to:\par}

\medskip
\noindent
\begin{minipage}[t]{0.54\textwidth}
\textbf{\footnotesize Declarations and policy $F$}
\begin{smtcode}
(set-logic QF_LIA)
(declare-const total Int)
(declare-const amount Int)
(declare-const approved Bool)
(declare-const credentials Bool)
(declare-const recurring Bool)
(declare-const allowed Bool)
(assert (= allowed
  (and (<= total 500) (<= amount 200) approved
       (not credentials) (not recurring))))
\end{smtcode}
\end{minipage}\hfill
\begin{minipage}[t]{0.42\textwidth}
\textbf{\footnotesize Facts, then $q_+$ and $q_-$}
\begin{smtcode}
(assert (= total 350))
(assert (= amount 150))
(assert approved)
(assert (not credentials))
(assert (not recurring))
(push 1)
(assert allowed)
(check-sat)
(pop 1)
(push 1)
(assert (not allowed))
(check-sat)
(pop 1)
\end{smtcode}
\end{minipage}
\medskip

\reviewchange{Extraction checks that the complete decoded response canonically equals the received \$150 purchase from \texttt{acme}, with no credential sharing or recurring commitment. It derives \texttt{total = 200 + 150} from prior reserved spending and checks \texttt{acme} in the authenticated vendor registry. Amounts are nonnegative whole US dollars, with \texttt{amount} at most \texttt{total}.\par}

\reviewchange{The first request yields $(\mathsf{SAT},\mathsf{UNSAT})$ and \textsf{VALID}. The receiver reserves funds and consumes the session in one transaction. The next fresh session at \texttt{total = 500} also approves. The third, at \texttt{total = 650}, yields $(\mathsf{UNSAT},\mathsf{SAT})$, \textsf{INVALID} and no reservation (Table~\ref{tab:stateful-receipts}). Authorization rejects stale extraction state.\par}

\subsection{\reviewchange{Solver Check Using Only Prohibitions}}

\reviewchange{\reviewchange{The solver experiment} (\S\ref{sec:bench-smt-pi-prime}) uses QF\_LRA and the five prohibitions without the permission equivalence. Its \$150 approved-vendor purchase supplies premise $P$ and conclusion $C=\texttt{isPurchaseAllowed}$. The policy and $q_+$ are:\par}

\medskip
\noindent
\begin{minipage}[t]{0.54\textwidth}
\textbf{\footnotesize Policy}
\begin{smtcode}
(set-logic QF_LRA)
(declare-const spendTotal Real)
(declare-const purchaseAmount Real)
(declare-const isApprovedVendor Bool)
(declare-const sharesPaymentCredentials Bool)
(declare-const isRecurringCommitment Bool)
(declare-const isPurchaseAllowed Bool)
(assert (=> (> spendTotal 500)
            (not isPurchaseAllowed)))
(assert (=> (> purchaseAmount 200)
            (not isPurchaseAllowed)))
(assert (=> (not isApprovedVendor)
            (not isPurchaseAllowed)))
(assert (=> sharesPaymentCredentials
            (not isPurchaseAllowed)))
(assert (=> isRecurringCommitment
            (not isPurchaseAllowed)))
\end{smtcode}
\end{minipage}\hfill
\begin{minipage}[t]{0.42\textwidth}
\textbf{\footnotesize Query $q_+$}
\begin{smtcode}
(assert (= purchaseAmount 150))
(assert (= spendTotal 150))
(assert isApprovedVendor)
(assert (not sharesPaymentCredentials))
(assert (not isRecurringCommitment))
(assert isPurchaseAllowed)
(check-sat)
\end{smtcode}
\end{minipage}
\medskip

\reviewchange{Replacing \texttt{(assert isPurchaseAllowed)} with \texttt{(assert (not isPurchaseAllowed))} gives $q_-$. Both queries are satisfiable: prohibitions force denial on violation but do not entail permission otherwise. The verdict is therefore \textsf{SATISFIABLE} and does not authorize dispatch. Permission would require a positive rule such as\par}
\begin{smtcode}
(assert (=> (and (<= spendTotal 500)
                 (<= purchaseAmount 200)
                 isApprovedVendor
                 (not sharesPaymentCredentials)
                 (not isRecurringCommitment))
            isPurchaseAllowed))
\end{smtcode}
\reviewchange{\reviewchange{The measured experiment} executes both queries with the five fixed prohibitions and omits this positive rule. It builds assertions from typed inputs, using Boolean equalities equivalent to the notation above, and does not prove AV extraction from action text (\S\ref{sec:bench-smt-pi-prime}, Table~\ref{tab:bench-components}).\par}

\section{BlindFold in Jolt Atlas}
\label{app:blindfold-jolt-atlas}

\reviewchange{Atlas BlindFold with HyperKZG~\cite{joltatlas2026} applies Vega's Succinct NovaBlindFold~\cite{vega}, based on HyperNova~\cite{hypernova2023}, to the ONNX sum-check verifier. Folding this small instance avoids a random witness as large as inference. Jolt applies the construction to RISC-V with Dory~\cite{jolt2023}. Measured native Atlas also uses hiding Dory and direct verification outside Jolt.\par}

\subsection{Setting}

\reviewchange{An Atlas proof has $S$ sum-check stages linked by polynomial-evaluation claims, with topology fixed by the preprocessed ONNX graph. Stage $s$ has $n_s$ rounds of degree-$d_s$ univariates; let $N=\sum_{s=1}^{S}n_s$ be the total round count.\par}

\reviewchange{BlindFold commits to round polynomials and cross-stage evaluations that would otherwise reveal information about the model inputs, weights, activations and policy.\par}

\subsection{Relaxed R1CS Encoding of the Verifier}

Following Nova~\cite{nova2022}, BlindFold operates over the relaxed R1CS relation, in which an instance is satisfied by a triple $(Z, u, E)$ with
\[
(A Z) \circ (B Z) = u \cdot (C Z) + E,
\]
\reviewchange{Standard R1CS has $u=1$, $E=\mathbf{0}$. BlindFold encodes the succinct sum-check verifier across all $N$ rounds:\par}
\begin{enumerate}
    \item \emph{Sum constraints.} For each round $i$ of stage $s$, the round polynomial $s_i$ must satisfy $s_i(0) + s_i(1)$ equal to the running sum-check claim from round $i-1$.
    \item \emph{Chain constraints.} The next claim equals $s_i(r_i)$, evaluated by Horner's rule against the verifier challenge $r_i$.
    \item \emph{Opening constraints.} The final claim of each stage is reduced to an evaluation claim on the committed witness polynomials at the appropriate point. Only this reduction is encoded here. In the target hiding configuration, the evaluation \emph{value} must not be sent in the clear: it occupies a private witness slot of $Z$ together with the blinding factor of its HyperKZG opening, the prover sends a commitment $\bar V$ to it, and an additional constraint asserts that the value bound by $\bar V$ equals that slot. This constraint is what ties the sum-check reductions to the polynomial-commitment proof. Without it a prover could satisfy every sum-check constraint with evaluation values inconsistent with the committed polynomials~\cite{joltatlas2026}. In the recursion layer, $\pi'$ certifies the non-pairing group computations of the cryptographic opening check, while $\widehat V_i$ evaluates its final pairing directly (\S\ref{sec:proof-composition}, Appendix~\ref{app:verifier-ops}).
\end{enumerate}
\reviewchange{Cross-stage constraints link final openings to subsequent initial claims. Fiat--Shamir challenges, batching coefficients and initial claims enter $A,B,C$, not $Z$, so the transcript and graph determine the R1CS. Bounded round degrees give $O(N)$ constraints: rounds are logarithmic per tensor domain, but their sum grows with the stage count. Fixed graphs and shapes fix the verifier R1CS regardless of which witness entries are private.\par}

The witness $Z$ is laid out as an $R' \times C$ grid following Hyrax~\cite{hyrax2018}, so that Pedersen~\cite{pedersen1992} commitments to each row enable batched openings at any evaluation point of the form $r_y = (r_{\text{row}}, r_{\text{col}})$.

\subsection{Hiding via NovaBlindFold}

Let $(Z_1, u_1 = 1, E_1 = \mathbf{0})$ denote the relaxed R1CS instance that encodes the prover's actual sum-check verification. BlindFold proceeds in four steps.

\textit{Step 1: Commit round polynomials.} In every sum-check round, the prover sends a Pedersen commitment to the round polynomial coefficients rather than the coefficients themselves:
\[
\mathsf{Com}(s_i) = \sum_{j=0}^{d_s} c_{i,j} \cdot G_j + \rho_i \cdot H,
\]
with independent generators $G_0, \dots, G_{d_s}, H$ and a fresh blinding scalar $\rho_i$. The commitment is appended to the Fiat-Shamir transcript in place of the polynomial. The verifier challenge $r_i$ is derived from the transcript exactly as in the cleartext case.

\reviewchange{\textit{Witness commitments and evaluations.} Transcript randomization alone does not hide the witness. The target also requires blinded HyperKZG commitments $C_i$, preventing guesses even for low-entropy witnesses of the same shape, and hidden evaluations committed as $\bar V$. Clear $W(r)$ values leak linear equations that independent challenges across proofs can make solvable. Blinding factors and evaluation values occupy private slots of $Z$~\cite{joltatlas2026revised}. Both commitments and openings must hide their values.\par}

\reviewchange{\textit{Step 2: Sample a random satisfying instance.} Sample $Z_2,u_2$ uniformly and set $E_2=(AZ_2)\circ(BZ_2)-u_2(CZ_2)$, giving a uniformly random satisfying relaxed instance. Commit $Z_2$ and $E_2$ by Hyrax rows. Their $O(N)$ verifier-instance size makes Succinct NovaBlindFold randomization a constant-factor cost, without a random witness as large as inference.\par}

\textit{Step 3: Fold.} The prover forms the Nova cross-term
\[
T = (A Z_1) \circ (B Z_2) + (A Z_2) \circ (B Z_1) - u_1 (C Z_2) - u_2 (C Z_1),
\]
commits each row of $T$ via Pedersen, and after Fiat-Shamir draws a folding challenge $r \in \mathbb{F}$. The folded instance is
\[
Z' = Z_1 + r Z_2,\qquad u' = u_1 + r u_2,\qquad E' = E_1 + r T + r^2 E_2,
\]
and a direct computation shows
\(
(A Z') \circ (B Z') = u' (C Z') + E',
\)
so $(Z', u', E')$ satisfies the relaxed R1CS relation.

\textit{Step 4: Spartan decider with Hyrax openings.} The prover proves
satisfaction of the folded instance via Spartan~\cite{spartan2020}. Pad the
row count $m$ to a power of two and derive a fresh verifier challenge
$\tau\in\mathbb F^{\log m}$ after binding the folded instance. Define
\[
g(x)=\widetilde{\mathsf{eq}}(\tau,x)
\left(\widetilde{AZ'}(x)\widetilde{BZ'}(x)
      -u'\widetilde{CZ'}(x)-\widetilde{E'}(x)\right).
\]
The outer sum-check starts from
$\sum_{x\in\{0,1\}^{\log m}}g(x)=0$ and uses $\log m$ rounds of
degree-$3$ univariates. At its random terminal point $r_x$, the verifier
checks that the final running claim equals $g(r_x)$. This terminal value
need not be zero: the row constraints hold at Boolean indices, whereas
products of multilinear extensions need not preserve those equalities at
arbitrary field points. An inner sum-check reduces a random linear
combination of the claimed $AZ'$, $BZ'$ and $CZ'$ evaluations to a witness
evaluation $W(r_y)$ and public matrix evaluations. The committed rows of
$Z'$ are opened at $r_y$ row by row using the homomorphism of Pedersen
commitments: the prover sends a combined row whose $j$-th entry is
$\sum_i\widetilde{\mathsf{eq}}(r_{\text{row}},i)Z'_{i,j}$ and the
corresponding combined blinding. The verifier checks the analogous linear
combination of row commitments. The error vector $E'$ is opened at $r_x$
by the same construction over its grid.

\subsection{Security and Complexity}

\reviewchange{Vega~\cite{vega}, building on HyperNova~\cite{hypernova2023}, establishes honest-verifier zero knowledge for NovaBlindFold. For fixed nonzero $r$, uniform $(Z_2,u_2)$ makes $(Z',u',E')$ uniform over satisfying instances and independent of $Z_1$: the affine map on $(Z_2,u_2)$ is bijective, and $E'$ is determined by the relaxed R1CS equation. At $r=0$, the fold equals the original instance. For a uniform honest-verifier challenge, this event has probability $1/|\mathbb F|$, contributing at most that much statistical error to this masking argument. Fiat--Shamir security additionally requires the transcript and simulation assumptions of \S\ref{sec:composed-security}. Witness and evaluation hiding require PCS blinding and $\bar V$. Knowledge soundness reduces to Nova folding and PCS openings, using BN254 HyperKZG assumptions here and hiding Dory assumptions for measured native Atlas.\par}

\reviewchange{The Fiat--Shamir protocol additionally needs a simulator in the programmable random-oracle model covering its commitments and openings. Theorem~\ref{thm:zkarc-composition} assumes adaptive multi-theorem zero knowledge; this property remains unproved for the implementation's intended hidden witnesses.\par}

\reviewchange{For fixed graph and tensor shapes, the random witness $Z_2$ and cross-term $T$ have fixed dimensions, and the decider adds one Spartan invocation. Thus the randomization work does not depend on which entries are private. Native measurements do not isolate hiding overhead or cover private model parameters. The graph and tensor shapes remain public; hiding architecture is outside scope.\par}

\section{Efficient Recursion: Subprotocols}
\label{app:efficient-recursion}
\label{app:ec-piops}

\reviewchange{Dao et al.'s~\cite{joltrecursion2026} three subprotocols certify Dory's curve operations over Grumpkin. Sum-check and zero-check soundness follows Schwartz--Zippel, with error linear in rounds and per-round degree over the $254$-bit Grumpkin scalar field $\mathbb F_q$. Independent wiring weights $\lambda_e$ and a union bound give the combined error used in \S\ref{sec:verifier-split}.\par}

\subsection{PIOPs for Elliptic-Curve Arithmetic}

The Dory verifier performs scalar multiplications in $G_1$ and $G_2$ and exponentiations in $G_T$. Each operation type is expressed as a polynomial constraint system and proved via a specialised sum-check PIOP.

\textit{$G_T$ exponentiation.} An element $g \in G_T$ is a $4$-variate MLE over $\mathbb{F}_q$ (12 coefficients padded to 16). To prove $h = g^\alpha$, the canonical integer representative of $\alpha \in \mathbb{F}_r$ is base-$b$ decomposed with most significant digit first as $\alpha = \sum_{i=0}^{n-1} d_i b^{n-1-i}$, with $0\le d_i<b$. The recurrence $\rho_{s+1} = \rho_s^b \cdot g^{d_s}$ for $0\le s<n$ ($\rho_0 = 1_{G_T}$, $\rho_n = h$) is encoded as a zero-check, with $\rho_{s+1}$ related to $\rho_s$ by a shift sum-check using EqPlusOne and a virtual (uncommitted) digit selector $\tilde{b}$. All $G_T$ exponentiations (up to 256 for large traces) batch into one zero-check over a stacked constraint index $c \in \{0,1\}^k$:
\[
\sum_{c \in \{0,1\}^k} \widetilde{\mathsf{eq}}(r_c, c) \sum_{x \in \{0,1\}^\ell} \widetilde{\mathsf{eq}}(r_x, x) \cdot \tilde{C}(x, c) = 0.
\]

\textit{$G_1$ and $G_2$ scalar multiplication.} For $Q = [\alpha]P$ with $P \in G_1$ (or $G_2$), the double-and-add chain $A_{i+1} = [2]A_i + b_i P$ is encoded in $m = \lceil \log_2 n \rceil$ variables over $n = \lceil \log_2 |\mathbb{F}_r| \rceil$ steps, committing accumulator and doubled-point coordinates and indicators. Constraints cover denominator-free doubling, conditional addition (branching on $\tilde{b}$ and infinity indicators), and boundary checks ($A_0 = \mathcal{O}$, $A_n = Q$). For $G_2$, coordinates lie in $\mathbb{F}_{q^2} = \mathbb{F}_q[u]/(u^2+1)$, so each coordinate MLE splits into two $\mathbb{F}_q$ polynomials, doubling the constraint count.

\subsection{Copy Constraints via Sum-Check}

\reviewchange{Wiring checks require each EC operation's output to equal the next operation's input. A port with $d_f$ base-field coordinates is compressed as $V_f=\sum_{j=0}^{d_f-1}\mu^jV_{f,j}$ using a fresh transcript challenge $\mu$. Unequal tuples collide with probability at most $(d_f-1)/|\mathbb F_q|$.\par}

\reviewchange{Each edge $e$ links source and destination ports at specified indices. For $f\in\{\mathsf{src},\mathsf{dst}\}$, let $k_{f(e)}$ be the index dimension of the endpoint's family and $\mathsf{idx}_{f(e)}\in\{0,1\}^{k_{f(e)}}$ its local index. Pad to a common dimension $k$. The coordinate projection $\mathsf{pr}_{f(e)}:\mathbb F_q^k\to\mathbb F_q^{k_{f(e)}}$ deletes the $k-k_{f(e)}$ dummy coordinates. Following Dao et al.~\cite{joltrecursion2026}, set $\beta_{f(e)}=2^{-(k-k_{f(e)})}$ and define\par}
\[
\reviewchange{W_e^f(x,c):=
\beta_{f(e)}\widetilde{\mathsf{eq}}\!\left(\mathsf{idx}_{f(e)},\mathsf{pr}_{f(e)}(c)\right)V_{f(e)}(x).}
\]
\reviewchange{Each local index has $2^{k-k_{f(e)}}$ preimages, so the normalized selector sums to one and $\sum_{c\in\{0,1\}^k}W_e^f(x,c)=V_{f(e)}(x)$.\par}
A single wiring sum-check per group proves all $\sim$1{,}000--2{,}000 edges:
\[
D(x,c):=\sum_{e\in E}\lambda_e
\bigl(W_e^{\mathsf{src}}(x,c)-W_e^{\mathsf{dst}}(x,c)\bigr),\qquad
\sum_{x\in\{0,1\}^\ell}\widetilde{\mathsf{eq}}(r_x,x)
\sum_{c\in\{0,1\}^k}D(x,c)=0,
\]
where \reviewchange{$\lambda_e$ are independent random edge-batching challenges sampled after the port polynomials are fixed}. Because $|E|$ is small, the verifier pays the $O(|E|)$ cost directly, avoiding the need to commit an accumulator polynomial.

\subsection{Prefix Packing}

\reviewchange{Dao et al.~\cite{joltrecursion2026} fit witnesses within $2^{21}$ coefficients over their analyzed trace range, not for arbitrary traces. A balanced Hyrax opening uses two $O(\sqrt{N})$ MSMs for $N$ coefficients, so smaller witnesses reduce opening cost. PIOPs, transcript processing and direct curve checks remain additional.\par}

The global-opening implementation pads each source polynomial to the
largest dimension $\ell_{\max}$ and then assigns equal-length prefixes. With
$p_j$ source polynomials in family $j$, its packed dimension is
\[
n_{\mathsf{pack}}=\ell_{\max}
 +\left\lceil\log_2\!\left(\sum_j p_j\right)\right\rceil.
\]
This can waste space when family dimensions differ. The implementation
does not use variable-length prefixes, so the reference construction's
size bound is not a measured bound for this uniformly padded layout.

\section{Partition of Verifier Operations}
\label{app:verifier-ops}

\reviewchange{The tables partition cryptographic verifier operations. Native sum-checks, claim chaining and Fiat--Shamir derivations remain direct. Auxiliary $\pi'_i$ certifies non-pairing rows using Dao et al.'s EC PIOPs, wiring and packing~\cite{joltrecursion2026}; $\widehat V_i$ directly evaluates the final pairing in $\mathcal C_i^{\mathsf{dir}}$.\par}

\subsection{Dory Verifier (Jolt SMT Solver)}
\label{app:verifier-ops-dory}

\reviewchange{Dory~\cite{lee2021dory} arranges $n$ coefficients in a $2^\nu\times2^\sigma$ matrix, with $\nu+\sigma=\log_2 n$. Reference counts assume a square layout, $\nu=\sigma=(\log_2 n)/2$, and even $\log_2 n$. The artifact reads dimensions from each opening. Dao et al.'s group computation~\cite{joltrecursion2026} has three phases.\par}

\textit{Phase 1: VMV initialisation.} The verifier reads $(C, D_2, E_1)$ from the proof and computes $E_2 = [v] H_2$, where $v \in \mathbb{F}_r$ is the claimed evaluation and $H_2 \in G_2$ is a setup generator. This contributes one $G_2$ scalar multiplication.

\textit{Phase 2: Reduce-and-fold ($\sigma$ rounds).} The verifier maintains five accumulators $C, D_1, D_2 \in G_T$, $E_1 \in G_1$, $E_2 \in G_2$. Each round draws Fiat-Shamir challenges $(\alpha_i, \beta_i)$ and updates every accumulator. The $G_T$ accumulator $C$ evolves as
\[
C \leftarrow C \cdot \chi_i \cdot D_2^{\beta_i} \cdot D_1^{\beta_i^{-1}} \cdot C_+^{\alpha_i} \cdot C_-^{\alpha_i^{-1}},
\]
where $\chi_i$ is a setup-correction term and $C_\pm$ are the round's left/right halves. Analogous updates apply to $D_1, D_2, E_1, E_2$.

\textit{Phase 3: Final check.} A single size-3 multi-pairing combines the post-fold accumulators with the final prover messages, the deferred VMV relation $D_2 \stackrel{?}{=} e(E_1, H_2)$ folded in by a $d^2$-weighted random linear combination so one Miller loop and one final exponentiation suffice.

\textit{Operation counts.} Table~\ref{tab:dory-verifier-ops} reproduces the counts from~\cite{joltrecursion2026} as a function of $\sigma$, with examples indexed by committed polynomial size, not CPU trace length.

\begin{table}[tbp]
\centering
\small
\setlength{\tabcolsep}{8pt}
\newcommand{\grp}[1]{\multicolumn{4}{@{}l}{\footnotesize\textbf{#1}}\\[1pt]}
\begin{tabular}{@{}l ccc@{}}
\toprule
& \multicolumn{3}{c}{\textbf{Operation count}} \\
\cmidrule(l){2-4}
\textbf{Group operation} & Symbolic & $n=2^{18}$, $\sigma=9$ & $n=2^{22}$, $\sigma=11$ \\
\midrule
\grp{Certified by $\pi'_{\mathsf{SMTsolver}}$}
\quad $G_1$ scalar multiplication & $3\sigma + 4$ & $31$ & $37$ \\
\quad $G_1$ addition & $3\sigma + 3$ & $30$ & $36$ \\
\quad $G_2$ scalar multiplication & $3\sigma + 4$ & $31$ & $37$ \\
\quad $G_2$ addition & $3\sigma + 2$ & $29$ & $35$ \\
\quad $G_T$ exponentiation & $10\sigma + 4$ & $94$ & $114$ \\
\quad $G_T$ multiplication & $11\sigma + 5$ & $104$ & $126$ \\
\addlinespace[3pt]
\grp{Checked directly}
\quad Multi-pairing & $1$ (size-$3$) & $1$ & $1$ \\
\bottomrule
\end{tabular}
\caption{Dory verifier operations under the direct-pairing split, as a
function of the reduce-and-fold round count $\sigma=(\log_2 n)/2$. The
examples assume a square commitment layout and do not identify $n$ with the
number of CPU cycles. The artifact's zero-knowledge opening adds checks, so
these reference counts are not its measured operation counts.}
\label{tab:dory-verifier-ops}
\end{table}

\reviewchange{\textit{What $\pi'_{\mathsf{SMTsolver}}$ proves.} Each non-pairing table row groups operations into one zero-check. The EC PIOPs (Appendix~\ref{app:ec-piops}) use two base-field limbs per $G_2$ coordinate in $\mathbb F_{q^2}$. Wiring links the groups. Their witnesses share one packed Hyrax-over-Grumpkin commitment, opened by two MSMs. The size-$3$ pairing remains direct.\par}

\subsection{HyperKZG Verifier (Jolt Atlas Inference)}
\label{app:verifier-ops-hyperkzg}

\begin{table}[tbp]
\centering
\small
\setlength{\tabcolsep}{8pt}
\newcommand{\grp}[1]{\multicolumn{4}{@{}l}{\footnotesize\textbf{#1}}\\[1pt]}
\begin{tabular}{@{}l l cc@{}}
\toprule
& & \multicolumn{2}{c}{\textbf{Operation count}} \\
\cmidrule(l){3-4}
\textbf{Phase} & \textbf{Group operation} & Symbolic & $C=120$, $\ell=10$ \\
\midrule
\grp{Certified by $\pi'_{\mathsf{Atlas}}$}
\quad Joint commit fold & $G_1$ scalar multiplication & $C - 1$ & $119$ \\
 & $G_1$ addition & $C - 1$ & $119$ \\
\addlinespace[2pt]
\quad KZG batch (MSM) & $G_1$ scalar multiplication & $\ell + 4$ & $14$ \\
 & $G_1$ addition & $\ell + 3$ & $13$ \\
\addlinespace[2pt]
\quad KZG batch ($R$) & $G_1$ scalar multiplication & $2$ & $2$ \\
 & $G_1$ addition & $2$ & $2$ \\
\addlinespace[3pt]
\grp{Checked directly}
\quad Multi-pairing & n/a & $1$ (size-$2$) & $1$ \\
\bottomrule
\end{tabular}
\caption{HyperKZG verifier operations under the direct-pairing split,
parameterised by the number of committed polynomials $C$ and joint dimension
$\ell=\max_i n_i$. The numerical column illustrates the stated dimensions. It
is not the measured Qwen trace.}
\label{tab:hyperkzg-verifier-ops}
\end{table}

\reviewchange{HyperKZG needs only $G_1$ linear combinations before a size-$2$ pairing, without Dory's $G_T$/$G_2$ accumulators or $\sigma$-round loop. Atlas commits sparse one-hot polynomials per ONNX operator, rather than over the whole trace (\S\ref{sec:proof-composition}).\par}

\subsubsection{Verifier algorithm (per Jolt Atlas proof)} Given the prover's commitments $\{C_0, \ldots, C_{C-1}\} \subset G_1$ (one per committed polynomial), the verifier: (i) forms the joint commitment $\bar C = \sum_i \gamma^i C_i$, a size-$C$ $G_1$ MSM with transcript scalar $\gamma$, (ii) reduces the $C$ evaluation claims to a single claim on $\bar C$ at a shared point $r \in \mathbb{F}_r^{\ell}$ ($\ell = \max_i n_i$) by an information-theoretic sum-check (not part of $\pi'$), and (iii) runs the HyperKZG inner check, which after $\ell$ native field folding-consistency equations reduces to one size-$(\ell + 4)$ $G_1$ MSM $L$ (over $\bar C$, the $\ell - 1$ round commitments, the three quotient commitments $W_0, W_1, W_2$, and the generator $[1]_1$), a size-$3$ combination $R = W_0 + d_0 W_1 + d_0^2 W_2$ with $d_0$ a transcript batching challenge (not an evaluation point), and a single size-$2$ multi-pairing $e(L, [1]_2)\, e(-R, [\tau]_2) = 1$.

\reviewchange{\textit{Operation counts and $\pi'_{\mathsf{Atlas}}$.} Table~\ref{tab:hyperkzg-verifier-ops} counts operations in $C$ and $\ell$, with fixed-SRS-base work treated as preprocessing. The target certifies $O(C+\ell)$ scalar multiplications and additions, wiring their outputs through each MSM, and checks the pairing directly. The implemented bundle omits the joint-commitment fold (\S\ref{sec:bench-atlas-pi-prime}), so its packing dimensions do not bound the target. Dimensions depend on operation counts, coordinates and padding (Appendix~\ref{app:ec-piops}). No model-independent small-language-model bound is established.\par}

\begingroup
\color{black}
\section{\reviewchange{Hash and Commitment Definitions}}
\label{app:protocol-encodings}
\label{sec:protocol-objects}

\reviewchange{These definitions expand $\langle\cdot\rangle$ in Figures~\ref{fig:protocol-setup}--\ref{fig:protocol-action}. The receiver checks them against the received action and authenticated context (\S\ref{sec:deployment-acceptance}).\par}

\reviewchange{Let $H$ be collision-resistant and $\mathsf{Com}(m;r)$ binding and hiding for byte strings, with private randomness $r$. The \reviewchange{configuration approved by the verifier} fixes parameters, canonical byte encodings and distinct usage tags. The injective tuple encoding $\mathsf{EncodeTuple}$ includes versions and types. A structured action $X$ contains a tool identifier and typed arguments under schema $\Sigma$. Its encoding $\mathsf{enc}_{\Sigma}(X)$ rejects missing, duplicate, ill-typed, unsupported and noncanonical fields. \reviewchange{State supplied by the receiver} $S$ has encoding $\mathsf{enc}_S$ and hash\par}
\[
d_S=H\!\left(\mathsf{EncodeTuple}(\mathsf{tag}_S,\mathsf{enc}_S(S))\right).
\]
\reviewchange{The spending extraction proof takes public $S$, containing prior spending and the approved-vendor registry, authenticated independently of the operator. With no external state, use fixed empty $S$. The receiver's $\mathsf{ctx}_X$ encodes application domain, counterparty, $\mathsf{sid}$, schema identifier, policy context, authorized policy commitment and $d_S$. It computes\par}
\[
d_X=H\!\left(\mathsf{EncodeTuple}(\mathsf{tag}_X,\mathsf{ctx}_X,\mathsf{enc}_{\Sigma}(X))\right).
\]
Thus a receipt for one application, recipient, session, schema, or action cannot be reinterpreted as a receipt for another.

\reviewchange{The receiver authenticates an expected commitment $c_Q^{\mathsf{exp}}$ to the session's full planner input $Q$. With encoding $\mathsf{enc}_Q$, define\par}
\[
c_Q=\mathsf{Com}\!\left(\mathsf{EncodeTuple}(\mathsf{tag}_Q,\mathsf{ctx}_X,\mathsf{enc}_Q(Q));r_Q\right).
\]
\reviewchange{Fresh independent $r_Q$ blinds the commitment. Provenance proves its opening to the planner's exact $Q$, including the specified text-to-token conversion. Acceptance requires $c_Q=c_Q^{\mathsf{exp}}$ using the receiver's independently authenticated value, never an expectation supplied by the receipt.\par}

\reviewchange{The cached-policy context $\mathsf{ctx}_F$ encodes owner, policy identifier, version and epoch. Its stable authorized commitment is\par}
\[
c_F^{\mathsf{root}}=\mathsf{Com}\!\left(\mathsf{EncodeTuple}(\mathsf{tag}_F,\mathsf{ctx}_F,\mathsf{enc}(F));r_F\right).
\]
\reviewchange{Omitting $\mathsf{sid}$ lets $\pi_{\mathsf{PMC}}$ be cached. A receipt may use a freshly randomized policy commitment with a proof that it binds the same $F$. This hides the online value but not the registration, still identified by the public stable commitment and cached proof. Per-action claims include the receipt context. Verifier policy authenticates all tuple encodings and relation, schema and circuit versions.\par}

\reviewchange{For a hidden value $z_e$ shared by two proofs, the abbreviated encoding is}
\[
\reviewchange{\langle\mathsf{ctx}_e,z_e\rangle_e=
\mathsf{EncodeTuple}(\mathsf{tag}_e,\mathsf{ctx}_e,\mathsf{enc}_e(z_e)).}
\]
\reviewchange{Both component relations open the shared commitment, using $c_F^{\mathsf{root}}$
for the cached policy. Any fresh policy commitment is also opened inside the SMT
relation to the same $F$ (Definition~\ref{def:zkarc-relation}). Per-action contexts
include $\mathsf{sid}$. The cached-policy context contains the owner, policy
identifier, version and epoch.}

\endgroup

\end{document}